\documentclass[traditabstract]{aa}

\usepackage{caption}
\usepackage{subcaption}
\usepackage{soul}
\usepackage{aecompl}

\usepackage{array}
\usepackage{amsmath}
\usepackage{graphicx}
\usepackage{amssymb}

\usepackage[round]{natbib}

\usepackage{calc}

\usepackage[colorinlistoftodos]{todonotes}

\usepackage{graphicx,bm}
\usepackage[colorlinks,citecolor=blue]{hyperref}
\usepackage{placeins}

\usepackage{xcolor}

\providecommand{\aap}{Astron.\ Astrophys.}
\providecommand{\apjs}{Astrophys.\ J.\ Suppl.}
\providecommand{\apjl}{Astrophys.\ J.}

\begin{document}

\title{Observational probes of the neutron star equation of state with hyperons, bosonic dark matter, and quark matter}

\author{
Mahboubeh~Shahrbaf\inst{1,2,3}\thanks{{\em email:} mahboubeh.shahrbafmotlagh@uwr.edu.pl}
\and Davood~Rafiei Karkevandi\inst{1}\thanks{{\em email:} Davood.Rafiei64@gmail.com}
\and Alexander~Ayriyan\inst{1,4}\thanks{{\em email:} alexander.ayriyan@uwr.edu.pl}
\and Stefan~Typel\inst{5,6}
}

\institute{
Institute for Theoretical Physics, University of Wroc{\l}aw, Plac Maksa Borna 9, 50-204 Wroc{\l}aw, Poland
\and Incubator of Scientific Excellence, Center for Simulations of Superdense Fluids, University of Wroc{\l}aw, Plac Maksa Borna 9, 50-204 Wroc{\l}aw, Poland
\and Frankfurt Institute for Advanced Studies, Giersch Science Center, D-60438 Frankfurt am Main, Germany
\and Alikhanyan National Science Laboratory (Yerevan Institute of Physics), Alikhanyan Brothers St 2, 0036 Yerevan, Armenia
\and Technische Universit\"{a}t Darmstadt, Fachbereich Physik, Institut f\"{u}r Kernphysik, Schlossgartenstra\ss{}e 9, D-64289 Darmstadt, Germany
\and GSI Helmholtzzentrum f\"{u}r Schwerionenforschung GmbH,
Theorie, Planckstra\ss{}e 1, D-64291 Darmstadt, Germany}

\authorrunning{M.~Shahrbaf et al.}

\date{Received date / Accepted date }

\abstract{

\textbf{Context.} The presence of dark matter in neutron stars is of growing interest due to its potential impact on the structure and observable properties of these objects. Among the various candidates, the hypothetical sexaquark has emerged as a promising bosonic dark matter particle, potentially forming under extreme conditions in neutron star cores.

\textbf{Aims.} We investigate whether a hybrid neutron star model that includes hyperons, bosonic dark matter (in the form of sexaquarks), and deconfined quark matter can satisfy all current observational constraints. We particularly focus on identifying the range of sexaquark masses consistent with mass-radius measurements and the tidal deformability limit.

\textbf{Methods.} We used the DD2Y-T model for the hadronic phase, which includes hyperons, and a nonlocal Nambu–Jona-Lasinio model for the deconfined quark phase. The phase transition was modeled as a smooth crossover using the replacement interpolation construction method. Sexaquark-baryon interactions were introduced via an effective mass shift representing repulsion. We incorporated the full set of current observational data, including NICER measurements of PSRs J0437–4715 and newly published J0614–3329 data, and performed a Bayesian analysis to constrain the sexaquark mass.

\textbf{Results.} Our results show that the presence of the sexaquark softens the equation of state, enabling the hybrid model to satisfy both the radius and tidal deformability constraints around the canonical $1.4\,M_\odot$ neutron stars. We find that hybrid EOSs with a sexaquark mass around $1900$ MeV are in agreement with all available constraints, including those from HESS J1731–347 and PSR J0952–0607, which represent the lowest and highest mass neutron stars observed to date. The Bayesian analysis favors a sexaquark mass range of $1885$–$1935$ MeV, supporting the potential relevance of this exotic particle in neutron star interiors.

}


\maketitle

\section{Introduction}
\label{sec:intro}
 
The accepted cosmological models that are presented and founded in general gelativity (GR) suggest that almost 27\% of the matter-energy composition of the Universe exists in the form of enigmatic dark matter \citep[DM;][]{Bertone:2016nfn,deMartino:2020gfi}. 
Multi-messenger astrophysics is actively engaged in the search for and examination of new ideas concerning DM particles. These candidates are believed to be non-relativistic particles that exist beyond the Standard Model of particle physics. They interact weakly with ordinary baryonic matter (BM) solely through gravitational forces \citep{Salucci:2020nlp, Arbey:2021gdg}. In this context,
weakly interacting massive particles are one of the notable DM candidates \citep{Bertone:2018krk, Arcadi:2017kky}. Moreover, hypothetical bosonic particles such as the axion \citep{Marsh:2015xka, Chadha-Day:2021szb} and the sexaquark (S; \citep{Farrar:2022mih, Doser:2023gls, Moore:2024mot}) also exhibit the primary characteristics required to be considered potential DM candidates. Their interaction with the BM must be very weak, which is why they have remained undetected thus far. The mass of candidate DM particles spans a wide range, from approximately $10^{-22}$ eV to $10^{10}$ eV, and is based on the particular type of candidate \citep{cirelli2024darkmatter}. Moreover, their self-interaction is another key parameter of the DM model. Indeed, self-interacting DM has the potential to address several issues in cosmology and astrophysics, including "core-cusp," "missing satellite," and "too big to fail" problems \citep{Spergel:1999mh, Peter:2012jh, Kaplinghat:2015aga}. Given these model parameters, DM particles must be incorporated into an equation of state (EOS), which is a fundamental tool for characterizing matter. 

It is believed that gravitationally stable astrophysical objects composed of DM could form in the Universe \citep{ Liebling:2012fv, Baryakhtar:2022hbu, Visinelli:2021uve}. In this context, they can exist as an object made entirely of DM, so-called dark stars \citep{Narain:2006kx, Maselli:2017vfi}. Such objects include axion stars, boson stars \citep{Eby:2015hsq, Visinelli:2017ooc, Pitz:2023ejc}, and stars that have a fraction of DM mixed with BM~\citep{Bramante:2023djs}. 
For the last type, compact objects such as black holes \citep{Shakeri:2022usk}, neutron stars (NSs; \citep{Karkevandi:2021ygv, Shakeri:2022dwg}), white dwarfs \citep{Ryan:2022hku}, and even hypothetical quark (strange) stars \citep{Ferreira:2022fjo,Yang:2024sxi,Sedaghat:2025dzt,Liu:2025vwm} offer a significant opportunity to probe DM indirectly.  These objects are considered promising astrophysical environments where DM may reside, either within them or in a surrounding halo.

Neutron stars, due to their extreme gravitational potential and high baryon density, have the potential to
accumulate DM particles and are widely regarded as astrophysical laboratories for studying DM  \citep{Grippa:2024ach}. In recent years, motivated by new and significant observational data on NS properties, numerous studies have investigated the possible presence of DM within these objects \citep{Ivanytskyi:2019wxd,RafieiKarkevandi:2021hcc, Dengler:2021qcq, Rutherford:2022xeb, Karkevandi:2024vov, Thakur:2023aqm, Konstantinou:2024ynd,Routaray:2024fcq,Biesdorf:2024dor,Arvikar:2025hej}. The influence of DM on observable features of NSs has been explored in order to place constraints on the parameter space of DM models \citep{Sedrakian:2015krq, Sedrakian:2018kdm, Shakeri:2022dwg, Giangrandi:2022wht, Diedrichs:2023trk, Avila:2023rzj,Issifu:2024htq,Cipriani:2025tga,Scordino:2024ehe,Mukherjee:2025omu}. Furthermore, unusual measurements of NS mass and radius have prompted attempts to explain them in terms of mixed configurations, hereafter referred to as DM admixed NSs. In this direction, HESS J1731-347 \citep{Sagun:2023rzp,Marzola:2024ame} and XTE J1814-338 ultracompact stars \citep{Pitz:2024xvh, Yang:2024ycl, Lopes:2024ixl} as well as very massive compact objects, such as the secondary ones detected in GW190814 and GW230529 events, are noteworthy cases \citep{Barbat:2024yvi,Vikiaris:2023vau}. 

Concerning the formation of DM admixed NSs, several scenarios have been considered, each leading to various DM fractions within the star \citep{Karkevandi:2021ygv}. Among these, DM production inside NSs or during supernova explosions is a notable mechanism \citep{Ellis:2018bkr, Nelson:2018xtr}.
An additional neutron decay channel into DM, potentially addressing the neutron decay anomaly, could result in the production of a significant fraction of fermionic DM inside NSs \citep{ Shirke:2023ktu, Shirke:2024ymc, Thakur:2024btu,Das:2025pjl}. Furthermore, as a novel concept, the production of S, a deeply bound bosonic DM candidate, offers a promising mechanism for forming these mixed objects \citep{Shahrbaf:2022upc, Blaschke:2022knl, Shahrbaf:2023uxy}. This mechanism is the focus of this paper.

Apart from the formation of these objects, modeling and investigation of DM admixed NSs are done based on the fact that DM and BM interact only through gravitational force or that there is a non-gravitational interaction between them. For the former, two-fluid Tolman-Oppenheimer-Volkoff equations describe the mixed object \citep{RafieiKarkevandi:2021hcc, Collier:2022cpr,Karkevandi:2024vov, Rutherford:2024bli, Shawqi:2024jmk,Dengler:2025ntz,Liu:2025cwy}. Depending on the DM parameters, it can reside as a core inside an NS, or it may form an extended halo around the NS. For more detail on this first approach, we refer to the works done by \citet{Karkevandi:2021ygv, Shakeri:2022dwg} and the references therein. However, when a non-gravitational interaction is considered between BM and DM, single fluid Tolman-Oppenheimer-Volkoff equations are employed to describe the hydrostatic equilibrium of the object \citep{Lenzi:2022ypb,Shahrbaf:2023uxy, Guha:2024pnn,Hajkarim:2024ecp, Kumar:2025ytm,Klangburam:2025rcb,Das:2025duq}. Therefore, DM production inside NSs, such as S, is investigated using the single-fluid approach, due to the non-gravitational interaction between DM particles and the baryonic medium \citep{Shahrbaf:2022upc}. In this paper, we demonstrate how the presence of S affects the EOS, leading to changes in the observable properties of DM admixed NSs.

It is worth mentioning that the EOS for strongly interacting
matter in NSs is tightly bound by observational constraints. Therefore, efficient boundaries on the features of DM candidates have been established using recent advancements in observational astronomy, 
 mainly facilitated by gravitational wave (GW) interferometers \citep{LIGOScientific:2018cki, Dietrich:2020efo,Pang:2022rzc,Ecker:2024uqv} and the Neutron Star Interior Composition ExploreR (NICER; \citep{Miller:2019cac, Miller:2021qha, Riley:2019yda, Riley:2021pdl,Vinciguerra:2023qxq,Salmi:2024aum, Salmi:2024bss,dittmann:2024:10215108,Choudhury:2024xbk,Mauviard:2025dmd}).  In this study, mass, radius, and tidal deformability serve as the fundamental constraints used to investigate these DM admixed objects.
 Thus, we focus on high-precision mass–radius (M-R) measurements obtained through X-ray pulse profile modeling by the NICER telescope. We include the recent result for PSR J0437-4715, which reports a mass of $M=1.418 \pm 0.037 M_{\odot}$ and a radius of $R=11.36^{+0.95}_{-0.63}$ km \cite{Choudhury:2024xbk}. Importantly, we also incorporate the latest NICER measurement of PSR J0614-3329, which infers a radius of $R=10.29^{+1.01}_{-0.86}$ km for a mass of $M=1.44^{+0.06}_{-0.07} M_{\odot}$ \citep{Mauviard:2025dmd}.
 These new NICER results are important, as they support a softer EOS and add valuable boundary conditions for constraining models of dense matter in NSs.

Moreover, the detection of GW signals arising from binary NS mergers, along with the subsequent determination of the tidal deformability parameter, have opened a new window for investigating the exotic internal structures of NSs \citep{LIGOScientific:2017vwq, LIGOScientific:2020aai}. Therefore, the tidal deformability constraint from the GW170817 event \citep{LIGOScientific:2018cki}, $\Lambda_{1.4}=190^{+390}_{-120}$, is also considered to obtain the allowed parameter space of the S bosonic DM model. 

The concept of a deeply bound S, an alternative term for the hexaquark family of hypothetical particles, was first introduced as a bosonic candidate for DM in references \citet{Farrar:2002ic, Farrar:2018hac, Farrar:2022mih, Doser:2023gls}. The quark composition of S is identical to that of the H-dibaryon \citep{Jaffe:1976yi}, with both consisting of uuddss quarks. However, S exhibits significantly stronger binding compared to the H-dibaryon. If S represents a molecular state such as $\Lambda\Lambda$, the binding of two $\Lambda$ particles can exclusively occur through the exchange of color-neutral entities, such as mesons. However, when S comprises a complex system consisting of three colored diquarks, these entities engage in interactions primarily governed by the color force. This force is notably stronger than the meson exchange force, particularly when considering short-range distances.

{It is important to note that the $S$ particle is a member of the wider family of hexaquark states. 
The first nontrivial hexaquark to be experimentally established was the $d^*(2380)$ dibaryon, 
discovered by the WASA-at-COSY Collaboration \citep{WASA-at-COSY:2011bjg}. 
This state has been investigated as an exotic degree of freedom in NSs. 
It has been shown that the $d^*(2380)$ could appear at baryon densities three to five times the saturation density, 
leading to a reduction in the maximum NS mass \citep{Vidana:2017qey, Celi:2023gtj}. 
More recent work \citet{Celi:2025wnc} has demonstrated that the inclusion of the $d^*(2380)$ softens the hadronic EOS 
and consequently delays the onset of deconfinement in a first-order phase transition. 
These studies highlight that an appropriate reparametrization of the hadronic models is required 
to remain consistent with astrophysical constraints when exotic dibaryons are included.
}

The proposed S is a boson at a spin-color-flavor singlet and even parity state, and it is characterized by a zero electric charge, a baryon number of two, and a strangeness value of -2. The potential of the 
S particle to serve as DM depends on several parameters, such as its mass and its stability against dissociation into two baryons, which requires a small dissociation amplitude. Other important factors include its elastic scattering cross section with nucleons and its annihilation with antibaryons (or antisexaquarks) into baryons or other Standard Model final states. S may manifest itself as a deeply bound state, displaying a mass sufficiently low to either be absolutely or effectively stable against decay. Our main focus in this study is the mass of the S. Notably, all measurements show that its mass is less than 2230 MeV, the threshold for producing two $\Lambda$ baryons. This suggests that the S may be a bound state \citep{Jaffe:1976yi, Kodama:1994np, Azizi:2019xla, Gross:2018ivp, Shahrbaf:2022upc, Evans:2023zde}.

However, Lattice QCD calculations suggest that H-dibaryon is an unbound particle \citep{Inoue:2010es, NPLQCD:2010ocs, Green:2021qol, Shanahan:2011su}, but \citet{Farrar:2022mih} assumes S as a deeply bound state that is difficult to detect in lattice calculations, making it different from the loosely bound H-dibaryon. For S to be a viable DM candidate, both its possible decay channels and any Standard Model particle decays involving the S must either be kinematically forbidden or highly suppressed.
If the S particle were to decay, the energetically most favorable channel would produce two protons and two electrons, with a total mass of approximately $1878$ MeV. Consequently, the S particle would be absolutely stable if its mass lies below this threshold, i.e., $m_\text{S} < 1878$ MeV. From an experimental point of view, the existence of S with a mass between $2054$ MeV and $2223.7$ MeV has been ruled out \citep{Takahashi:2001nm}. Additionally, \citet{Farrar:2023wta} argues that if the S mass equals the combined mass of a lambda hyperon, a proton, and an electron ($m_\text{S} = 2054$ MeV), its decay into two nucleons would involve a doubly weak interaction, resulting in a lifetime longer than the age of the Universe. The constraints on the mass of the S have been extensively discussed in \citet{Farrar:2023wta} and the references therein. Meanwhile, \citet{Gal:2024nbr} have demonstrated that the observation of double-$\Lambda$ hypernuclei decaying into single-$\Lambda$ hypernuclei does not exclude the existence of a deeply bound S within the mass range of $2m_n$ to $m_\Lambda+m_n$. However, its $\Delta S = 2$ decay lifetime ($\tau(H \rightarrow nn)$) is estimated to be shorter than the timescale required for it to be a viable DM candidate. Moreover, the ratio of densities between S DM and ordinary BM in the Universe, which can be estimated through considerations of statistical equilibrium within the quark-gluon plasma, aligns with observational data \citep{Farrar:2022mih}. Under the above-mentioned conditions, we assume that the S has the potential to be a DM candidate and examine its mass based on observational constraints from NSs. A mechanism for producing such an uuddss state at rest through the formation of antiprotonic atoms has been proposed in \citet{Doser:2023gls}. Ultimately, lattice QCD can be used to study the interactions and decay channels of the S, offering deeper insight into its properties.

Complementary to the S DM hypothesis, heavy-ion collision experiments have demonstrated that hyperons can be abundantly produced in high-density environments. 
These findings are significant for understanding the behavior of dense nuclear matter, 
as they provide empirical evidence supporting the inclusion of hyperons in the EOS of matter under extreme conditions \citep{STAR:2010gyg, Rappold:2015una, ALargeIonColliderExperiment:2021puh, Leung:2022flt, ALICE:2022sco}. Therefore, including hyperons in NS EOS 
models is essential for a comprehensive understanding of their structure, 
stability, and evolution, aligning theoretical predictions with observational data. Moreover, given that our DM candidate carries double strangeness, it is natural to include hyperons in the hadronic phase as well.
In this study, the S production is established using a generalized relativistic density functional approach known as DD2Y-T
\citep{Stone:2019blq}. This hadronic model contains all $J=1/2$ octet baryons, namely, the neutron, proton, $\Lambda$, $\Sigma$, and $\Xi$ hyperons, giving a more reliable description of the high-density matter inside NSs. The EOS for the entire NS, including both the core and the crust, is constructed consistently using the DD2 crust model as in \citet{Typel:2024myq}, which ensures a smooth and thermodynamically consistent matching between the outer layers and the dense core. In addition to the strongly interacting baryonic components, the EOS must also account for leptonic degrees of freedom - specifically, electrons and muons - due to the requirements of charge neutrality and beta equilibrium in NS matter. These leptons are treated as non-interacting, relativistic Fermi gases and play a crucial role in balancing the electric charge introduced by the presence of protons and charged hyperons.

Recent studies on NS and the nuclear matter properties, as well as thermodynamic stability, have highlighted a rapid stiffening of quantum chromodynamics (QCD) matter above saturation density. This stiffening is observed to be faster compared to purely hadronic models as the density increases \citep{Kojo:2024ejq}. Although a clear signature of deconfined quark matter (QM) within a realistic EOS is still to be confirmed, it has been shown that post-merger GWs may distinguish between scenarios with and without deconfinement \citep{Fujimoto:2022xhv}. Moreover, various studies suggest that GW signals, along with signals from supernova explosions, NS mergers, and accreting NSs, could be impacted by a phase transition to QM \citep{Bauswein:2018bma, Most:2018eaw, Weih:2019xvw, Bauswein:2022vtq, Largani:2023kjx}.

Furthermore, from a phenomenological point of view, a peak in the square of the speed of sound in QCD matter has been observed, which is associated with the percolation threshold and the phase transition to deconfined QM \citep{Satz:1998kg, Magas:2003wi, Castorina:2008vu, Braun-Munzinger:2014lba, Fukushima:2020cmk, Marczenko:2022jhl}. These findings support the idea that the core of NSs may consist not only of hadronic matter but also deconfined QM.
As a result, it is highly possible to encounter a phase transition to QM in the core of NS \citep{Li:2023zty, Gholami:2024ety, Li:2024sft, Lindblom:2025wme}. Therefore, our model is connected to an advanced QM model named the nonlocal Nambu-Jona-Lasinio (nlNJL) model, which is based on the crossover phase transition through a novel method. 

Consequently, we modeled hybrid stars with a dense core of deconfined QM surrounded by layers of bosonic S DM, hypernuclear matter, and ordinary nuclear matter. Our aim is to examine whether a unified EOS incorporating these exotic components can satisfy current multi-messenger constraints on NSs {while also placing constraints on the upper bound of the bosonic DM mass from astrophysical observations.}

In our prior work, \citet{Shahrbaf:2022upc}, we established the lower-mass limit for S, ensuring that the possible
emergence of S as a DM candidate remains consistent with
observational constraints derived from NSs. Our findings disclosed the feasibility of S appearance in the hadronic phase of NS, demonstrating that a minimum mass of $m_\texttt{S} = 1885$ MeV fulfills the necessary conditions that S does not appear at sub-saturation densities. In the present work,
we adopt the same approach to ascertain the upper limit for the mass of S, assuming that the interaction strength with BM is fixed at its lowest value. This provides a reliable assumption about the DM-ordinary matter interaction.

Furthermore, we performed a Bayesian analysis (BA) to systematically determine the posterior probability distribution of the allowed mass range of S, accounting for both observational uncertainties and model assumptions. By applying multiple constraints, the Bayesian framework provides more precise limits on the allowed bosonic DM parameter space within the context of NS physics. For our analysis, we considered fundamental constraints on NSs, mass, radius, and tidal deformability along with two extreme cases: HESS J1731-347 \citep{Doroshenko:2022nwp}, a light and ultracompact NS, and PSR J0952–0607~\citep{Romani:2022jhd}, which is a black widow (BW) pulsar and the most massive confirmed NS. The constraints are first applied independently to infer the respective posterior distributions of the S mass. Subsequently, these are combined to yield the final posterior distribution.

The paper is organized as follows: 
In Section~\ref{sec:EOS}, we introduce our model to describe the hadronic phase with hyperons and DM as well as the QM phase. Our approach for combining these two phases and constructing a crossover phase transition is also discussed in this section. Our findings and in-depth discussions concerning the thermodynamic properties of the resulting hybrid stars are presented in Section~\ref{sec:Thermodynamics}.
In Section~\ref{sec:M_R_L}, we investigate the impact of NS constraints to find the upper limit of the mass of DM particles. Our BA approach based on the NS observations is presented in Section~\ref{BA}. Finally, we summarize our work and draw conclusions in Section~\ref{sec:summary}.

\section{
Unified modeling of dense matter: Hyperonic, dark, and quark phases}
\label{sec:EOS}

\subsection{Hadronic matter equation of state including hyperons and dark matter}
\label{sec:setup1}

 In the DD2Y-T model, all particles are treated as quasi-particles with effective mass and effective chemical potential. The interactions in this model exclusively entail meson exchange interactions, and the couplings to $\sigma$, $\omega$, $\rho$, and $\phi$ mesons depend on the density of the medium. The density-dependent couplings are accurately adjusted to faithfully reproduce the characteristics of atomic nuclei, as established by earlier works \citep{Typel:1999yq, Typel:2005ba, Typel:2009sy}.

 The DD2Y-T model has been employed to assess the properties of NSs in both high-temperature (or "hot") and low-temperature (or "cold") hyperonic configurations \citep{Stone:2019blq}. Notably, it is essential to highlight that this model successfully addresses the 'hyperon puzzle', a longstanding problem that has been extensively deliberated in the literature \citep{Sedrakian:2020kbi, Choi:2023qtk, Tolos:2024adu, Fujimoto:2024doc, Tong:2025sui}. It is considered one of the most challenging issues within the domain of hypernuclear physics, and various works have tried to address the solutions to this puzzle \citep{Maslov_2015, Shahrbaf:2019wex, Shahrbaf:2019vtf, Shahrbaf:2020uau, Li:2023vso, Tsiopelas:2024ksy, Jangal:2025maa}.

In the following, we refer to the DD2Y model incorporating S DM as DD2Y-T+S. In this model, the S particle serves as an additional baryonic degree of freedom, featuring a mass shift that depends on the baryonic density, which captures the impact of its interaction with the surrounding medium. In principle, the interaction between the S particle and the surrounding medium can be theoretically elucidated by establishing couplings to mesons, analogous to the approach employed for other baryonic degrees of freedom. However, this requires introducing several meson–sexaquark couplings, the precise values of which remain undetermined.

As established in \citet{Shahrbaf:2022upc}, we have demonstrated that a constant mass assumption for the S particle is incompatible with the structural stability of NS. Consequently, we introduced a linear density-dependent mass shift for the S particle, with a slope parameter harmonized with the corresponding mass shifts observed in other baryons. Detailed constraints regarding the lower threshold of the mass and the slope parameter have been thoughtfully presented and illustrated in Figure 8 of \citet{Shahrbaf:2022upc}. In our approach,
\begin{equation}
  \label{eq:S_S}
  S_{S} =   -\Delta m_{S}
\end{equation}
is the scalar potential for the S particle, and $\Delta m_\text{S}$ is the mass shift of S. The effective mass for S is thus defined as 
\begin{equation}
\label{eq:m_S}
 m^*_{S} = m_{S} - S_{S} = m_{S}+ \Delta m_{S},
\end{equation}
with the vacuum mass $m_{S}$ of S. The mass shift is given by
\begin{equation}
\label{eq:Deltam}
    \Delta m_{S} = m_{S} x_{S} \frac{n_{B}}{n_{0}},
\end{equation}
where $n_B$ is the baryonic density, $x_{S}$ is a slope parameter, and $n_0 = 0.15$~ fm$^{-3}$ is the nuclear saturation density. {For the S particle, we adopted the minimal value of the slope parameter. This choice, while simplified, produces a significant mass shift at high densities of NSs and ensures that the S remains a deeply
bound but weakly interacting state. Such a treatment is consistent with the
expectation that a viable DM candidate should interact only weakly
with BM, and it serves to mimic the possible interactions of the
S with the medium without introducing additional free parameters.}
A finite density of S exclusively emerges at zero temperature when the condition $m^{*}_S = \mu^{*}_S$ is met, leading to the occurrence of Bose-Einstein condensation. The scalar density for S is then equal to the vector density and reads
\begin{eqnarray}
    n^{(s)}_{S} = n^{(v)}_{S} &=& n_{b}-n_{n}^{(v)}-n_{p}^{(v)}
    -n_{\Lambda}^{(v)}-n_{\Sigma^{+}}^{(v)}-n_{\Sigma^{0}}^{(v)}- \nonumber \\ 
    &&
n_{\Sigma^{-}}^{(v)}-n_{\Xi^{0}}^{(v)}-n_{\Xi^{-}}^{(v)}.
\end{eqnarray}
Here, $n_{i}^{(v)}$ signifies the vector density of the fermions.
All relevant information pertaining to the thermodynamic properties of the system can be deduced from the grand canonical thermodynamic potential density, denoted as $\Omega({\mu_{i}})$, wherein $\mu_{i}$ signifies the chemical potential of each particle at zero temperature. The condensate contribution of the bosonic S in grand canonical thermodynamic potential density is formally given by
\begin{equation}
  \Omega_{S} = n^{(s)}_{S}(m_{S}^{\ast}-\mu_{S}^{\ast}),
\end{equation} 
where $\mu_{S}^{\ast} = \mu_S - V_S$ is the effective chemical potential and $\mu_S = 2\mu_B$ is the chemical potential of the S particle. $V_S$ is the vector potential for the S particle and reads
\begin{equation}
  V_S = n_{S}^{(s)} \frac{\partial\Delta m_{S}}{\partial n_{S}^{(v)}}.
\end{equation}
It appears as a rearrangement contribution due to the density dependence of the mass shift and is required for thermodynamic consistency.
For a more comprehensive exploration of the topic, we refer the reader to the extensive discussion presented in \citet{Shahrbaf:2022upc}. 

The pressure ($P$) can be straightforwardly expressed as $P = -\Omega$. Moreover, the free energy density ($f$) coincides with the internal energy density ($\varepsilon$) and can be readily computed using the following relation:
\begin{equation}
f = \varepsilon = \Omega + \sum_{i}\mu_{i}n_{i}^{(v)}.
\end{equation}

\subsection{Quark-matter equation of state}
\label{sec:setup2}
 
One of the best candidates for describing the color superconducting QM phase is a covariant nlNJL model \citep{Radzhabov:2010dd}. This effective model successfully accounts for the key features of low-energy QCD, such as dynamical chiral symmetry breaking, and also reasonably fulfills NS conditions. 

It is worth noting that the precise values of the vector meson coupling ($\eta_{V}$) and the diquark coupling ($\eta_{D}$) remain undetermined in this effective model. In one of our previous works, we determined optimal parameters for those couplings in the nlNJL model \citep{Shahrbaf:2021cjz}. This was achieved through BA, specifically in the context of a first-order phase transition from hadronic matter to deconfined QM within a systematic study of hybrid NS EOS.

{ In this study, the employed QM EOS is the microscopic nlNJL model. 
The coupling constants $\eta_V$ and $\eta_D$ are taken from the BA of the nlNJL model performed in \citet{Shahrbaf:2021cjz}, while the squared speed of sound $c_s^2$, which emerges as a property determined by the chosen parametrizations, is obtained through the established mapping between the nlNJL and CSS parametrizations in the same paper. 
Thus, each quark EOS in our study is characterized by the parameter set $(\eta_D, \eta_V)$. The chosen parameters fall within the range of $0.10 < \eta_V < 0.12$ and $0.70 < \eta_D < 0.74$. For the Bayesian approach in reference \citet{Shahrbaf:2021cjz}, two constraints derived from observations of NSs have been employed. The first is based on the mass measurement of the NS component within the binary system PSR J0740+6620, determined using the relativistic Shapiro time delay effect. The gravitational mass of this NS is estimated to be approximately ${2.08}_{-0.07}^{+0.07}$$M_{\odot}$ with a 68\% credibility interval, as reported in references \cite{Fonseca:2021wxt}. The NS's radius has been inferred through fitting rotating hot spot patterns to data from the NICER and X-ray Multi-Mirror (XMM-Newton) X-ray observations, resulting in an estimated radius of approximately ${13.7}_{-1.5}^{+2.6}$~km at the 68\% confidence level, as reported in reference \citep{Miller:2021qha}. The second constraint utilized in our analysis pertains to the radius of a NS with a mass of 1.4$M_{\odot}$, estimated to be approximately ${11.75}^{+0.86}_{-0.81}$ km at a 90\% confidence level. This constraint was derived through a combined analysis of the GW event GW170817 with its electromagnetic counterparts AT2017gfo and GRB170817A, as well as the GW event GW190425, both originating from NS mergers, as described in reference \citet{Dietrich:2020efo}.

\subsection{Hybrid equation of state}
\label{sec:hybridEOS}

In the context of constructing a phase transition to QM, traditional Maxwell constructions typically result in a first-order phase transition \citep{Christian:2023hez, Huang:2025vfl, Counsell:2025hcv}. However, recent theoretical studies suggest that a continuous transition, or crossover, could be a more accurate description in the QCD framework \citep{Schafer:1998ef, Hirono:2018fjr, Fujimoto:2019sxg, Fujimoto:2024ymt}. Percolation theory, in particular, supports the idea that the transition from hadronic to QM need not be abrupt. Instead, it could occur as a smooth crossover, consistent with lattice QCD results at high temperatures. The geometric and statistical principles of percolation theory provide valuable insight into quark-hadron continuity, suggesting that for baryon densities in the range $2n_0 <n_B < (4-7)~n_0$, the composition of matter should change smoothly from hadronic to QM \citep{Fukushima:2020cmk, Baym:2017whm}.

Recent observations of NSs show similar radii for stars with masses of $2.1$ and $1.4$ solar masses. This finding challenges the onset of a sharp first-order phase transition in the density range slightly above saturation, extending up to the baryon densities found in the cores of two-solar-mass NSs ($n_B \approx 4-7~n_0$) \citep{Kojo:2021hqh}. This observation supports the hypothesis of hadron-quark continuity, where a weak first-order phase transition may still exist but is not dominant, allowing for a more accurate description of the QCD matter EOS inside NSs \citep{Kojo:2020krb}.

It is also worth noting that \citet{Hensh:2024onv} recently analyzed both crossover and strong first-order phase transitions in GW signals. They found that the post-merger main frequency $f_2$ is lower than predicted by hadronic models with the same tidal deformability ($\Lambda$). They concluded that the detection of both inspiral and post-merger signals could provide evidence of the presence of a crossover phase transition \citep{Hensh:2024onv}.

Moreover, the determination of surface tension at the interface between the hadronic phase and quark phase in strongly interacting matter remains a topic of ongoing investigations. An infinite surface tension implies the applicability of a Maxwell construction, whereas a vanishing surface tension necessitates the use of a Glendenning construction. In scenarios where the surface tension is variable, a crossover construction approach can be considered.

{In light of the aforementioned points, we adopt the replacement interpolation construction (RIC) method \citep{Ayriyan:2021prr, Ayriyan:2017nby, Ayriyan:2017tvl, Shahrbaf:2020uau} to model the phase transition. Importantly, two distinct interpolation philosophies exist in the literature. In the earlier works of \citet{Ayriyan:2017nby, Ayriyan:2017tvl}, the interpolated EOS was considered as a thermodynamically meaningful mixed phase, in which the $\Delta_P$ parameter reflects the physical surface tension at the hadron--quark interface. In contrast, in the later formulation of \citet{Ayriyan:2021prr}, the interpolation is introduced purely as a mathematical bridge between two trustworthy regions, namely the hadronic EOS at $\mu_B < \mu_H$ and the quark EOS at $\mu_B > \mu_Q$. In this latter interpretation, which we explicitly follow in the present work, the interpolated EOS does not represent a thermodynamically favored state. Instead, it is a pragmatic construction enabling us to explore the astrophysical consequences of a smooth crossover rather than a sharp first-order transition. It is worth mentioning that both the baryonic-QM and the leptonic sectors are taken into account in the interpolation procedure.

The hadronic EOS is considered appropriate at lower densities, particularly in the vicinity of nuclear saturation density, where the BM does not reveal its mixed structure. We establish the upper limit of applicability for our hadronic EOS as $n_H(\mu_B)$. In regions of relatively high densities, where quarks exist independently of specific baryons, the deconfined QM EOS is invoked. The lower threshold of validity for the color superconducting QM EOS is defined as $n_Q(\mu_B)$. In the intermediate density range, denoted by $n_H(\mu_B) < n(\mu_B) < n_Q(\mu_B)$, neither the hadronic EOS nor the QM EOS is suitable, and the interpolated EOS is applied as a mathematical connection between them.

The EOS for the hadronic and QM phases are expressed in terms of the pressure versus chemical potential, denoted as $P_H(\mu_B)$ and $P_Q(\mu_B)$, respectively, at $T = 0$, which is relevant for NSs. Under these conditions, the effective crossover EOS, $P_M(\mu_B)$, can be conveniently represented as an interpolation function between the two phases. This interpolation requires that the pressure function connects smoothly the values characteristic of hadronic and quark phases at their lower and upper limits. Moreover, it must satisfy fundamental thermodynamic constraints, including a positive slope of density versus chemical potential, i.e., $\frac{\partial n_M}{\partial \mu_B} = \frac{\partial^2 P_M}{\partial \mu_B^2} > 0$. Additionally, the causality condition, which ensures that the adiabatic speed of sound at zero frequency, $c_s^2 = \frac{\partial P}{\partial \epsilon}$, does not exceed the speed of light, must be obeyed too. One effective and practical approach for achieving this interpolation is through the utilization of a polynomial function, which smoothly connects the pressure curves of the two phases.

The critical chemical potential, denoted as $\mu_c$, is defined as the mathematical crossing point between the hadronic and quark EOSs. While in the Maxwell construction, this point has a thermodynamic meaning, in our RIC implementation, it serves only as a reference point for the polynomial interpolation. This crossing is expressed as
\begin{equation}
P_Q(\mu_c) = P_H(\mu_c) = P_c.
\label{eq:mechequi}
\end{equation}
To model the pressure of the crossover phase, we employed a polynomial ansatz:
\begin{equation}
P_M(\mu_B) = \sum_{q=1}^{N}\alpha_q(\mu_B - \mu_c)^q + (1 + \Delta_P)P_c,
\label{eq:RIMpressure}
\end{equation}
with the free parameter, $\Delta_P$, quantifying the offset of the interpolated curve at $\mu_c$:
\begin{equation}
P_M(\mu_c) = P_c + \Delta P = P_M, \text{ where } \Delta_P = \frac{\Delta P}{P_c}.
\label{eq:addpressure}
\end{equation}
In this work, we fixed $\Delta_P = -0.02$ for simplicity. Given the choice of well-reliable EOSs for both the hadronic and quark phases and assuming the existence of a phase transition, but taking into account that these EOSs by themselves do not provide a reasonable Maxwell-type phase transition, the RIC method with negative $\Delta_P$ could be employed. Typically, within the framework of Eq.~\eqref{eq:RIMpressure}, it is common to consider $N=2$.

At the matching points $\mu_{H}$ and $\mu_{Q}$, the pressure of the cross-over phase aligns with the pressure of the hadronic and QM phases, respectively. This requires that both pressures and their first derivatives satisfy continuity conditions, as outlined below:
\begin{eqnarray}
P_H(\mu_{H}) &=&  P_M(\mu_{H})~,\label{eq:continuity1}\\
P_Q(\mu_{Q}) &=&  P_M(\mu_{Q})~,\\
\frac{\partial}{\partial\mu_B}P_H(\mu_{H}) &=&  \frac{\partial}{\partial\mu_B}P_M(\mu_{H})~,\\
\frac{\partial}{\partial\mu_B}P_Q(\mu_{Q}) &=&  \frac{\partial}{\partial\mu_B}P_M(\mu_{Q})~.
\label{eq:continuity}
\end{eqnarray}
By considering the relationship $n(\mu_B) = \frac{\partial P(\mu_B)}{\partial \mu_B}$, we undertake a numerical solution of the continuity equations, i.e., Eqs. \eqref{eq:continuity1}–\eqref{eq:continuity}, for baryon density. This approach enables us to deduce the unknown variables ($\alpha_1$, $\alpha_2$, $\mu_H$, $\mu_Q$) in the aforementioned equations.

Finally, we emphasize that for the EOSs employed here, the first crossing between the hadronic and quark curves takes place at very low densities. This crossing is discarded in our construction. Instead, the second crossing, corresponding to the so-called reconfinement transition at higher densities, is adopted as the boundary for the RIC interpolation. This procedure ensures that the hadronic EOS remains applicable at lower densities, while the QM EOS governs the high-density regime, and the interpolated EOS provides a smooth mathematical connection between them.}

\subsection{Parameter choices and ranges}
For completeness, we outline here the chosen parameter ranges and fixed inputs. These serve as the basis for constructing the hadronic, quark, and hybrid phases of our model.

{We scanned the S mass over the interval $m_S = 1890$--$2054~\mathrm{MeV}$. The mass-shift slope is fixed at the minimal value $x_S = 0.03$. Quark matter is modeled using two fixed nlNJL parametrizations: the first with $(\eta_D, \eta_V) = (0.70, 0.10)$ and the second with $(0.70, 0.11)$. For both parametrizations, the same squared speed of sound, $c_s^2 = 0.44$, is obtained through the established mapping between the nlNJL and CSS parametrizations, i.e., Eq.~(C4) in Ref.~\cite{Shahrbaf:2021cjz}. The hadron-quark crossover is constructed using the quadratic RIC; for each combination of a hadronic EOS (with a specific S-particle mass) and a quark EOS set, we determine numerically the matching chemical potentials $(\mu_H,\mu_Q)$ and ($\alpha_1, \alpha_2$) in Eq. \eqref{eq:RIMpressure} which control the interpolation shape. The pressure offset is fixed to $\Delta_P = -0.02$ and no additional free parameters are introduced. Therefore, the four unknowns $(\alpha_1, \alpha_2, \mu_H,\mu_Q)$ are determined by solving the four continuity conditions Eqs. \eqref{eq:continuity1}-\eqref{eq:continuity}. For readability, curves in multi-$m_S$ figures are labeled in increasing $m_S$, and some plots show subsets of the range.}

\section{Equation of state characterization across phases and thermodynamic behavior}
\label{sec:Thermodynamics}
We have determined specific values for the slope parameter, denoted as $x_\text{S}$, which corresponds to the coupling strength, and for the vacuum mass of S, $m_\text{S}$, in Eq.~\eqref{eq:Deltam}, by ensuring they satisfy the constraints arising from the interplay between these two parameters. The resulting constraint region, illustrating $x_\text{S}$ as a function of $m_\text{S}$, is presented in Figure 8 of our previous study \citep{Shahrbaf:2022upc}.

It is noteworthy that the DD2Y-T model, whose parameters are fine-tuned to reproduce the properties of atomic nuclei near saturation density and the experimentally constrained hyperon potentials at saturation, provides an extrapolation of the EOS beyond this regime. However, its predictions at higher densities are subject to increasing uncertainty due to the lack of direct experimental constraints. As a result, the constraints on $m_\text{S}$ and $x_\text{S}$ derived from saturation properties are notably more robust. In particular, the lower bounds on the mass and coupling strength of the $S$ particles are found to be $m_{\text{S}}^{\text{min}} = 1885$ MeV and $x_{\text{S}}^{\text{min}} = 0.03$, respectively \citep{Shahrbaf:2022upc}. Conversely, other constraints are based on the observations of NSs at high densities and may exhibit a degree of model dependency. Therefore, the forthcoming work will focus on investigating the model-dependent aspects of the current results.

\begin{figure}[h!]
   \centering
  \includegraphics[width=1.0\columnwidth]{p-mu_dd2_high_density.eps}

	\caption{Pressure as a function of baryonic chemical potential for both hadronic phase and quark phase. The solid lines correspond to the hadronic phase with S for which the coupling strength of S remains constant ($x_\text{S}=0.03$), while different values of the mass of S are supposed. The QM EOSs shown with dotted lines are characterized by two key parameters: the diquark coupling ($\eta_D$) and the vector meson coupling ($\eta_V$), respectively. The hadronic EOS without S is also shown with the dashed blue line as DD2Y-T for comparison. The colored dots show the critical chemical potential in Eq. (\ref{eq:mechequi}) at which the RIC has been applied. \label{fig:all_p_mu}}
	\end{figure}

This study aims to explore the upper limit for the mass of S as a candidate for DM within the framework of the DD2Y-T model. To determine the allowed range for the mass of S, it is important to explore the behavior of matter at high densities and make sure the results agree with the observational constraints from NSs. It is essential to note that a mass shift of S particles, $x_{\text{S}}$, serves the purpose of mimicking the interaction with the surrounding medium. Given the postulation that S is a candidate for DM, it is reasonable to anticipate a predominantly weak interaction with the medium. \citet{Moore:2024mot} emphasized that for S particles to be a major DM candidate, their interactions must be extremely suppressed.  Consequently, even though increasing the mass of S would inherently enable us to increase the value of $x_{\text{S}}$ based on our previous study \citep{Shahrbaf:2022upc}, it is a deliberate choice to maintain $x_{\text{S}}$ as the coupling strength of the DM interaction with BM at its minimal value, specifically $x_{\text{S}} = 0.03$ throughout the subsequent analyses. {Although the chosen value of $x_{\text{S}}$ is small, Eq.~\eqref{eq:Deltam} shows that increasing the baryon density still leads to a significant $\Delta m_{\text{S}}$, effectively mimicking the interactions between DM and BM in the dense cores of NSs. This simplifying approach allows us to systematically vary the mass of the DM candidate and examine its impact on the consistency of the hybrid star EOS, which includes hyperons in the hadronic phase, with current observational constraints.} 

Therefore, we applied RIC, as outlined in Section~\ref{sec:hybridEOS}, to model the phase transition from hadronic matter to deconfined QM. Within this approach, we studied the influence of the mass of S particles on the hybrid EOS and aim to determine the maximum allowed value for the mass of S.

Figure~\ref{fig:all_p_mu} illustrates the pressure as a function of the baryonic chemical potential, showcasing both the hadronic phase and the quark phase. In the context of the hadronic phase, the coupling strength of S remains constant, as previously discussed, while different values of the mass of S with an increasing trend are supposed. One should notice that in comparison to the hadronic EOS without S (as represented by DD2Y-T, which includes nucleons and hyperons), the inclusion of S as a DM candidate results in an obvious softening of the EOS, although a repulsive interaction between DM and BM is considered. 
In the quark phase, we depict two various EOSs with specific parameters determined through BA, as outlined in reference \citet{Shahrbaf:2021cjz}. Among the posterior-favored nlNJL parametrizations, we specifically select those that yield the earliest possible phase transition when combined with each hadronic EOS. This conservative choice ensures that deconfinement sets in as early as allowed by the model. Without this criterion, each hadronic EOS could, in principle, intersect with several different QM curves, leading to multiple possible transition points. 
 
The color dots show the crossing point between hadronic and quark lines, where the RIC with a negative value of $\Delta_P$ in Eq. \eqref{eq:RIMpressure} has been applied to correct the mischaracterization of a transition from hadronic matter to deconfined QM. As a result, at lower densities, the hadronic phase dominates, while at high densities, the quark phase remains a valid description. One should also point out that the RIC construction with $\Delta_P<0$ generates a peak in the squared speed of sound (as shown in Fig. 11 of \cite{Shahrbaf:2022upc}), consistent with the phenomenologically established behavior observed in QCD-based studies.

It is important to emphasize again that, for each hadronic EOS, we have selected the first intersection point with the QM EOS as the basis for implementing the phase transition construction. Although the two QM EOSs used in this study differ only slightly in stiffness, the softest three hadronic EOSs, corresponding to $m_\text{S}=$ 1890, 1895, and 1900 MeV, do not intersect with the stiffer of the two quark EOSs. Therefore, in the next section, where we perform a detailed scan over the mass of the S particle, these three soft hadronic models are combined with the first QM EOS (characterized by 
$\eta_D=0.70$, $\eta_V=0.10$). The remaining hadronic models are matched with the second, slightly stiffer QM EOS (with 
$\eta_D=0.70$, $\eta_V=0.11$). For both parametrizations, the squared speed of sound is inferred to be $c_s^2 = 0.44$.
 
It can be seen that increasing the mass of S leads to a noticeable stiffening of the hadronic EOS. As a result, the crossing point between the hadronic and QM EOS curves shifts toward higher values of the baryonic chemical potential, and consequently, to higher baryonic densities. This shift in the onset of the deconfinement transition has important implications for the internal structure of the star in the hadronic phase, influencing both the density profile and the composition of matter, including the presence and population of hadronic and exotic particles.

\begin{figure}[h!]
   \centering
	\includegraphics[width=1.0\columnwidth]{p-mu_dd2_RIC1.eps}
  
	\caption{Pressure as a function of baryonic chemical potential for the hadronic phase, QM phase, and the hybrid EOS constructed within the framework of RIC for $m_\text{S}=1900$ MeV.
      }
		\label{fig:threeRIC}  
	\end{figure}

    In Fig.~\ref{fig:threeRIC}, we illustrate how a RIC construction is performed between the hadronic EOS - represented by DD2Y-T+$\text{S}_{1900}$, which includes nucleons, hyperons, and a S with $m_\text{S}=1900$ MeV - and the QM EOS described by the nlNJL model with parameters $\eta_D=0.70$, $\eta_V=0.10$. The crossover phase transition between the two phases is characterized by a negative value of $\Delta_P$. The matching points defined earlier as $\mu_{H}$ and $\mu_{Q}$, representing the chemical potentials at which the pressure of the crossover phase is in accordance with the pressures of the hadronic and QM phases, are clearly illustrated in the figure. This figure shows how a negative value for $\Delta_ P$ in the context of RIC effectively addresses the issue of a nonphysical intersection point, often referred to as the reconfinement point. This adjustment restores the hadronic phase at lower densities and the QM phase at higher densities. {In contrast to our assumption, the inverted or cross hybrid star scenarios \citep{Zhang:2022pse, Zhang:2023zth, Negreiros:2024cvr} assume that QM can be absolutely stable at low densities (low pressure), allowing for hadronic cores surrounded by QM layers. These approaches represent complementary assumptions, and future high-precision astrophysical observations will be crucial to determine which scenario, if either, is realized in nature.
}

    \begin{figure}[h!]
   \centering
  \includegraphics[width=1.0\columnwidth]{p-e_new.eps}

	\caption{Pressure as a function of energy density for the baseline hadronic EOSs (DD2 and DD2Y-T) and the constructed hybrid EOSs. The yellow shaded region and dashed green line indicate the constraints from \citet{Hebeler:2013nza} and \citet{Miller:2021qha}, respectively. \label{fig:p-e}}
	\end{figure}

    {Figure~\ref{fig:p-e}, shows the pressure as a function of energy density for the baseline hadronic EOSs (DD2 and DD2Y-T) and the constructed hybrid EOSs. All employed and constructed EOSs are consistent with the Hebeler et al. \citep{Hebeler:2013nza} constraint (yellow region) and largely fall within the narrower Miller et al. \citep{Miller:2021qha} region (between dashed green lines), with only slight deviations at higher S-particle masses. At low energy densities, where hadronic matter dominates, the hybrid EOSs with DM are softer than the reference hadronic lines (DD2: black dash-dotted, DD2Y-T: light red lines) and also the hybrid RIC-DD2Y-T (violet) line, with stiffness increasing for higher S masses. At high energy densities, where QM governs, all hybrid EOSs converge, as both quark EOS sets share the same stiffness and speed of sound, rendering the high-density behavior almost insensitive to the S DM mass. At this region, the hybrid EOSs with DM (RIC-DD2Y-T+S) remain softer than DD2 but stiffer than DD2Y-T and RIC-DD2Y-T.}

    \begin{figure}[h]
   \centering
	\includegraphics[width=1.0\columnwidth]{S_fraction.eps}
	\caption{Fraction of S particles as a DM candidate in the hadronic phase calculated using the DD2Y-T model for various S particle masses. The fractions are shown from the respective onset densities up to densities beyond the central density of NSs, illustrating how the S fraction evolves within the purely hadronic phase. The black and red dash-dotted lines indicate the onset densities for the transition to deconfined QM, corresponding to $\mu_H$ for $m_{S}=1890$ MeV and $m_{S}=2054$ MeV, respectively.
		\label{fig:sfrac}
	}
	\end{figure}

Regarding the production of S particles inside the hybrid star modeled by the approach described above, Fig. \ref{fig:sfrac} shows the variation of the DM particle fraction as a function of baryon number density for different 
S-particle masses. In general, lighter S particles result in a higher DM fraction at a given density. Increasing the mass of S shifts the DM onset to higher densities, leading to a smaller fraction of heavier S particles being produced in stellar matter. For a fixed S mass, the DM fraction increases as the system becomes denser. However, at high densities, around $n_{B}\approx1 fm^{-3}$, the DM fraction becomes roughly the same across all considered S masses shown in the legend. The vertical black and dash-dotted red lines indicate the deconfinement onset densities, corresponding to $\mu_H$ for $m_{S}=1890$ MeV and $m_{S}=2054$ MeV, respectively, marking the chemical potentials at which some degrees of phase transition to QM occur. It is worth noting that the microscopic evolution of particle fractions within the mixed phase is not explicitly determined in our approach. The onset of the transition is specified by $\mu_H$, and the maximum DM fraction is reached at this threshold density for each stellar mass. Consequently, the maximum DM fraction allowed in our models varies from about $12\%$ for the highest S-particle mass to $15\%$ for the lightest one.
 Although the DM fractions vary along the curves, from zero up to about $30\%$ at very high densities within the pure hadronic phase, the maximum S fraction inside hybrid stars changes only within a narrow range of $12$-$15\%$. This behavior indicates that increasing the DM particle mass not only shifts the DM onset to higher densities but also delays the deconfinement transition, pushing it to even higher densities, as also seen in Fig.~\ref{fig:all_p_mu}.

\section{Neutron star constraints on the mass of dark matter particles}
\label{sec:M_R_L}

\begin{figure}[h!]
   \centering
	\includegraphics[width=1.0\columnwidth]{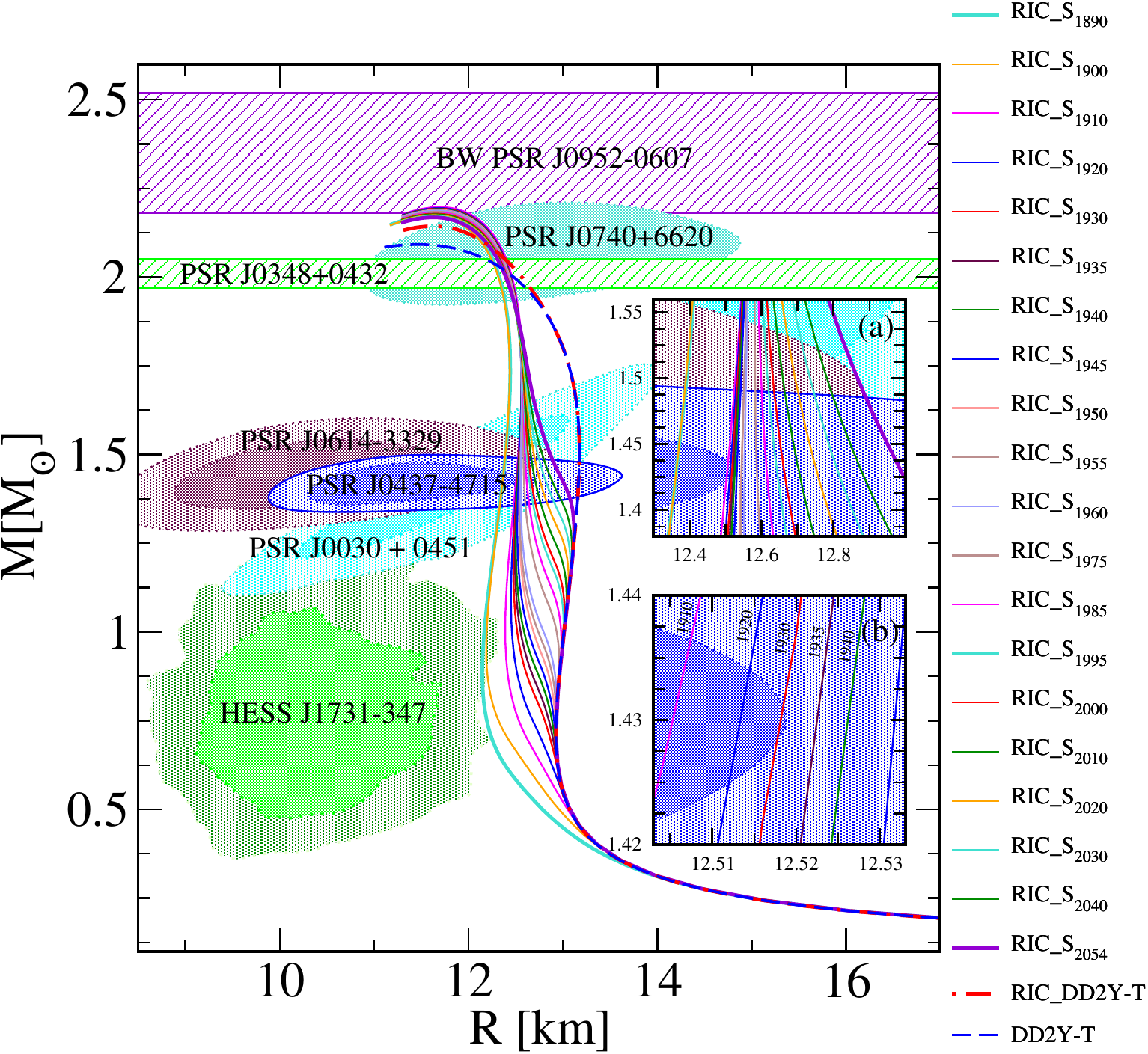}
	\caption{Mass-radius relations for NSs with a purely hadronic core (dashed blue line), hybrid stars without DM (dash-dotted red line), and hybrid stars including DM for different S particle masses (solid colored lines). The colored regions represent observational constraints for comparison. The 1-$\sigma$ M-R constraint from the NICER analysis of PSR J0740+6620 is shown in turquoise \citep{dittmann:2024:10215108}. The updated $95\%$ and $68\%$ credible NICER results for PSR J0030+0451 are displayed as nested light and dark cyan regions \citep{vinciguerra_2023_8239000}. The hatched magenta and green regions show the high-mass BW pulsar PSR J0952-0607 \citep{Romani:2022jhd} and PSR J0348+0432 \citep{Antoniadis:2013pzd}, respectively. The credible intervals for HESS J1731-347 are shown as nested green regions, representing the $68\%$ (inner) and $90\%$ (outer) levels based on X-ray spectral modeling \citep{doroshenko:2023:8232233}. The M-R constraints for PSR J0437-4715 are represented by overlapping dark and light blue regions, corresponding to the $68\%$ (inner) and $90\%$ (outer) credible intervals \citep{choudhury:2024:13766753}. The nested light and dark maroon regions show the most recent NICER analysis for the mass and radius of PSR J0614-3329, $95\%$ and $68\%$ credible results, respectively \citep{mauviard_2025_15603406}. The zoomed-in region (a) highlights that all RIC EOSs including S DM pass through the $95\%$ credible region of PSR J0614–3329, whereas the RIC\_DD2Y-T and DD2Y-T curves do not satisfy this constraint. The zoomed-in region (b) illustrates that for S masses larger than $1930$ MeV, the NS radius lies outside the $68\%$ credible region for PSR J0437-4715.
		\label{fig:M-R}
        }
	\end{figure}

 As explained in the previous sections, recent multimessenger observations of NSs offer a valuable chance to explore the structure of high-density matter, which may contain exotic components such as hyperons, QM, or even DM. In this section, based on the astrophysical constraints of NSs, we investigate our hybrid EOS and determine the allowed maximum mass for the S particle. For this purpose, the latest M-R measurements by NICER telescope have been applied, which provide reliable constraints for the radius of NSs with masses around $2M_\odot$ and $1.4M_\odot$ \citep{Miller:2019cac, Riley:2019yda, Miller:2021qha, Riley:2021pdl, choudhury:2024:13766753, Mauviard:2025dmd}.
 Since all of our hybrid EOSs pass well through the center of the NICER constraint region for PSR J0740+6620 with a mass of $2M_\odot$, we now focus primarily on the most recent results for the pulsars PSR J0614-3329 which has a measured mass of $M = 1.44^{+0.06}_{-0.07}\,M_{\odot}$ and radius $R = 10.29^{+1.01}_{-0.86}~\mathrm{km}$ \citep{Mauviard:2025dmd}, and PSR J0437-4715 ($M = 1.418 \pm 0.037\,M_\odot$, $R = 11.36^{+0.95}_{-0.63}\,\mathrm{km}$) \citep{choudhury:2024:13766753}.

 Furthermore, tidal deformability, which describes how a NS deforms under the gravitational influence of its companion in a binary system, is also taken into account. This parameter is highly sensitive to the EOS and the compactness of the star \citep{Hinderer:2007mb, Hinderer:2009ca}. We therefore examined whether our hybrid model aligns with the tidal deformability constraint for a $1.4M_\odot$ NS, given as $\Lambda_{1.4} = 190^{+390}_{-120}$ \citep{LIGOScientific:2018cki}.

In addition, since our model incorporates several important degrees of freedom in dense matter, including hyperons, DM, and the phase transition to deconfined QM, we include recent exotic mass measurements of NSs for a more comprehensive analysis. Specifically, we consider HESS J1731–347 as the lowest-mass known NS \citep{doroshenko:2023:8232233}, and PSR J0952–0607, the BW pulsar, as the highest-mass candidate \citep{Romani:2022jhd}.

To evaluate the reliability of our model and explore how the DM parameters align with observational constraints on NS properties, both the pure hadronic and hybrid EOSs are used as inputs to the Tolman–Oppenheimer–Volkoff equations \citep{PhysRev.55.364, PhysRev.55.374}. This allows us to obtain the M–R relations corresponding to each EOS. The results are displayed in Fig.~\ref{fig:M-R}, where the observational M–R constraints are shown as color-coded regions for comparison.

In Fig.~\ref{fig:M-R}, the M–R curve for the pure hadronic EOS, which includes hyperons (DD2Y-T), is shown with a dashed blue line. The dash-dotted red line represents the hybrid EOS without any DM (S) particles (RIC\_DD2Y-T). The solid colored lines show the full hybrid model that includes hyperons, S particles, and QM, where each color corresponds to a different S mass as indicated in the legend.

From the M–R curves, we can see that increasing the mass of S particles shifts the phase transition to QM to higher densities, meaning it happens at higher NS masses. This can be observed from the point where the curves start to deviate from the pure hadronic one. Because of the production of DM particles and the resulting softening, the radius of the star changes compared to the model without S, bringing the results in better agreement with all the observational constraints. The softening induced by DM acts mainly in the hadronic phase, which reduces radii at $\sim1.4M_\odot$. By contrast, the maximum mass is governed predominantly by the high-density quark EOS, which is essentially unchanged across our cases because both quark-matter parametrizations result in the same $c_s^2$. Consequently, the hybrid EOSs with DM yield comparable $M_{\max}$. The onset density of the phase transition also matters. For example, the RIC-DD2Y-T EOS (without DM) has a late transition, retains a larger hadronic fraction at high density, and therefore attains a smaller $M_{\max}$.  While all S masses satisfy the high-mass limit, the region corresponding to HESS J1731-347 
is only consistent with the lowest S masses ($m_\text{S} \lesssim1900$ MeV), due to the earlier deconfinement transition. The latest NICER M-R results for PSR J0614-3329, presented with maroon regions, indicate a measured mass of $\sim 1.4\,M_\odot$ having a radius comparable to HESS J1731-347. As also shown in the zoomed-in panel (a), all hybrid EOSs constructed using the RIC method and including S DM are found to intersect the $95\%$ credible region of PSR J0614-3329. In contrast, both the RIC\_DD2Y-T and DD2Y-T models, which either exclude DM or represent purely hadronic matter, fail to meet this observational constraint. This highlights that the consistency of a model with such a compact object requires additional softening beyond what is provided by hyperons or QM alone. Therefore, the presence of an additional degree of freedom, such as S DM, can supply the necessary softening and make our hybrid model consistent with the most recent NICER results.

In addition, considering the NICER results for PSR J0437-4715, 
which also gives the allowed radius range for a $1.4M_\odot$ NS, we find that including S particles shifts the curves toward the center of the nested blue regions, making the model more consistent with observations. In fact, as shown in the zoomed-in panel (b), for $m_\text{S} \leq 1930$ MeV, the hybrid EOSs even fall within the $68\%$ credible region from NICER, demonstrating consistency with the highest-probability region of the observational data.

We have also used the NICER measurements for PSR J0030+0451 from \citet{Vinciguerra:2023qxq}, which represent a recent update to the analysis by \citet{Riley:2019yda}. Examining the cyan regions, the impact of the S becomes evident. In contrast to the RIC\_DD2Y-T and DD2Y-T models, the EOSs that include S DM successfully pass through the $68\%$ credible region for PSR J0030+0451, indicating strong agreement with the most likely observational results. {We note that our hybrid EOSs remain consistent with the NICER measurement of PSR J1231-1411 \citep{Salmi:2024bss}, although this object is still under debate due to its complex pulse profile; for this reason, we did not include it in our M-R diagram.}

It is important to note that the presence of DM affects the onset of the phase transition compared to the RIC\_DD2Y-T EOS without S. This leads to a change in the radius of the star around $1.4M_\odot$, which is a key role of S particle production in making the model consistent with tidal deformability constraints, as we will discuss in the following.

Figures~\ref{fig:LM} and \ref{fig:L1L2} show how the presence of S particles affects the tidal deformability of NSs. In Fig.~\ref{fig:LM}, we show how the dimensionless tidal deformability, $\Lambda$, changes with the mass of the star. The vertical green line shows the observational limit for a $1.4M_\odot$ NS from the GW170817 event \citep{LIGOScientific:2018cki}. Different lines in the plot correspond to different S particle masses, as shown in the legend.

As seen in the figure, without DM, both the pure hadronic model (DD2Y-T) and the hybrid model without S (RIC\_DD2Y-T) do not meet the tidal deformability limit at $1.4M_\odot$. However, adding S particles helps soften the EOS near $1.4M_\odot$, which allows most hybrid models with S particles to satisfy the constraint. The zoomed-in part of the figure shows that for S masses less than about 2040 MeV, the value of $\Lambda_{1.4}$ stays below the upper limit of 580.

For completeness, Fig.~\ref{fig:L1L2} shows the allowed region in the $\Lambda_1$–$\Lambda_2$ plane based on the GW170817 observation \citep{LIGOScientific:2017vwq}. This figure also supports the upper limit of around 2040 MeV for the S particle mass. It can be seen that for $m_\text{S} \lesssim 2040$ MeV, the curves fall within the $90\%$ confidence region. In contrast, for heavier S particles, as well as for the pure hadronic and hybrid models without S, the curves fall outside the allowed region.

In conclusion, based on all the available multi-messenger data from NSs, we place a limit on the maximum allowed mass of the S particle. We find that the S mass should lie between 1885 and 2040 MeV to satisfy both nuclear and astrophysical constraints. Within this range, models that include hyperons, DM, and deconfined QM can all exist in the NS core. Among them, lighter S masses around 1900 MeV are especially favored, considering the M-R data from HESS J1731–347 and PSR J0437-4715. 

\begin{figure}[h!]
   \centering
	\includegraphics[width=1.0\columnwidth]{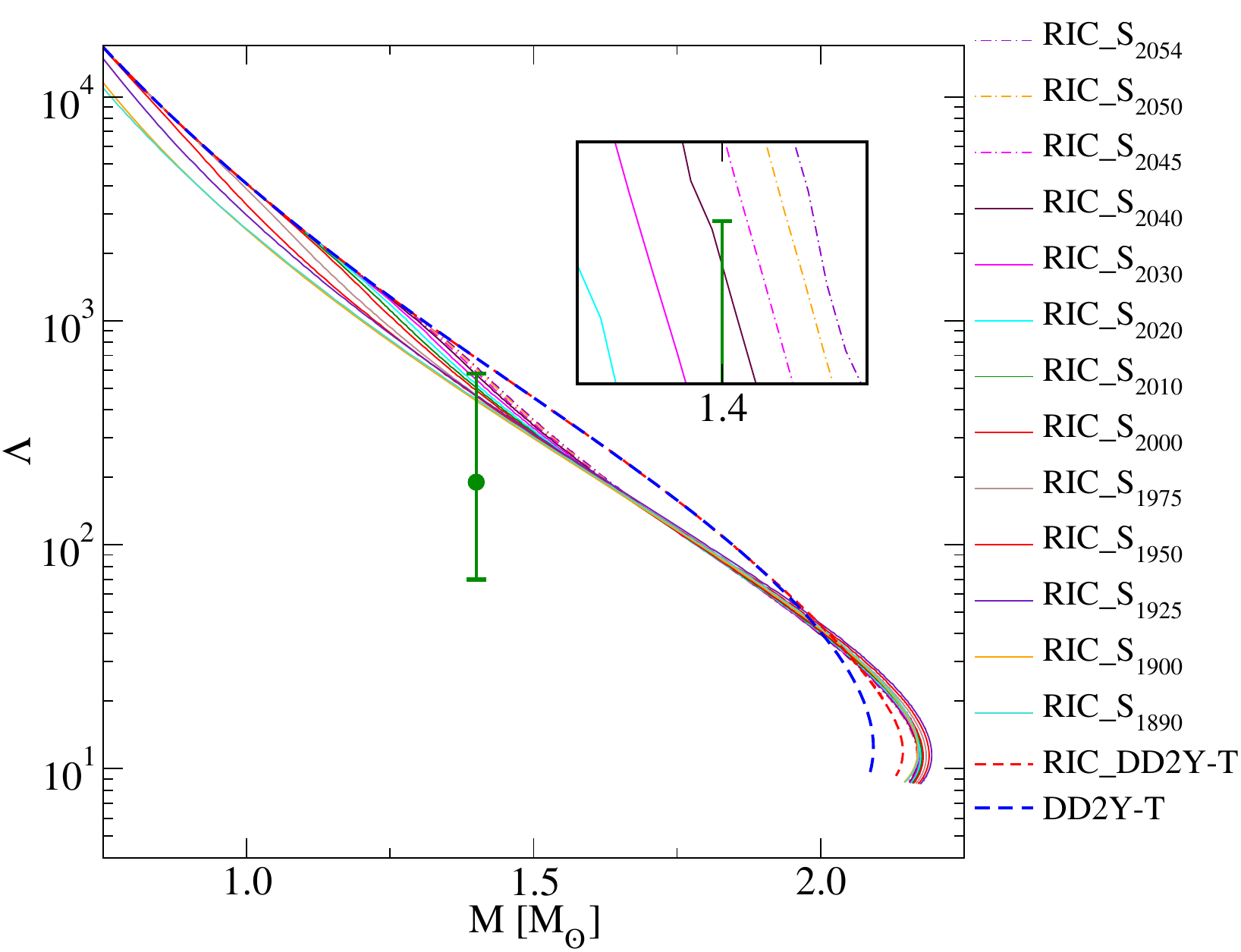}
	\caption{Dimensionless tidal deformability, denoted as $\Lambda$, presented as a function of stellar mass for the ensemble of EOSs illustrated in Fig.~\ref{fig:M-R}. The vertical green line in the figure signifies the $\Lambda_{1.4}$ constraint derived from the low-spin prior analysis of GW170817, as reported in \citep{LIGOScientific:2018cki}.
		\label{fig:LM}
	}
	\end{figure}

\begin{figure}[h!]
   \centering
	\includegraphics[width=1.0\columnwidth]{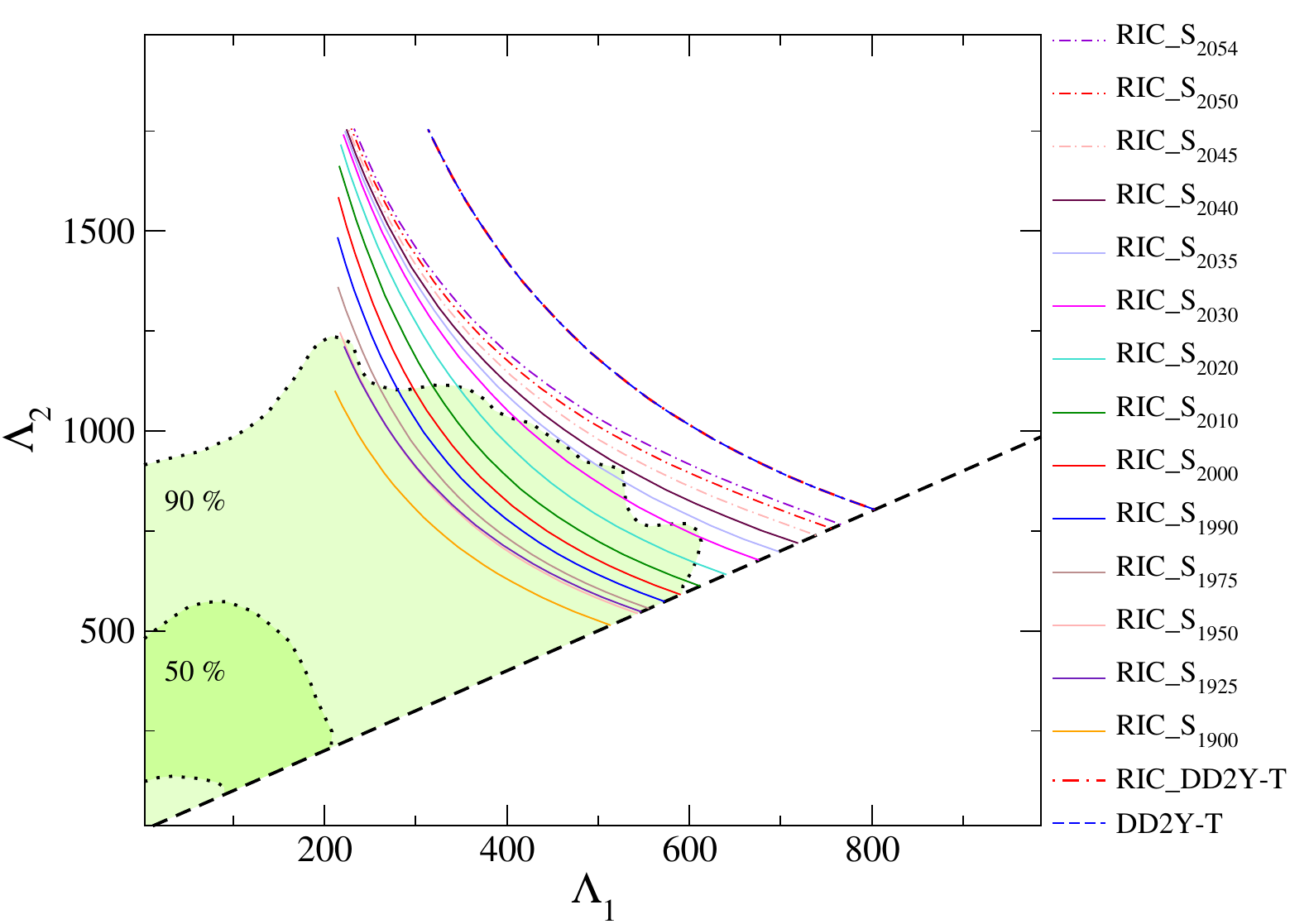}
	\caption{Tidal deformability parameters $\Lambda_1$ and $\Lambda_2$ of the high- and low-mass components of the binary merger. The green region is the placed constraint on the tidal effects by the LIGO and Virgo collaboration from the GW170817 event. 
 	\citep{LIGOScientific:2018cki}.
		\label{fig:L1L2}
	}
	\end{figure}

\section{Bayesian inference from neutron star observations}
\label{BA}

To constrain the value of the S mass in our BA,  we incorporated a set of astrophysical measurements leading to M-R and tidal deformability constraints presented in the previous section. Additionally, we examined the influence of individual subsets of these constraints on our EOS model estimates to assess the impact of the S mass.

\subsection{Basic observational dataset}

The basic dataset consists of the following

\begin{itemize}
    \item[(I)] Precise mass measurement of PSR J0348+0432 in a white dwarf binary~\citep{Antoniadis:2013pzd}:
    $$
        M = 2.01^{+0.04}_{-0.04}~M_{\odot}.
    $$
    This provides a robust lower bound on the maximum mass of NSs.

    \item[(II)]  The NICER measurements for mass and radius for PSR~J0740+6620 from \citet{Dittmann:2024mbo}, which is a recent refinement of~\citet{Miller:2021qha}:
    $$
        M = 2.08^{+0.07}_{-0.07}~M_{\odot}, \quad R = 12.92^{+2.09}_{-1.13}~\mathrm{km}.
    $$

    \item[(III)] The NICER measurements for PSR~J0030+0451 from \citet{Vinciguerra:2023qxq}, which is a recent refinement of~\citet{Riley:2019yda}:
    $$
        M = 1.44^{+0.15}_{-0.14}~M_{\odot}, \quad R = 13.02^{+1.24}_{-1.06}~\mathrm{km}.
    $$

    \item[(IV)] The mass and radius of PSR J0437-4715~\citep{Choudhury:2024xbk}:
    $$
        M = 1.418^{+0.037}_{-0.037}~M_{\odot}, \quad R = 11.36^{+0.95}_{-0.63}~\mathrm{km}.
    $$
    \item[(V)] Additionally, we considered the tidal deformability constraint from the GW170817 event~\citep{LIGOScientific:2018cki}:
    $$
        \Lambda_{1.4} = 190^{+390}_{-120}.
    $$
\end{itemize}

These entries constitute the basic dataset employed for Bayesian inference.

\subsection{Extended dataset}
Additionally, we tested the results by including two extreme measurements:

\begin{itemize}
    \item[(VI)] The high-mass BW pulsar PSR J0952-0607~\citep{Romani:2022jhd}:
    $$
        M = 2.35^{+0.17}_{-0.17}~M_{\odot}.
    $$
    \item[(VII)] The low-mass ultra compact object HESS J1731-347~\citep{Doroshenko:2022nwp}:
    $$
        M = 0.77^{+0.20}_{-0.17}~M_{\odot}, \quad R = 10.04^{+0.86}_{-0.78}~\mathrm{km}.
    $$
\end{itemize}

Furthermore, we examined EOSs in light of the newly reported mass and radius measurements for the rotation-powered millisecond pulsar PSR J0614-3329, based on NICER observations, which indicate that it is a moderately massive and relatively compact object, possessing the same radius as HESS J1731-347 but nearly twice the mass:

\begin{itemize}
    \item[(VIII)] Moderate-mass compact millisecond pulsar PSR~J0614-3329~\citep{Mauviard:2025dmd}:
    $$
        M = 1.44^{+0.06}_{-0.07}~M_{\odot}, \quad R = 10.29^{+1.01}_{-0.86}~\mathrm{km}.
    $$
\end{itemize}

\subsection{Bayesian analysis framework}

To determine the posterior distribution of the S mass  \(m_\text{S}\) given the data \(\mathcal{D}\), we apply Bayes' theorem:
\begin{equation}
    p(m_\text{S} | \mathcal{D}) = \frac{p(\mathcal{D} | m_\text{S}) \, p(m_\text{S})}{p(\mathcal{D})},
\end{equation}
assuming a uniform prior.

The total likelihood is a product over independent constraints:

\begin{equation}
    p(\mathcal{D} | m_\text{S}) = \prod_{\alpha} p(D_{\alpha} | m_\text{S}).
\end{equation}

The likelihood for maximum mass constraints uses the normal cumulative distribution function:
\begin{equation}
    p(D_{M} | m_\text{S}) = F_{\mathcal{N}}(M_{\max}(m_\text{S}); \mu_M,\sigma_M),
\end{equation}
with \(\mu_M\) and \(\sigma_M\) taken from measurements of the heaviest pulsars. Here $F_{\mathcal{N}}$ is the cumulative normal distribution function.

For M-R constraints, we evaluated the maximum likelihood approach:
\begin{equation}
    p(D_{MR}|m_\text{S}) = \max_{\varepsilon_c}\left[ f_{MR}\left(M(\varepsilon_c), R(\varepsilon_c)\right)\right].
\end{equation}
The density functions \(f_{MR}\) are obtained via kernel density estimation (KDE) using public datasets from Zenodo  \citep{vinciguerra_2023_8239000, dittmann:2024:10215108, choudhury:2024:13766753, doroshenko:2023:8232233, mauviard_2025_15603406}.

Similarly, the GW170817 constraint was evaluated as
\begin{equation}
    p(D_{GW} | m_\text{S}) =  f_{GW}\left(\Lambda(M=1.4; m_\text{S})\right),
\end{equation}
where \(f_{GW}\) is the non-symmetric split normal distribution of the tidal deformability for mass $1.4\,M_{\odot}$.

Finally, the Bayesian evidence was computed as
\begin{equation}
    p(\mathcal{D}) = \sum_{m_\text{S}} p(\mathcal{D} | m_\text{S} )\, p(m_\text{S}).
\end{equation}
Bayesian evidence plays two roles: a normalization factor in Bayesian inference, a measure of how well data can be explained by a model on average, considering all the parameter values it allows. We decided to provide the values of the evidences in the legends of the figures presenting our BA results, so that others can use them to calculate Bayes factors and compare their models with ours.

Note that we prefer the maximum likelihood approach over numerical marginalization, as employed, for example, in~\citet{Alvarez-Castillo:2016oln, Shahrbaf:2021cjz, Ayriyan:2024zfw}, because marginalizing over central density can introduce noticeable numerical inaccuracies when the EOS stiffness varies significantly within the target region. A fixed step in central density may lead to significant uneven sampling in observable space (e.g., mass and radius) across different EOSs, potentially biasing the results. We observed this effect in the region of the most probable NS masses, where our EOS exhibits substantial variation in stiffness. In this regime, numerical marginalization using M-R constraints from~\citet{vinciguerra_2023_8239000}, \citet{Choudhury:2024xbk} and \citet{Mauviard:2025dmd} becomes less reliable for our case. In contrast, the maximum likelihood method directly identifies the best-fitting models without relying on sensitive numerical integration.

\subsection{Bayesian analysis results}
 
The resulting posterior distributions, shown in Fig.~\ref{fig:basic_constraints}, reveal varying sensitivity to the different observational datasets. As seen in~Fig.~\ref{fig:basic_constraints:a}, the 2$M_\odot$ constraints from the precise mass measurement of PSR J0348+0432~\citep{Antoniadis:2013pzd} and the NICER measurement of PSR J0740+6620~\citep{Dittmann:2024mbo} yield a nearly uniform posterior, indicating that all considered S masses support sufficiently massive NSs. Similarly,~Fig.~\ref{fig:basic_constraints:c} shows that the tidal deformability constraint from GW170817, calibrated at 1.4$M_\odot$, only mildly disfavors the highest S masses. These two results are consistent with the fact that the physical parameter space for the S mass was pre-selected to satisfy both the 2$M_\odot$ mass requirement and the tidal deformability bound $\Lambda_{1.4}$. Consequently, the extremely massive BW pulsar PSR J0952-0607~\citep{Romani:2022jhd} (see~Fig.~\ref{fig:additional_constraints:a}) has a negligible impact on the posterior distribution as well. 

More discriminating behavior is observed in Fig.~\ref{fig:basic_constraints:b}, where NICER M-R measurements for PSR J0030+0451~\citep{Vinciguerra:2023qxq} and PSR J0437-4715~\citep{Choudhury:2024xbk}, both representative of canonical 1.4$M_\odot$ NSs, clearly favor lighter S masses (below 1930 MeV) while disfavoring the upper end of the tested range. This trend persists in Fig.~\ref{fig:basic_constraints:d} when combining all standard M-R and tidal deformability constraints (I)-(V). The preference for lower masses in this region reflects the requirement for an earlier onset of the phase transition to remain consistent with NICER radius measurements.

Incorporating the exotic compact objects HESS J1731–347 and PSR J0952–0607 in~Fig.~\ref{fig:additional_constraints} further reinforces this trend. In particular, the lightness and high compactness of HESS J1731–347 strongly favor S masses $m_\text{S}\lesssim 1900$~MeV(Fig.~\ref{fig:additional_constraints:b}), while our model intersects only the higher right edge of the 95\% confidence region inferred for this object by~\citet{Doroshenko:2022nwp}. This happens because the constraint yields a higher weighted likelihood difference for different values of $m_S$. 

It is worth noting that, due to the unexpectedly high compactness of this source, we expect that future refinements may lead to significant changes in its estimated mass and radius in favor of less compactness, despite this, various hypotheses are now being put forward about the nature of this ultracompact object. Several studies have proposed the nature of such light and ultracompact objects, it could be strange stars~\citep{DiClemente:2022wqp, Horvath:2023uwl, Yuan:2025mmn}, hybrid stars with a deconfined quark core~\citep{Ayriyan:2024zfw}, or even third family compact stars~\citep{Ayriyan:2024zfw, Alvarez-Castillo:2025yzu}. On the basis of the obtained results, we alternatively suggest that the presence of DM in such objects is a plausible hypothesis.

The final Fig.~\ref{fig:additional_constraints:c} illustrates the results of the analysis including all the constraints on mass and radius (I)-(VII). The dominant contribution to the posterior distribution arises from constraints in the region of canonical masses, as well as from the ultracompact object, resulting in an even stronger preference for lower masses of S.

In the final part of the analysis, we included a recently appeared estimate of the mass and radius of the pulsar PSR J0614-3329~\citep{Mauviard:2025dmd}, labeled as constraint (VIII). This object has a canonical mass of 1.44$M_{\odot}$ and a relatively small radius, making it a notably compact object in contrast to constraints (III) and (IV). Since these objects are assumed to have approximately the same mass, the M-R curve is expected to pass through the region where the uncertainties of these constraints overlap, this is satisfied primarily for lower S masses. The results shown in Fig.\ref{fig:new_constraint:a} indicate that the influence of this new constraint on the posterior distribution of the DM particle mass is similar to that of constraints (III) and (IV). Its combination with the baseline constraints (I)–(V) clearly enhances the preference for low-mass S particles (see fig.~\ref{fig:new_constraint:b}). When all constraints are taken into account, this preference becomes extremely dominant, as seen in Fig.~\ref{fig:new_constraint:c}.

\begin{figure*}[htbp]
    \centering
    \begin{tabular}{cc}
        \begin{subfigure}[t]{0.44\textwidth}
            \centering
            \includegraphics[width=\linewidth,height=0.75\linewidth]{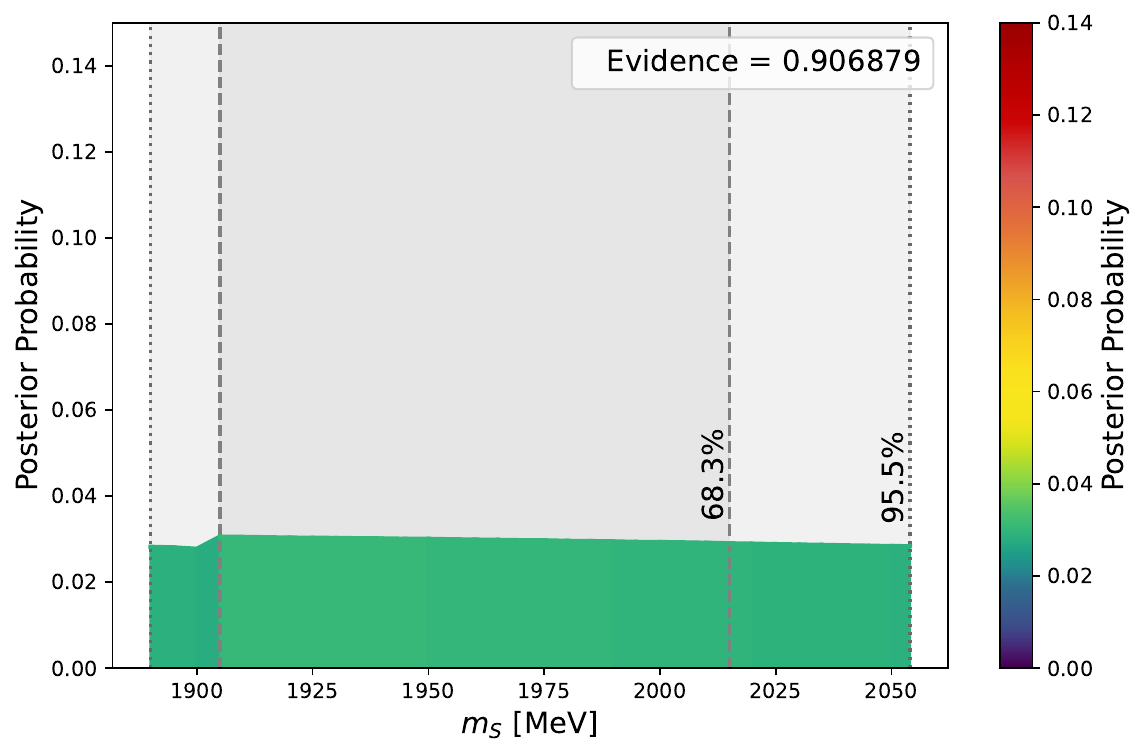}
            \caption{(I) and (II)}
            \label{fig:basic_constraints:a}
        \end{subfigure} &
        \begin{subfigure}[t]{0.44\textwidth}
            \centering
            \includegraphics[width=\linewidth,height=0.75\linewidth]{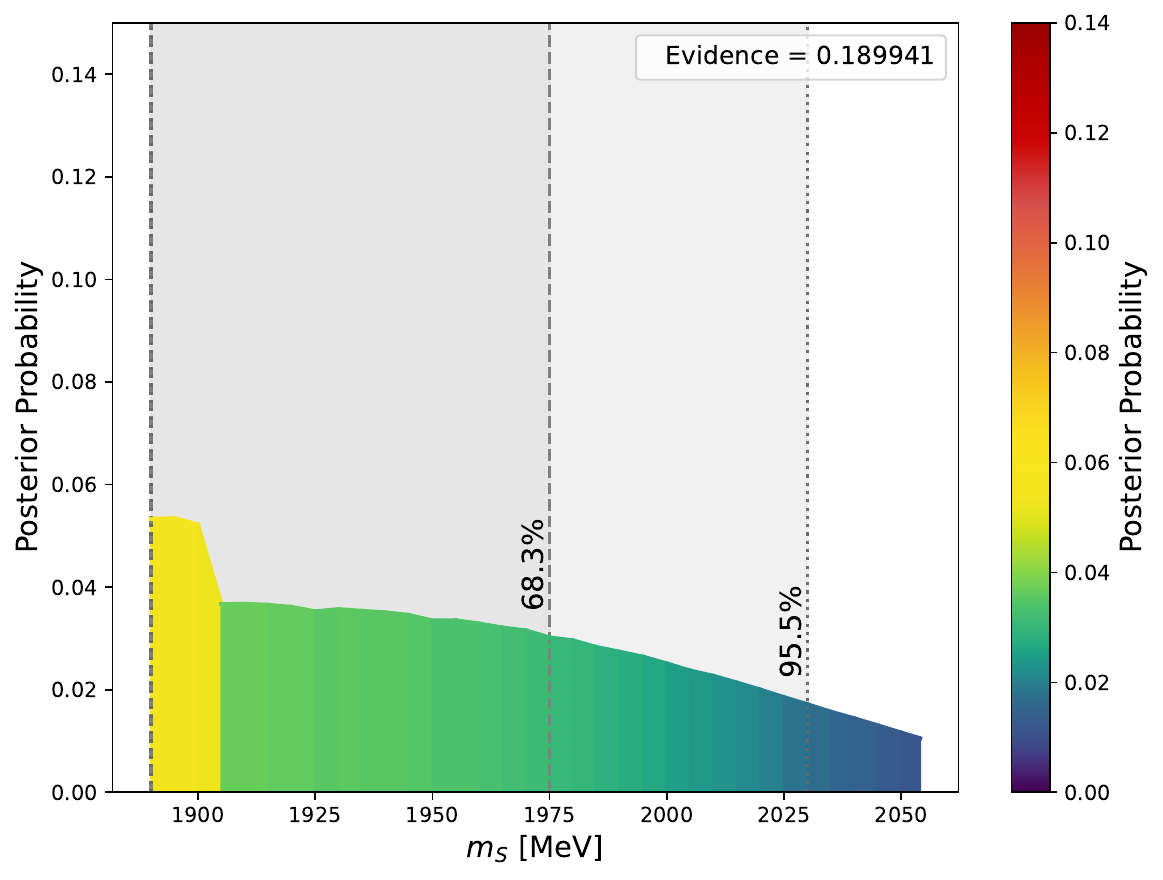}
            \caption{(III) and (IV)}
            \label{fig:basic_constraints:b}
        \end{subfigure} \\
        \begin{subfigure}[t]{0.44\textwidth}
            \centering
            \includegraphics[width=\linewidth,height=0.75\linewidth]{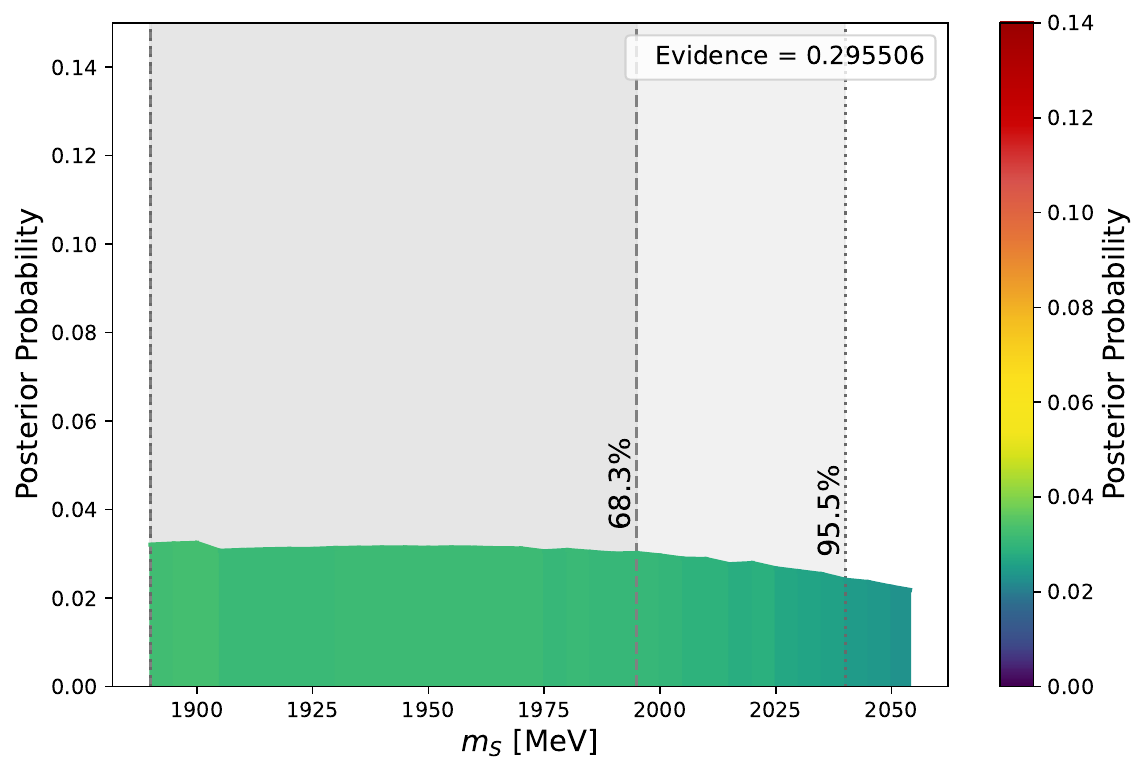}
            \caption{(V)}
            \label{fig:basic_constraints:c}
        \end{subfigure} &
        \begin{subfigure}[t]{0.44\textwidth}
            \centering
            \includegraphics[width=\linewidth,height=0.75\linewidth]{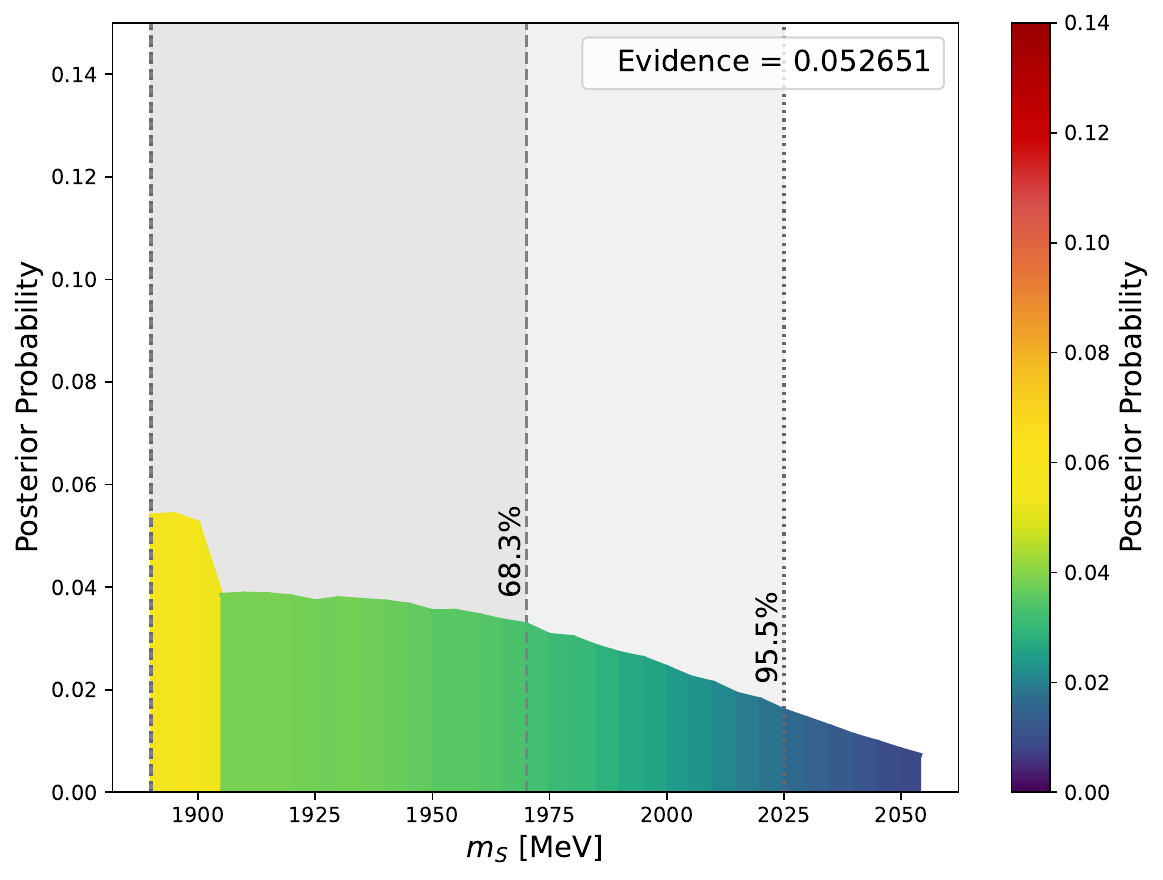}
            \caption{(I)--(V)}
            \label{fig:basic_constraints:d}
        \end{subfigure} \\
    \end{tabular}
    \caption{Posterior distribution for different sets of constraints.}
    \label{fig:basic_constraints}
\end{figure*}

\begin{figure}[htbp]
    \centering
    \begin{tabular}{cc}
        \begin{subfigure}[t]{0.42\textwidth}
            \centering
            \includegraphics[width=\linewidth,height=0.75\linewidth]{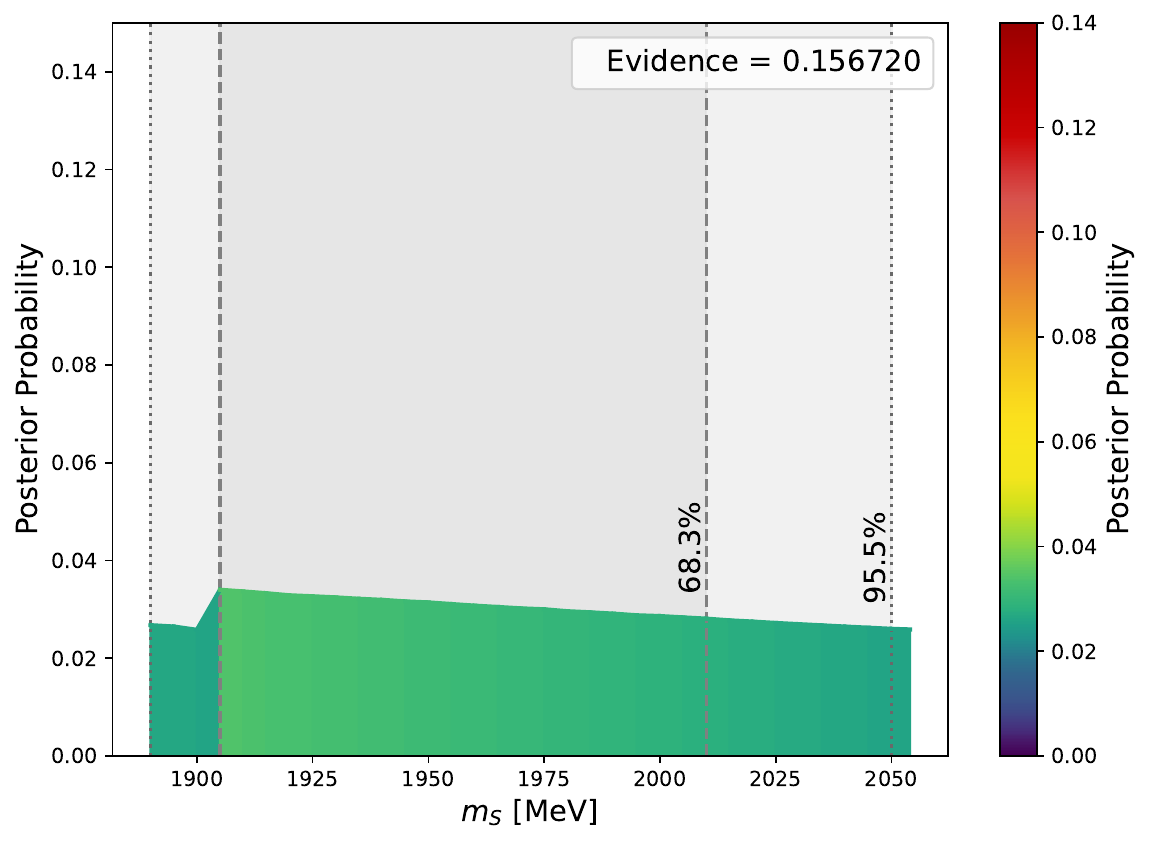}
            \caption{(VI)}
            \label{fig:additional_constraints:a}
        \end{subfigure} \\
        \begin{subfigure}[t]{0.42\textwidth}
            \centering
            \includegraphics[width=\linewidth,height=0.75\linewidth]{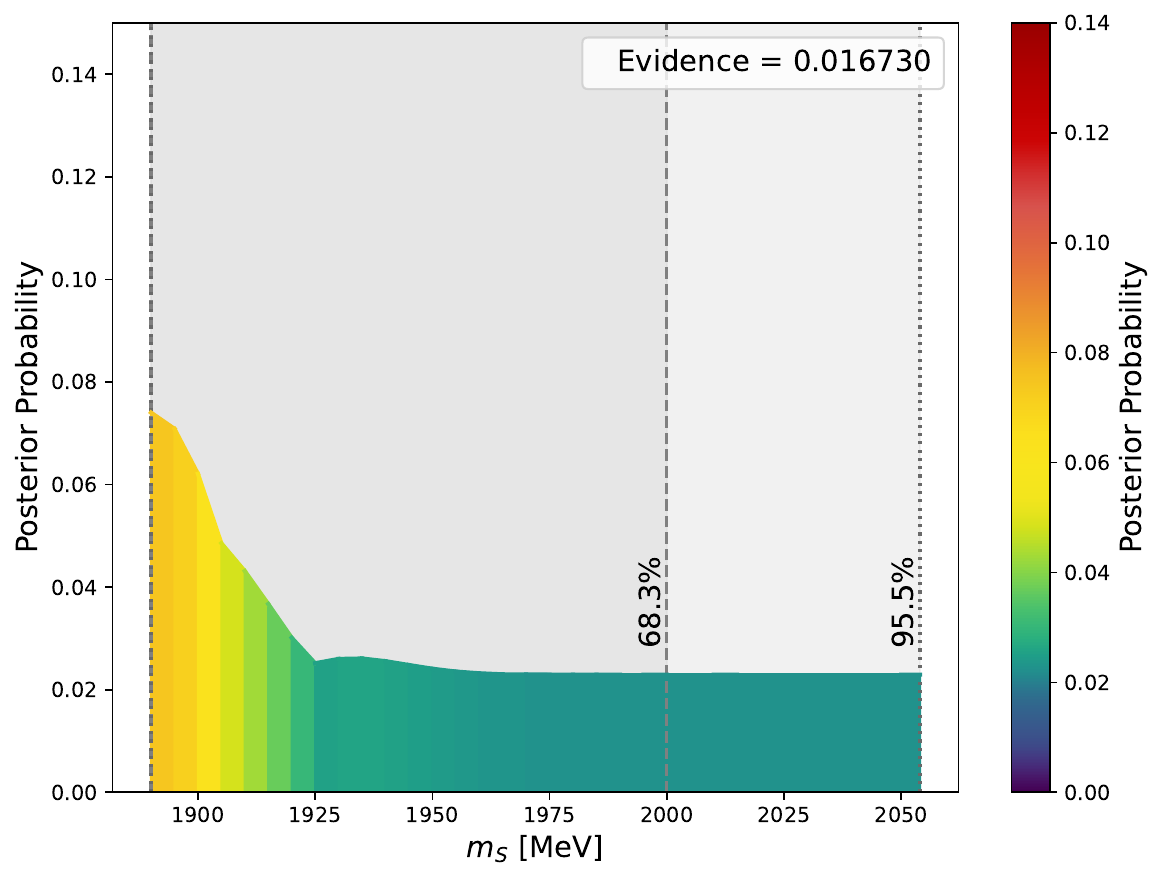}
            \caption{(VII)}
            \label{fig:additional_constraints:b}
        \end{subfigure} \\
        \begin{subfigure}[t]{0.42\textwidth}
            \centering
            \includegraphics[width=\linewidth,height=0.75\linewidth]{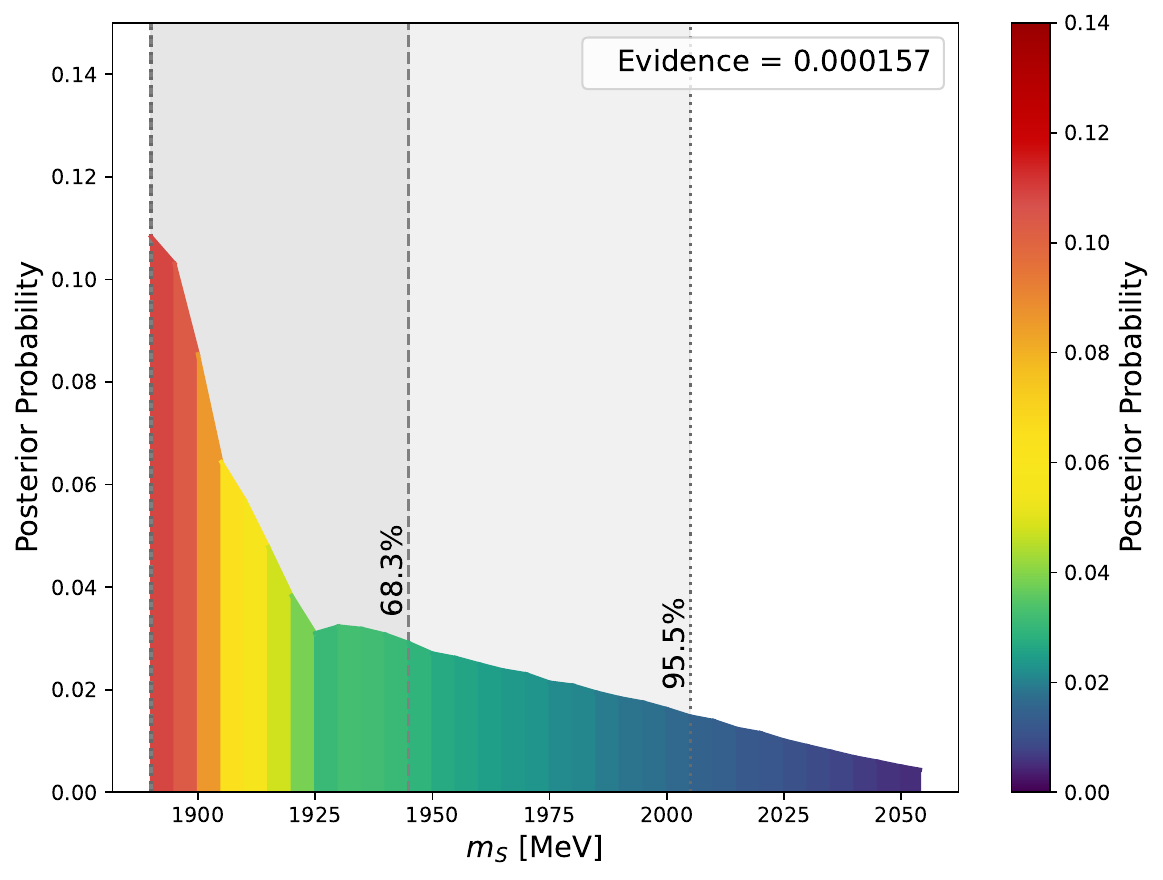}
            \caption{(I)--(VII)}
             \label{fig:additional_constraints:c}
        \end{subfigure}
    \end{tabular}
    \caption{Posterior for an additional set of constraints: (VI)~high-mass BW pulsar PSR J0952-0607~\citep{Romani:2022jhd} and (VII)~ultracompact HESS J1731-347~\citep{Doroshenko:2022nwp}.}
    \label{fig:additional_constraints}
\end{figure}

\begin{figure}[htbp]
    \centering
    \begin{tabular}{cc}
        \begin{subfigure}[t]{0.42\textwidth}
            \centering
            \includegraphics[width=\linewidth,height=0.75\linewidth]{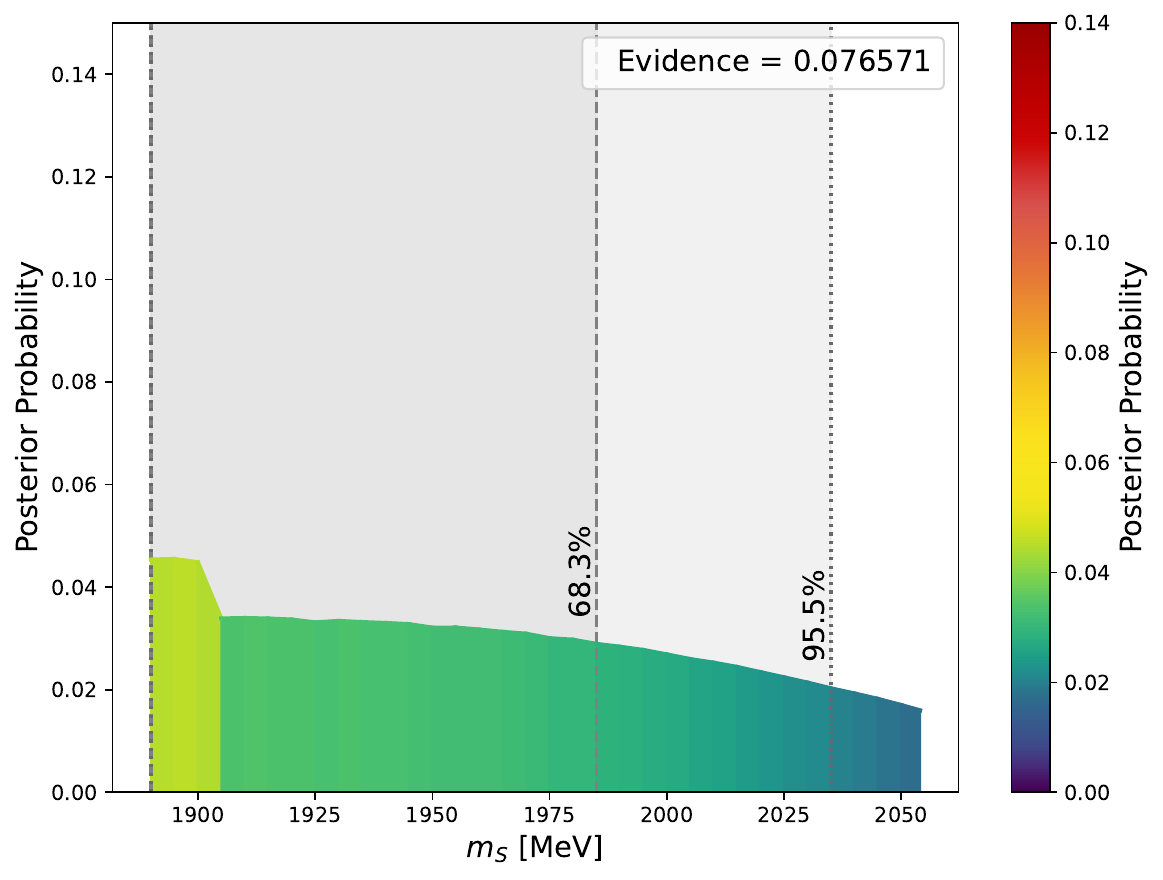}
            \caption{(VIII)}
            \label{fig:new_constraint:a}
        \end{subfigure} \\
        \begin{subfigure}[t]{0.42\textwidth}
            \centering
            \includegraphics[width=\linewidth,height=0.75\linewidth]{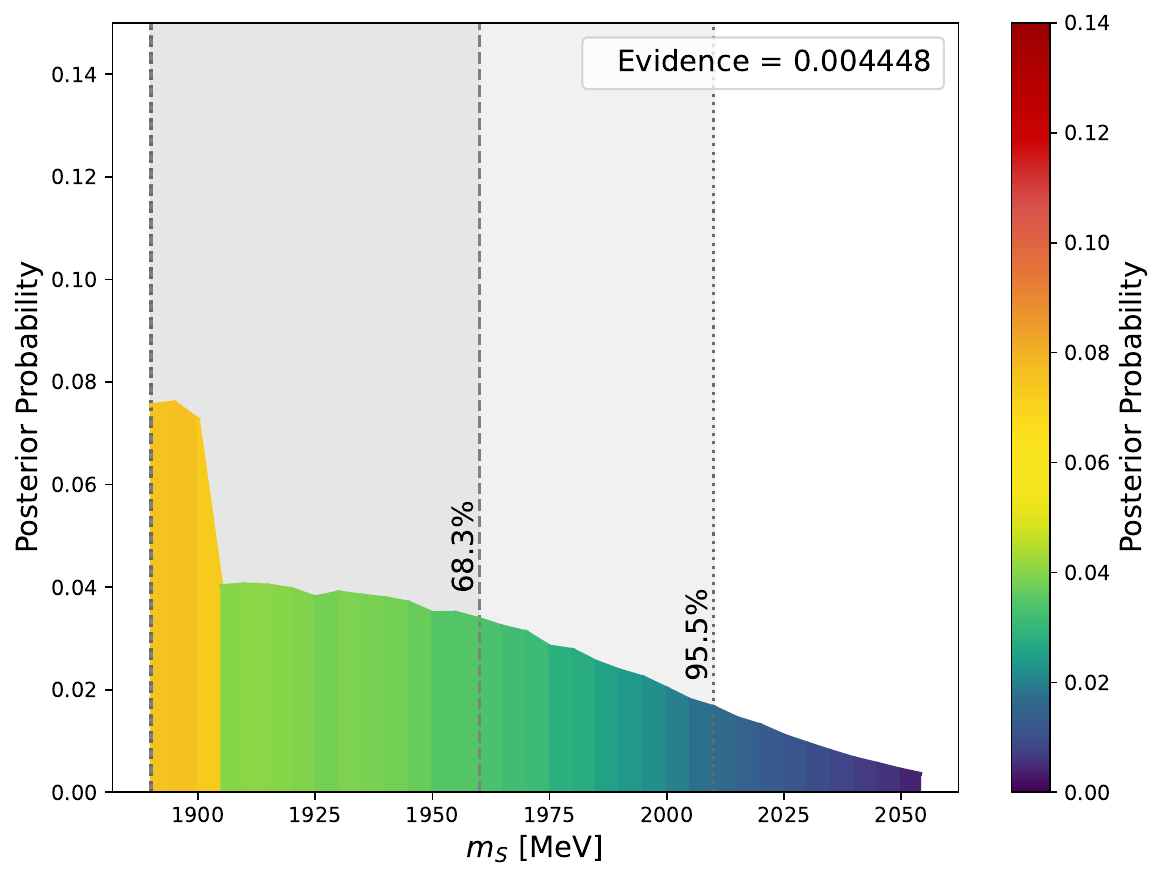}
            \caption{(I)--(V) and (VIII)}
            \label{fig:new_constraint:b}
        \end{subfigure} \\
        \begin{subfigure}[t]{0.42\textwidth}
            \centering
            \includegraphics[width=\linewidth,height=0.75\linewidth]{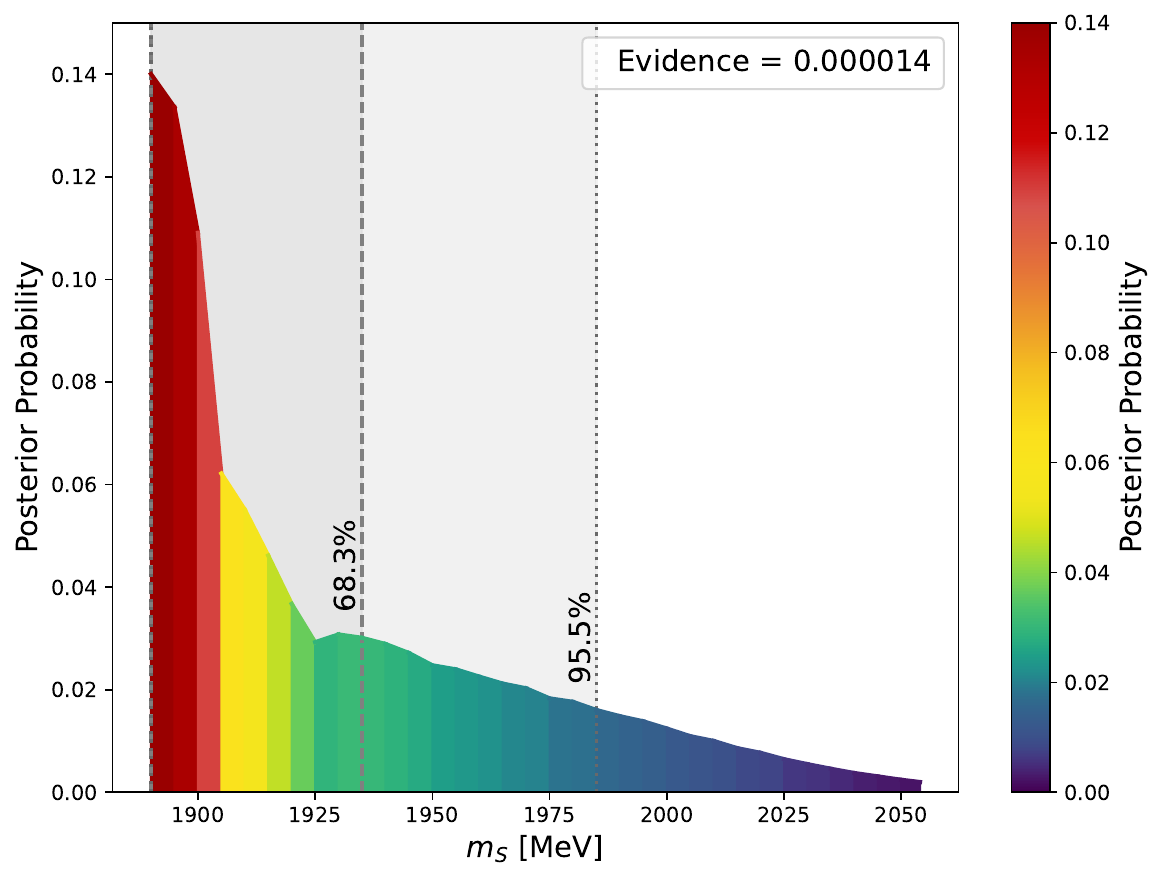}
            \caption{(I)--(VIII)}
             \label{fig:new_constraint:c}
        \end{subfigure}
    \end{tabular}
    \caption{Posterior distribution with new  NICER constraints (VIII) from the millisecond pulsar PSR~J0614--3329~\citep{Mauviard:2025dmd}}
    \label{fig:new_constraint}
\end{figure}

\section{Summary and conclusions}
\label{sec:summary}

Motivated by the theoretical existence of a stable sexaquark configuration, S=uuddss, with a mass below $2054$ MeV, we have explored its role as a bosonic DM candidate within the core of NSs. Using the DD2Y-T model to describe the hadronic phase, incorporating both hyperons and the sexaquark (S) as a bosonic DM component, we modeled the superconducting QM phase with the nlNJL model. A smooth crossover transition between these two phases was constructed using the RIC method. We find that hybrid configurations reach a maximum S DM fraction of about $12$-$15\%$.

Our results show that the inclusion of S particles with a weak repulsive interaction, modeled via a positive mass shift, leads to a softening of the EOS. {This effect is consistent with the role of the first established hexaquark, 
the $d^*(2380)$, whose inclusion in the NS EOS has also been shown 
to soften the EOS and alter the deconfinement threshold.} This softening plays a critical role in addressing the tension between stiff hadronic EOSs and tidal deformability constraints, particularly around the canonical $1.4 M_{\odot}$ mass range. In contrast to purely hadronic or hybrid models without DM, the resulting hybrid stars, comprising a deconfined QM core and a hypernuclear outer layer enriched with S DM, satisfy all current observational constraints on NS properties.

Incorporating recent astrophysical observations, we find that for $m_{\text{S}} \leq 1935$ MeV, the M-R relations of the resulting hybrid stars not only remain consistent with existing measurements but also lie within the $68\%$ credible region from NICER for PSR J0437-4715 around $1.4 M_{\odot}$. Moreover, the hybrid star configurations with the two lowest S masses used in this work agree with the constraints inferred from the low-mass ultracompact object HESS J1731-347. For $1885 < m_{\text{S}} < 2040$ MeV, the tidal deformability $\Lambda_{1.4}$ remains below the observational upper limit of 580, further supporting the viability of our EOS. {For comparison, we note that axion-like DM is expected in the $\mu$eV-MeV range, 
while weakly interacting massive particles are typically considered in the GeV-TeV range. 
Our results constrain the $S$ particle mass to the $\sim 2$ GeV scale, placing it in a distinct region among DM candidates.} 

It is worth noting that without the presence of DM, the considered hadronic (DD2Y-T) and hybrid (RIC\_DD2Y-T) models are not consistent with  $\Lambda_{1.4}$. 
In addition, by employing the recently updated NICER M-R results for PSR J0030+0451, we showed that the EOSs including S DM are compatible with the most probable observational values, i.e., the $68\%$ credible region, while the hybrid model without S and the pure hadronic EOS are only consistent with the $95\%$ region.

Furthermore, to place the most stringent constraint on the S mass while accounting for all available observational data, we have incorporated the latest NICER M-R measurements of PSR J0614-3329, which became available prior to the submission of this study. We find that, unlike the EOSs without the S component, all models that include DM favor the $95\%$ credible region, with lighter S particles being particularly preferred. This noteworthy result provides strong support for our model in which S particles are produced inside NSs and thus coexist with other possible internal structures, including hyperons and QM.

We conclude that including S DM as an additional degree of freedom alongside hyperons and deconfined QM plays a crucial role in ensuring the consistency of the model with all available observational data. In particular, it makes the model more favorable around 1.4$M_{\odot}$, reflecting close alignment with the most robustly inferred NS properties.

To investigate the parameter space of S and constrain its mass more precisely, we performed a BA that incorporates all key observational constraints within a consistent statistical framework. This includes all available M-R measurements by the NICER telescope and the tidal deformability constraint from the GW170817 event by LIGO/Virgo detectors. Additionally, we considered two astrophysical extremes: HESS J1731–347, representing the most compact (lightest) known NS, and PSR J0952–0607, the most massive NS observed to date. These sources provide critical boundary conditions for constraining the EOS across the full density range. 

To the best of our knowledge, this study incorporates all currently available and relevant observational data to evaluate the viability of the proposed model. We find that our hybrid model, comprising ordinary nuclear matter, hyperons, S bosonic DM, and deconfined QM, satisfies all observational constraints only when S DM is included. This highlights the essential role of the additional softening and thus the smaller radii around $1.4M_{\odot}$ introduced by the S component. The Bayesian inference yields a posterior distribution that disfavors S particle masses higher than $2000~\text{MeV}$ (placing them outside the 95.5\% credibility region) and strongly favors  $m_{S}\lesssim1935~\text{MeV}$ (within the 68.3\% credibility region), suggesting that such a DM candidate may contribute meaningfully to the dense matter composition of NSs.

While our study does not claim that the presence of S particles is required to explain NS properties, it provides compelling evidence that their inclusion leads to the improved consistency of our model with current observational data. Future observations, particularly those sensitive to the thermal and cooling behavior of NSs, may offer further insight into the role of such exotic particles in the interiors of compact stars.

{We emphasize that in this work, we have
focused on testing the astrophysical viability of the S particle as a DM candidate inside NSs. Our assumption of a small fixed value of $x_S$ effectively models a weak coupling to the
medium and is consistent with current literature. 
A DM $S$ state remains viable provided its vacuum couplings to baryons are small, while in-medium effects at sufficiently high densities can still shift its effective mass to influence the EOS. Our present results should be interpreted as the limit of equilibrium configurations. We note that a definitive assessment requires future work that (i) computes weak reaction rates and equilibration timescales involving $S$, (ii) implements a dynamical vector interaction for $S$ beyond a linear mass shift, and (iii) treats possible deconfinement-driven dissociation consistently. These improvements will determine whether the equilibrium compositions assumed here are achieved on astrophysical timescales, but they lie
beyond the scope of the present study.

In summary, our analysis shows that the inclusion of DM provides a viable mechanism to soften the stiff hadronic EOSs, such as DD2Y-T, enabling them to remain consistent with modern astrophysical constraints. We stress, however, that this conclusion is model dependent, as softer EOSs may not require, and could even be disfavored by, the presence of DM. In this sense, DM should be regarded as one possible solution among several proposed in the literature to reconcile current tensions between the precise measurements of high mass NSs and the existence of very compact NSs, alongside scenarios such as two families of compact stars \citep{Drago:2015cea, Drago:2015dea, Drago:2013fsa}, slowly stable hybrid stars \citep{Mariani:2024gqi, Laskos-Patkos:2024fdp}, or scenarios involving self-bound hybrid stars and hybrid stars that remain radially stable in the context of slow (or even fast) phase transitions \citep{Zhang:2025rnf}.

As a final remark, we note that our analysis is not in contradiction with halo-like DM distributions on galactic scales. Rather, our model describes the distribution of S particles inside NSs under the assumption of non-gravitational BM-DM coupling. The shell-like configuration of S DM at the stellar level can, in principle, coexist with extended DM halo structures on larger astrophysical scales. Thus, our results are not in tension with the DM galactic halo distribution but are complementary to it, providing an additional way to probe this model through multi-messenger observations of NSs.

\section*{Acknowledgements}
M. Sh. would like to thank G. Farrar and D. Blaschke, her co-authors on a previous work, for valuable discussions and correspondence that significantly contributed to her understanding of the sexaquark concept and its implications for neutron star physics. M. Sh. has been supported by NCN under SONATINA 7 grant NO. 2023/48/C/ST2/00297 and also by the program Excellence Initiative-Research University of the University of Wroclaw of the Ministry of Education and Science. D.R.K was supported by SONATINA 7 grant NO. 2023/48/C/ST2/00297. D.R.K was, in part,  supported by Polish NCN Grant No. 2023/51/B/ST9/02798. A.A. was supported by NCN under grant No. 2021/43/P/ST2/03319.

\bibliographystyle{aa}
\bibliography{references}

@article{LIGOScientific:2020aai,
    author = "Abbott, B. P. and others",
    collaboration = "LIGO Scientific, Virgo",
    title = "{GW190425: Observation of a Compact Binary Coalescence with Total Mass $\sim 3.4 M_{\odot}$}",
    eprint = "2001.01761",
    archivePrefix = "arXiv",
    primaryClass = "astro-ph.HE",
    reportNumber = "LIGO-P190425",
    doi = "10.3847/2041-8213/ab75f5",
    journal = "Astrophys. J. Lett.",
    volume = "892",
    number = "1",
    pages = "L3",
    year = "2020"
}

@article{LIGOScientific:2017vwq,
    author = "Abbott, B. P. and others",
    collaboration = "LIGO Scientific, Virgo",
    title = "{GW170817: Observation of Gravitational Waves from a Binary Neutron Star Inspiral}",
    eprint = "1710.05832",
    archivePrefix = "arXiv",
    primaryClass = "gr-qc",
    reportNumber = "LIGO-P170817",
    doi = "10.1103/PhysRevLett.119.161101",
    journal = "Phys. Rev. Lett.",
    volume = "119",
    number = "16",
    pages = "161101",
    year = "2017"
}

@article{Hinderer:2007mb,
    author = "Hinderer, Tanja",
    title = "{Tidal Love numbers of neutron stars}",
    eprint = "0711.2420",
    archivePrefix = "arXiv",
    primaryClass = "astro-ph",
    doi = "10.1086/533487",
    journal = "Astrophys. J.",
    volume = "677",
    pages = "1216--1220",
    year = "2008",
    note = "[Erratum: Astrophys.J. 697, 964 (2009)]"
}

@article{Hinderer:2009ca,
    author = "Hinderer, Tanja and Lackey, Benjamin D. and Lang, Ryan N. and Read, Jocelyn S.",
    title = "{Tidal deformability of neutron stars with realistic equations of state and their gravitational wave signatures in binary inspiral}",
    eprint = "0911.3535",
    archivePrefix = "arXiv",
    primaryClass = "astro-ph.HE",
    doi = "10.1103/PhysRevD.81.123016",
    journal = "Phys. Rev. D",
    volume = "81",
    pages = "123016",
    year = "2010"
}

@article{deMartino:2020gfi,
    author = "de Martino, Ivan and Chakrabarty, Sankha S. and Cesare, Valentina and Gallo, Arianna and Ostorero, Luisa and Diaferio, Antonaldo",
    title = "{Dark matters on the scale of galaxies}",
    eprint = "2007.15539",
    archivePrefix = "arXiv",
    primaryClass = "astro-ph.CO",
    doi = "10.3390/universe6080107",
    journal = "Universe",
    volume = "6",
    number = "8",
    pages = "107",
    year = "2020"
}

@article{Salucci:2020nlp,
    author = "Salucci, Paolo and others",
    title = "{Einstein, Planck and Vera Rubin: Relevant Encounters Between the Cosmological and the Quantum Worlds}",
    eprint = "2011.09278",
    archivePrefix = "arXiv",
    primaryClass = "gr-qc",
    doi = "10.3389/fphy.2020.603190",
    journal = "Front. in Phys.",
    volume = "8",
    pages = "603190",
    year = "2021"
}

@article{Bertone:2018krk,
    author = "Bertone, Gianfranco and Tait, M. P., Tim",
    title = "{A new era in the search for dark matter}",
    eprint = "1810.01668",
    archivePrefix = "arXiv",
    primaryClass = "astro-ph.CO",
    doi = "10.1038/s41586-018-0542-z",
    journal = "Nature",
    volume = "562",
    number = "7725",
    pages = "51--56",
    year = "2018"
}

@article{Marsh:2015xka,
    author = "Marsh, David J. E.",
    title = "{Axion Cosmology}",
    eprint = "1510.07633",
    archivePrefix = "arXiv",
    primaryClass = "astro-ph.CO",
    reportNumber = "KCL-PH-TH-2015-50",
    doi = "10.1016/j.physrep.2016.06.005",
    journal = "Phys. Rept.",
    volume = "643",
    pages = "1--79",
    year = "2016"
}

@article{Farrar:2022mih,
  author        = {Farrar, Glennys R.},
  title         = {A Stable Sexaquark: Overview and Discovery Strategies},
  year          = {2022},
  month         = {January},
  archivePrefix = {arXiv},
  eprint        = {2201.01334},
  primaryClass  = {hep-ph},
  note          = {arXiv:2201.01334},
  url           = {https://arxiv.org/abs/2201.01334}
}

@article{Doser:2023gls,
    author = "Doser, Michael and Farrar, Glennys and Kornakov, Georgy",
    title = "{Searching for a dark matter particle with anti-protonic atoms}",
    eprint = "2302.00759",
    archivePrefix = "arXiv",
    primaryClass = "hep-ph",
    doi = "10.1140/epjc/s10052-023-12319-8",
    journal = "Eur. Phys. J. C",
    volume = "83",
    number = "12",
    pages = "1149",
    year = "2023"
}

@article{LIGOScientific:2018cki,
    author = "Abbott, B. P. and others",
    collaboration = "LIGO Scientific, Virgo",
    title = "{GW170817: Measurements of neutron star radii and equation of state}",
    eprint = "1805.11581",
    archivePrefix = "arXiv",
    primaryClass = "gr-qc",
    reportNumber = "LIGO-P1800115",
    doi = "10.1103/PhysRevLett.121.161101",
    journal = "Phys. Rev. Lett.",
    volume = "121",
    number = "16",
    pages = "161101",
    year = "2018"
}

@article{Drago:2013fsa,
    author = "Drago, Alessandro and Lavagno, Andrea and Pagliara, Giuseppe",
    title = "{Can very compact and very massive neutron stars both exist?}",
    eprint = "1309.7263",
    archivePrefix = "arXiv",
    primaryClass = "nucl-th",
    doi = "10.1103/PhysRevD.89.043014",
    journal = "Phys. Rev. D",
    volume = "89",
    number = "4",
    pages = "043014",
    year = "2014"
}

@article{Drago:2015cea,
    author = "Drago, Alessandro and Lavagno, Andrea and Pagliara, Giuseppe and Pigato, Daniele",
    title = "{The scenario of two families of compact stars}: {1. Equations of state, mass-radius relations and binary systems}",
    eprint = "1509.02131",
    archivePrefix = "arXiv",
    primaryClass = "astro-ph.SR",
    doi = "10.1140/epja/i2016-16040-3",
    journal = "Eur. Phys. J. A",
    volume = "52",
    number = "2",
    pages = "40",
    year = "2016"
}

@article{Drago:2015dea,
    author = "Drago, Alessandro and Pagliara, Giuseppe",
    title = "{The scenario of two families of compact stars}: {2. Transition from hadronic to quark matter and explosive phenomena}",
    eprint = "1509.02134",
    archivePrefix = "arXiv",
    primaryClass = "astro-ph.SR",
    doi = "10.1140/epja/i2016-16041-2",
    journal = "Eur. Phys. J. A",
    volume = "52",
    number = "2",
    pages = "41",
    year = "2016"
}

@article{Mariani:2024gqi,
    author = "Mariani, Mauro and Ranea-Sandoval, Ignacio F. and Lugones, Germ{\'a}n and Orsaria, Milva G.",
    title = "{Could a slow stable hybrid star explain the central compact object in HESS J1731-347?}",
    eprint = "2407.06347",
    archivePrefix = "arXiv",
    primaryClass = "astro-ph.HE",
    doi = "10.1103/PhysRevD.110.043026",
    journal = "Phys. Rev. D",
    volume = "110",
    number = "4",
    pages = "043026",
    year = "2024"
}

@article{Laskos-Patkos:2024fdp,
    author = "Laskos-Patkos, P. and Moustakidis, Ch. C.",
    title = "{XTE J1814-338: A potential hybrid star candidate}",
    eprint = "2410.18498",
    archivePrefix = "arXiv",
    primaryClass = "astro-ph.HE",
    doi = "10.1103/PhysRevD.111.063058",
    journal = "Phys. Rev. D",
    volume = "111",
    number = "6",
    pages = "063058",
    year = "2025"
}

@article{Zhang:2025rnf,
    author = "Zhang, Chen and Pretel, Juan M. Z. and Xu, Renxin",
    title = "{Slow Stable Self-bound Hybrid Star Can Relieve All Tensions}",
    eprint = "2507.01371",
    archivePrefix = "arXiv",
    primaryClass = "astro-ph.HE",
    month = "7",
    year = "2025"
}

@article{Miller:2019cac,
    author = "Miller, M. C. and others",
    title = "{PSR J0030+0451 Mass and Radius from $NICER$ Data and Implications for the Properties of Neutron Star Matter}",
    eprint = "1912.05705",
    archivePrefix = "arXiv",
    primaryClass = "astro-ph.HE",
    doi = "10.3847/2041-8213/ab50c5",
    journal = "Astrophys. J. Lett.",
    volume = "887",
    number = "1",
    pages = "L24",
    year = "2019"
}

@article{Hebeler:2013nza,
    author = "Hebeler, K. and Lattimer, J. M. and Pethick, C. J. and Schwenk, A.",
    title = "{Equation of state and neutron star properties constrained by nuclear physics and observation}",
    eprint = "1303.4662",
    archivePrefix = "arXiv",
    primaryClass = "astro-ph.SR",
    doi = "10.1088/0004-637X/773/1/11",
    journal = "Astrophys. J.",
    volume = "773",
    pages = "11",
    year = "2013"
}

@article{Miller:2021qha,
    author = "Miller, M. C. and others",
    title = "{The Radius of PSR J0740+6620 from NICER and XMM-Newton Data}",
    eprint = "2105.06979",
    archivePrefix = "arXiv",
    primaryClass = "astro-ph.HE",
    doi = "10.3847/2041-8213/ac089b",
    journal = "Astrophys. J. Lett.",
    volume = "918",
    number = "2",
    pages = "L28",
    year = "2021"
}

@article{Riley:2019yda,
    author = "Riley, Thomas E. and others",
    title = "{A $NICER$ View of PSR J0030+0451: Millisecond Pulsar Parameter Estimation}",
    eprint = "1912.05702",
    archivePrefix = "arXiv",
    primaryClass = "astro-ph.HE",
    doi = "10.3847/2041-8213/ab481c",
    journal = "Astrophys. J. Lett.",
    volume = "887",
    number = "1",
    pages = "L21",
    year = "2019"
}

@article{Riley:2021pdl,
    author = "Riley, Thomas E. and others",
    title = "{A NICER View of the Massive Pulsar PSR J0740+6620 Informed by Radio Timing and XMM-Newton Spectroscopy}",
    eprint = "2105.06980",
    archivePrefix = "arXiv",
    primaryClass = "astro-ph.HE",
    doi = "10.3847/2041-8213/ac0a81",
    journal = "Astrophys. J. Lett.",
    volume = "918",
    number = "2",
    pages = "L27",
    year = "2021"
}

@article{Fonseca:2021wxt,
    author = "Fonseca, E. and others",
    title = "{Refined Mass and Geometric Measurements of the High-mass PSR J0740+6620}",
    eprint = "2104.00880",
    archivePrefix = "arXiv",
    primaryClass = "astro-ph.HE",
    doi = "10.3847/2041-8213/ac03b8",
    journal = "Astrophys. J. Lett.",
    volume = "915",
    number = "1",
    pages = "L12",
    year = "2021"
}

@article{Dietrich:2020efo,
    author = "Dietrich, Tim and Coughlin, Michael W. and Pang, Peter T. H. and Bulla, Mattia and Heinzel, Jack and Issa, Lina and Tews, Ingo and Antier, Sarah",
    title = "{Multimessenger constraints on the neutron-star equation of state and the Hubble constant}",
    eprint = "2002.11355",
    archivePrefix = "arXiv",
    primaryClass = "astro-ph.HE",
    reportNumber = "LA-UR-20-21470",
    doi = "10.1126/science.abb4317",
    journal = "Science",
    volume = "370",
    number = "6523",
    pages = "1450--1453",
    year = "2020"
}

@article{Karkevandi:2021ygv,
    author = "Karkevandi, Davood Rafiei and Shakeri, Soroush and Sagun, Violetta and Ivanytskyi, Oleksii",
    title = "{Bosonic dark matter in neutron stars and its effect on gravitational wave signal}",
    eprint = "2109.03801",
    archivePrefix = "arXiv",
    primaryClass = "astro-ph.HE",
    doi = "10.1103/PhysRevD.105.023001",
    journal = "Phys. Rev. D",
    volume = "105",
    number = "2",
    pages = "023001",
    year = "2022"
}

@inproceedings{RafieiKarkevandi:2021hcc,
    author = "Rafiei Karkevandi, Davood and Shakeri, Soroush and Sagun, Violetta and Ivanytskyi, Oleksii",
    title = "{Tidal deformability as a probe of dark matter in neutron stars}",
    booktitle = "{16th Marcel Grossmann Meeting on~Recent Developments in Theoretical and Experimental General Relativity, Astrophysics and Relativistic Field Theories}",
    eprint = "2112.14231",
    archivePrefix = "arXiv",
    primaryClass = "astro-ph.HE",
    doi = "10.1142/9789811269776_0307",
    month = "12",
    year = "2021"
}

@article{Shakeri:2022dwg,
    author = "Shakeri, Soroush and Karkevandi, Davood Rafiei",
    title = "{Bosonic dark matter in light of the NICER precise mass-radius measurements}",
    eprint = "2210.17308",
    archivePrefix = "arXiv",
    primaryClass = "astro-ph.HE",
    doi = "10.1103/PhysRevD.109.043029",
    journal = "Phys. Rev. D",
    volume = "109",
    number = "4",
    pages = "043029",
    year = "2024"
}

@article{Ellis:2018bkr,
    author = {Ellis, John and H\"utsi, Gert and Kannike, Kristjan and Marzola, Luca and Raidal, Martti and Vaskonen, Ville},
    title = "{Dark Matter Effects On Neutron Star Properties}",
    eprint = "1804.01418",
    archivePrefix = "arXiv",
    primaryClass = "astro-ph.CO",
    reportNumber = "CERN-TH-2018-072, KCL-PH-TH/2018-13, KCL-PH-TH-2018-13",
    doi = "10.1103/PhysRevD.97.123007",
    journal = "Phys. Rev. D",
    volume = "97",
    number = "12",
    pages = "123007",
    year = "2018"
}

@article{Nelson:2018xtr,
    author = "Nelson, Ann and Reddy, Sanjay and Zhou, Dake",
    title = "{Dark halos around neutron stars and gravitational waves}",
    eprint = "1803.03266",
    archivePrefix = "arXiv",
    primaryClass = "hep-ph",
    reportNumber = "INT-PUB-18-010",
    doi = "10.1088/1475-7516/2019/07/012",
    journal = "JCAP",
    volume = "07",
    pages = "012",
    year = "2019"
}

@article{Ivanytskyi:2019wxd,
    author = "Ivanytskyi, O. and Sagun, V. and Lopes, I.",
    title = "{Neutron stars: New constraints on asymmetric dark matter}",
    eprint = "1910.09925",
    archivePrefix = "arXiv",
    primaryClass = "astro-ph.HE",
    doi = "10.1103/PhysRevD.102.063028",
    journal = "Phys. Rev. D",
    volume = "102",
    number = "6",
    pages = "063028",
    year = "2020"
}

@article{Shahrbaf:2023uxy,
    author = "Shahrbaf, Mahboubeh",
    title = "{Appearance of sexaquark in the core of neutron stars as a candidate of dark matter}",
    doi = "10.1088/1742-6596/2536/1/012001",
    journal = "J. Phys. Conf. Ser.",
    volume = "2536",
    number = "1",
    pages = "012001",
    year = "2023"
}

@article{Rutherford:2022xeb,
    author = "Rutherford, Nathan and Raaijmakers, Geert and Prescod-Weinstein, Chanda and Watts, Anna",
    title = "{Constraining bosonic asymmetric dark matter with neutron star mass-radius measurements}",
    eprint = "2208.03282",
    archivePrefix = "arXiv",
    primaryClass = "astro-ph.HE",
    doi = "10.1103/PhysRevD.107.103051",
    journal = "Phys. Rev. D",
    volume = "107",
    number = "10",
    pages = "103051",
    year = "2023"
}

@article{Shirke:2023ktu,
    author = {Shirke, Swarnim and Ghosh, Suprovo and Chatterjee, Debarati and Sagunski, Laura and Schaffner-Bielich, J\"urgen},
    title = "{R-modes as a new probe of dark matter in neutron stars}",
    eprint = "2305.05664",
    archivePrefix = "arXiv",
    primaryClass = "astro-ph.HE",
    reportNumber = "LIGO-P2300140",
    doi = "10.1088/1475-7516/2023/12/008",
    journal = "JCAP",
    volume = "12",
    pages = "008",
    year = "2023"
}

@article{Farrar:2002ic,
    author = "Farrar, G. R.",
    editor = "Dvali, G. and Gunzig, E. and Verdaguer, E.",
    title = "{A stable H dibaryon: Dark matter candidate within QCD?}",
    doi = "10.1023/A:1025702431127",
    journal = "Int. J. Theor. Phys.",
    volume = "42",
    pages = "1211--1218",
    year = "2003"
}

@article{Farrar:2018hac,
  author        = {Farrar, Glennys R.},
  title         = {A precision test of the nature of Dark Matter and a probe of the QCD phase transition},
  year          = {2018},
  month         = {May},
  archivePrefix = {arXiv},
  eprint        = {1805.03723},
  primaryClass  = {hep-ph},
  note          = {arXiv:1805.03723},
  url           = {https://arxiv.org/abs/1805.03723}
}

@article{Jaffe:1976yi,
    author = "Jaffe, Robert L.",
    title = "{Perhaps a Stable Dihyperon}",
    reportNumber = "SLAC-PUB-1828",
    doi = "10.1103/PhysRevLett.38.195",
    journal = "Phys. Rev. Lett.",
    volume = "38",
    pages = "195--198",
    year = "1977",
    note = "[Erratum: Phys.Rev.Lett. 38, 617 (1977)]"
}

@article{Shahrbaf:2022upc,
    author = "Shahrbaf, M. and Blaschke, D. and Typel, S. and Farrar, G. R. and Alvarez-Castillo, D. E.",
    title = "{Sexaquark dilemma in neutron stars and its solution by quark deconfinement}",
    eprint = "2202.00652",
    archivePrefix = "arXiv",
    primaryClass = "nucl-th",
    doi = "10.1103/PhysRevD.105.103005",
    journal = "Phys. Rev. D",
    volume = "105",
    number = "10",
    pages = "103005",
    year = "2022"
}

@article{Typel:1999yq,
    author = "Typel, S. and Wolter, H. H.",
    title = "{Relativistic mean field calculations with density dependent meson nucleon coupling}",
    doi = "10.1016/S0375-9474(99)00310-3",
    journal = "Nucl. Phys. A",
    volume = "656",
    pages = "331--364",
    year = "1999"
}

@article{Typel:2005ba,
    author = "Typel, S.",
    title = "{Relativistic model for nuclear matter and atomic nuclei with momentum-dependent self-energies}",
    eprint = "nucl-th/0501056",
    archivePrefix = "arXiv",
    doi = "10.1103/PhysRevC.71.064301",
    journal = "Phys. Rev. C",
    volume = "71",
    pages = "064301",
    year = "2005"
}

@article{Typel:2009sy,
    author = "Typel, S. and Ropke, G. and Klahn, T. and Blaschke, D. and Wolter, H. H.",
    title = "{Composition and thermodynamics of nuclear matter with light clusters}",
    eprint = "0908.2344",
    archivePrefix = "arXiv",
    primaryClass = "nucl-th",
    doi = "10.1103/PhysRevC.81.015803",
    journal = "Phys. Rev. C",
    volume = "81",
    pages = "015803",
    year = "2010"
}

@article{Stone:2019blq,
    author = "Stone, J. R. and Dexheimer, V. and Guichon, P. A M. and Thomas, A. W. and Typel, S.",
    title = "{Equation of state of hot dense hyperonic matter in the Quark\textendash{}Meson-Coupling (QMC-A) model}",
    eprint = "1906.11100",
    archivePrefix = "arXiv",
    primaryClass = "nucl-th",
    reportNumber = "ADP-19-13/T1093",
    doi = "10.1093/mnras/staa4006",
    journal = "Mon. Not. Roy. Astron. Soc.",
    volume = "502",
    number = "3",
    pages = "3476--3490",
    year = "2021"
}

@article{Shahrbaf:2019wex,
    author = "Shahrbaf, M. and Moshfegh, H. R.",
    title = "{Appearance of hyperons in neutron stars within LOCV method}",
    doi = "10.1016/j.aop.2019.01.008",
    journal = "Annals Phys.",
    volume = "402",
    pages = "66--77",
    year = "2019"
}

@article{Shahrbaf:2019vtf,
    author = "Shahrbaf, M. and Blaschke, D. and Grunfeld, A. G. and Moshfegh, H. R.",
    title = "{First-order phase transition from hypernuclear matter to deconfined quark matter obeying new constraints from compact star observations}",
    eprint = "1908.04740",
    archivePrefix = "arXiv",
    primaryClass = "nucl-th",
    reportNumber = "JINR report No. E2-2019-41",
    doi = "10.1103/PhysRevC.101.025807",
    journal = "Phys. Rev. C",
    volume = "101",
    number = "2",
    pages = "025807",
    year = "2020"
}

@article{Shahrbaf:2020uau,
    author = "Shahrbaf, M. and Blaschke, D. and Khanmohamadi, S.",
    title = "{Mixed phase transition from hypernuclear matter to deconfined quark matter fulfilling mass-radius constraints of neutron stars}",
    eprint = "2004.14377",
    archivePrefix = "arXiv",
    primaryClass = "nucl-th",
    doi = "10.1088/1361-6471/abaa9a",
    journal = "J. Phys. G",
    volume = "47",
    number = "11",
    pages = "115201",
    year = "2020"
}

@article{Radzhabov:2010dd,
    author = "Radzhabov, A. E. and Blaschke, D. and Buballa, M. and Volkov, M. K.",
    title = "{Nonlocal PNJL model beyond mean field and the QCD phase transition}",
    eprint = "1012.0664",
    archivePrefix = "arXiv",
    primaryClass = "hep-ph",
    doi = "10.1103/PhysRevD.83.116004",
    journal = "Phys. Rev. D",
    volume = "83",
    pages = "116004",
    year = "2011"
}

@article{Shahrbaf:2021cjz,
    author = "Shahrbaf, Mahboubeh and Anti\'c, Sofija and Ayriyan, A. and Blaschke, David and Grunfeld, Ana Gabriela",
    title = "{Constraining free parameters of a color superconducting nonlocal Nambu\textendash{}Jona-Lasinio model using Bayesian analysis of neutron stars mass and radius measurements}",
    eprint = "2105.00029",
    archivePrefix = "arXiv",
    primaryClass = "nucl-th",
    doi = "10.1103/PhysRevD.107.054011",
    journal = "Phys. Rev. D",
    volume = "107",
    number = "5",
    pages = "054011",
    year = "2023"
}

@article{Bertone:2016nfn,
    author = "Bertone, Gianfranco and Hooper, Dan",
    title = "{History of dark matter}",
    eprint = "1605.04909",
    archivePrefix = "arXiv",
    primaryClass = "astro-ph.CO",
    reportNumber = "FERMILAB-PUB-16-157-A",
    doi = "10.1103/RevModPhys.90.045002",
    journal = "Rev. Mod. Phys.",
    volume = "90",
    number = "4",
    pages = "045002",
    year = "2018"
}

@article{Choudhury:2024xbk,
    author = "Choudhury, Devarshi and others",
    title = "{A NICER View of the Nearest and Brightest Millisecond Pulsar: PSR J0437\textendash{}4715}",
    eprint = "2407.06789",
    archivePrefix = "arXiv",
    primaryClass = "astro-ph.HE",
    doi = "10.3847/2041-8213/ad5a6f",
    journal = "Astrophys. J. Lett.",
    volume = "971",
    number = "1",
    pages = "L20",
    year = "2024"
}

@article{Salmi:2024bss,
    author = "Salmi, Tuomo and others",
    title = "{A NICER View of PSR J1231\ensuremath{-}1411: A Complex Case}",
    eprint = "2409.14923",
    archivePrefix = "arXiv",
    primaryClass = "astro-ph.HE",
    doi = "10.3847/1538-4357/ad81d2",
    journal = "Astrophys. J.",
    volume = "976",
    number = "1",
    pages = "58",
    year = "2024"
}

@misc{dittmann:2024:10215108,
  doi = {10.5281/ZENODO.10215108},
  url = {https://zenodo.org/doi/10.5281/zenodo.10215108},
  author = {Dittmann,  A. J. and Miller,  M. C. and Lamb,  F. K. and Holt,  I. and Chirenti,  C. and Wolff,  M. T. and Bogdanov,  S. and Guillot,  S. and Ho,  W. C. G. and Morsink,  S. M. and Arzoumanian,  Z. and Gendreau,  K. C.},
  title = {Updated NICER PSR J0740+6620 Illinois-Maryland MCMC Samples},
  publisher = {Zenodo},
  year = {2024},
  copyright = {Creative Commons Attribution 4.0 International}
}

@article{choudhury:2024:13766753,
  author       = {Choudhury, Devarshi and
                  Salmi, Tuomo and
                  Serena, Vinciguerra and
                  Riley, Thomas and
                  Kini, Yves and
                  Watts, Anna L. and
                  Dorsman, Bas and
                  Bogdanov, Slavko and
                  Guillot, Sebastien and
                  Ray, Paul S. and
                  Reardon, Daniel and
                  Remillard, Ronald A. and
                  Bilous, Anna and
                  Huppenkothen, Daniela and
                  Lattimer, James and
                  Rutherford, Nathan and
                  Arzoumanian, Zaven and
                  Gendreau, Keith and
                  Morsink, Sharon and
                  Ho, Wynn C. G.},
  title        = {Reproduction package for:  'A NICER View of the
                   Nearest and Brightest Millisecond Pulsar: PSR
                   J0437–4715'
                  },
  month        = {9},
  year         = {2024},
  publisher    = {Zenodo},
  doi          = {10.5281/zenodo.13766753},
  url          = {https://doi.org/10.5281/zenodo.13766753},
}

@article{doroshenko:2023:8232233,
  author       = {Doroshenko, Victor and
                  Suleimanov, Valery F. and
                  Pühlhofer, Gerd and
                  Santangelo, Andrea},
  title        = {MCMC samples for X-ray spectra fits summarised in
                   the paper "A strangely light neutron star"
                  },
  month        = {8},
  year         = {2023},
  publisher    = {Zenodo},
  version      = {V2},
  doi          = {10.5281/zenodo.8232233},
  url          = {https://doi.org/10.5281/zenodo.8232233},
}

@article{Dittmann:2024mbo,
    author = "Dittmann, Alexander J. and others",
    title = "{A More Precise Measurement of the Radius of PSR J0740+6620 Using Updated NICER Data}",
    eprint = "2406.14467",
    archivePrefix = "arXiv",
    primaryClass = "astro-ph.HE",
    doi = "10.3847/1538-4357/ad5f1e",
    journal = "Astrophys. J.",
    volume = "974",
    number = "2",
    pages = "295",
    year = "2024"
}

@article{Ayriyan:2024zfw,
    author = "Ayriyan, Alexander and Blaschke, David and Carlomagno, Juan Pablo and Contrera, Gustavo A. and Grunfeld, Ana Gabriela",
    title = "{Bayesian analysis of hybrid neutron star EOS constraints within an instantaneous nonlocal chiral quark matter model}",
    eprint = "2501.00115",
    archivePrefix = "arXiv",
    primaryClass = "nucl-th",
    doi = "10.3390/universe11050141",
    journal = "Universe",
    volume = "11",
    pages = "141",
    year = "2025"
}

@article{Doroshenko:2022nwp,
    author = {Doroshenko, Victor and Suleimanov, Valery and P\"uhlhofer, Gerd and Santangelo, Andrea},
    title = "{A strangely light neutron star within a supernova remnant}",
    doi = "10.1038/s41550-022-01800-1",
    journal = "Nature Astron.",
    volume = "6",
    number = "12",
    pages = "1444--1451",
    year = "2022"
}

@article{Romani:2022jhd,
    author = "Romani, Roger W. and Kandel, D. and Filippenko, Alexei V. and Brink, Thomas G. and Zheng, WeiKang",
    title = "{PSR J0952\ensuremath{-}0607: The Fastest and Heaviest Known Galactic Neutron Star}",
    eprint = "2207.05124",
    archivePrefix = "arXiv",
    primaryClass = "astro-ph.HE",
    doi = "10.3847/2041-8213/ac8007",
    journal = "Astrophys. J. Lett.",
    volume = "934",
    number = "2",
    pages = "L17",
    year = "2022"
}

@article{Moore:2024mot,
    author = "Moore, Marianne and Slatyer, Tracy R.",
    title = "{Cosmology and terrestrial signals of sexaquark dark matter}",
    eprint = "2403.03972",
    archivePrefix = "arXiv",
    primaryClass = "hep-ph",
    reportNumber = "MIT-CTP/5685",
    doi = "10.1103/PhysRevD.110.023515",
    journal = "Phys. Rev. D",
    volume = "110",
    number = "2",
    pages = "023515",
    year = "2024"
}

@article{Arbey:2021gdg,
    author = "Arbey, A. and Mahmoudi, F.",
    title = "{Dark matter and the early Universe: a review}",
    eprint = "2104.11488",
    archivePrefix = "arXiv",
    primaryClass = "hep-ph",
    reportNumber = "CERN-TH-2021-066",
    doi = "10.1016/j.ppnp.2021.103865",
    journal = "Prog. Part. Nucl. Phys.",
    volume = "119",
    pages = "103865",
    year = "2021"
}

@article{Arcadi:2017kky,
    author = "Arcadi, Giorgio and Dutra, Ma\'\i{}ra and Ghosh, Pradipta and Lindner, Manfred and Mambrini, Yann and Pierre, Mathias and Profumo, Stefano and Queiroz, Farinaldo S.",
    title = "{The waning of the WIMP? A review of models, searches, and constraints}",
    eprint = "1703.07364",
    archivePrefix = "arXiv",
    primaryClass = "hep-ph",
    doi = "10.1140/epjc/s10052-018-5662-y",
    journal = "Eur. Phys. J. C",
    volume = "78",
    number = "3",
    pages = "203",
    year = "2018"
}

@article{Chadha-Day:2021szb,
    author = "Chadha-Day, Francesca and Ellis, John and Marsh, David J. E.",
    title = "{Axion dark matter: What is it and why now?}",
    eprint = "2105.01406",
    archivePrefix = "arXiv",
    primaryClass = "hep-ph",
    reportNumber = "KCL-PH-TH/2021-20, CERN-TH-2021-045, IPPP/20/91",
    doi = "10.1126/sciadv.abj3618",
    journal = "Sci. Adv.",
    volume = "8",
    number = "8",
    pages = "abj3618",
    year = "2022"
}

@article{Spergel:1999mh,
    author = "Spergel, David N. and Steinhardt, Paul J.",
    title = "{Observational evidence for selfinteracting cold dark matter}",
    eprint = "astro-ph/9909386",
    archivePrefix = "arXiv",
    doi = "10.1103/PhysRevLett.84.3760",
    journal = "Phys. Rev. Lett.",
    volume = "84",
    pages = "3760--3763",
    year = "2000"
}

@article{Peter:2012jh,
    author = "Peter, Annika H. G. and Rocha, Miguel and Bullock, James S. and Kaplinghat, Manoj",
    title = "{Cosmological Simulations with Self-Interacting Dark Matter II: Halo Shapes vs. Observations}",
    eprint = "1208.3026",
    archivePrefix = "arXiv",
    primaryClass = "astro-ph.CO",
    reportNumber = "NSF-KITP-12-147",
    doi = "10.1093/mnras/sts535",
    journal = "Mon. Not. Roy. Astron. Soc.",
    volume = "430",
    pages = "105",
    year = "2013"
}

@article{Kaplinghat:2015aga,
    author = "Kaplinghat, Manoj and Tulin, Sean and Yu, Hai-Bo",
    title = "{Dark Matter Halos as Particle Colliders: Unified Solution to Small-Scale Structure Puzzles from Dwarfs to Clusters}",
    eprint = "1508.03339",
    archivePrefix = "arXiv",
    primaryClass = "astro-ph.CO",
    doi = "10.1103/PhysRevLett.116.041302",
    journal = "Phys. Rev. Lett.",
    volume = "116",
    number = "4",
    pages = "041302",
    year = "2016"
}

@article{Bramante:2023djs,
    author = "Bramante, Joseph and Raj, Nirmal",
    title = "{Dark matter in compact stars}",
    eprint = "2307.14435",
    archivePrefix = "arXiv",
    primaryClass = "hep-ph",
    doi = "10.1016/j.physrep.2023.12.001",
    journal = "Phys. Rept.",
    volume = "1052",
    pages = "1--48",
    year = "2024"
}

@article{Grippa:2024ach,
    author = "Grippa, Francesco and Lambiase, Gaetano and Poddar, Tanmay Kumar",
    title = "{Searching for New Physics in an Ultradense Environment: A Review on Dark Matter Admixed Neutron Stars}",
    eprint = "2412.09381",
    archivePrefix = "arXiv",
    primaryClass = "astro-ph.HE",
    doi = "10.3390/universe11030074",
    journal = "Universe",
    volume = "11",
    number = "3",
    pages = "74",
    year = "2025"
}

@article{Liebling:2012fv,
    author = "Liebling, Steven L. and Palenzuela, Carlos",
    title = "{Dynamical boson stars}",
    eprint = "1202.5809",
    archivePrefix = "arXiv",
    primaryClass = "gr-qc",
    doi = "10.1007/s41114-023-00043-4",
    journal = "Living Rev. Rel.",
    volume = "26",
    number = "1",
    pages = "1",
    year = "2023"
}

@article{Visinelli:2021uve,
    author = "Visinelli, Luca",
    title = "{Boson stars and oscillatons: A review}",
    eprint = "2109.05481",
    archivePrefix = "arXiv",
    primaryClass = "gr-qc",
    doi = "10.1142/S0218271821300068",
    journal = "Int. J. Mod. Phys. D",
    volume = "30",
    number = "15",
    pages = "2130006",
    year = "2021"
}

@inproceedings{Baryakhtar:2022hbu,
    author = "Baryakhtar, Masha and others",
    title = "{Dark Matter In Extreme Astrophysical Environments}",
    booktitle = "{Snowmass 2021}",
    eprint = "2203.07984",
    archivePrefix = "arXiv",
    primaryClass = "hep-ph",
    month = "3",
    year = "2022"
}

@article{Narain:2006kx,
    author = "Narain, Gaurav and Schaffner-Bielich, Jurgen and Mishustin, Igor N.",
    title = "{Compact stars made of fermionic dark matter}",
    eprint = "astro-ph/0605724",
    archivePrefix = "arXiv",
    doi = "10.1103/PhysRevD.74.063003",
    journal = "Phys. Rev. D",
    volume = "74",
    pages = "063003",
    year = "2006"
}

@article{Maselli:2017vfi,
    author = "Maselli, Andrea and Pnigouras, Pantelis and Nielsen, Niklas Gr\o{}nlund and Kouvaris, Chris and Kokkotas, Kostas D.",
    title = "{Dark stars: gravitational and electromagnetic observables}",
    eprint = "1704.07286",
    archivePrefix = "arXiv",
    primaryClass = "astro-ph.HE",
    doi = "10.1103/PhysRevD.96.023005",
    journal = "Phys. Rev. D",
    volume = "96",
    number = "2",
    pages = "023005",
    year = "2017"
}

@article{Eby:2015hsq,
    author = "Eby, Joshua and Kouvaris, Chris and Nielsen, Niklas Gr\o{}nlund and Wijewardhana, L. C. R.",
    title = "{Boson Stars from Self-Interacting Dark Matter}",
    eprint = "1511.04474",
    archivePrefix = "arXiv",
    primaryClass = "hep-ph",
    doi = "10.1007/JHEP02(2016)028",
    journal = "JHEP",
    volume = "02",
    pages = "028",
    year = "2016"
}

@article{Pitz:2023ejc,
    author = {Pitz, Sarah Louisa and Schaffner-Bielich, J\"urgen},
    title = "{Generating ultracompact boson stars with modified scalar potentials}",
    eprint = "2308.01254",
    archivePrefix = "arXiv",
    primaryClass = "astro-ph.HE",
    doi = "10.1103/PhysRevD.108.103043",
    journal = "Phys. Rev. D",
    volume = "108",
    number = "10",
    pages = "103043",
    year = "2023"
}

@article{Shakeri:2022usk,
    author = "Shakeri, Soroush and Hajkarim, Fazlollah",
    title = "{Probing axions via light circular polarization and event horizon telescope}",
    eprint = "2209.13572",
    archivePrefix = "arXiv",
    primaryClass = "hep-ph",
    doi = "10.1088/1475-7516/2023/04/017",
    journal = "JCAP",
    volume = "04",
    pages = "017",
    year = "2023"
}

@article{Karkevandi:2024vov,
    author = "Karkevandi, Davood Rafiei and Shahrbaf, Mahboubeh and Shakeri, Soroush and Typel, Stefan",
    title = "{Exploring the Distribution and Impact of Bosonic Dark Matter in Neutron Stars}",
    eprint = "2402.18696",
    archivePrefix = "arXiv",
    primaryClass = "astro-ph.HE",
    doi = "10.3390/particles7010011",
    journal = "Particles",
    volume = "7",
    number = "1",
    pages = "201--213",
    year = "2024"
}

@article{Ryan:2022hku,
    author = "Ryan, Michael and Radice, David",
    title = "{Exotic compact objects: The dark white dwarf}",
    eprint = "2201.05626",
    archivePrefix = "arXiv",
    primaryClass = "astro-ph.HE",
    doi = "10.1103/PhysRevD.105.115034",
    journal = "Phys. Rev. D",
    volume = "105",
    number = "11",
    pages = "115034",
    year = "2022"
}

@article{Sedaghat:2025dzt,
    author = "Sedaghat, J. and Bordbar, G. H. and Haghighat, M. and Zebarjad, S. M.",
    title = "{Influence of dark matter on the structure of strange quark stars in one-fluid model}",
    eprint = "2502.16317",
    archivePrefix = "arXiv",
    primaryClass = "hep-ph",
    doi = "10.1140/epjc/s10052-025-14023-1",
    journal = "Eur. Phys. J. C",
    volume = "85",
    number = "3",
    pages = "283",
    year = "2025"
}

@article{Dengler:2021qcq,
    author = {Dengler, Yannick and Schaffner-Bielich, J\"urgen and Tolos, Laura},
    title = "{Second Love number of dark compact planets and neutron stars with dark matter}",
    eprint = "2111.06197",
    archivePrefix = "arXiv",
    primaryClass = "astro-ph.HE",
    doi = "10.1103/PhysRevD.105.043013",
    journal = "Phys. Rev. D",
    volume = "105",
    number = "4",
    pages = "043013",
    year = "2022"
}

@article{Thakur:2023aqm,
    author = "Thakur, Prashant and Malik, Tuhin and Das, Arpan and Jha, T. K. and Provid\^encia, Constan\c{c}a",
    title = "{Exploring robust correlations between fermionic dark matter model parameters and neutron star properties: A two-fluid perspective}",
    eprint = "2308.00650",
    archivePrefix = "arXiv",
    primaryClass = "hep-ph",
    doi = "10.1103/PhysRevD.109.043030",
    journal = "Phys. Rev. D",
    volume = "109",
    number = "4",
    pages = "043030",
    year = "2024"
}

@article{Konstantinou:2024ynd,
    author = "Konstantinou, Andreas",
    title = "{The Effect of a Dark Matter Core on the Structure of a Rotating Neutron Star}",
    eprint = "2405.01487",
    archivePrefix = "arXiv",
    primaryClass = "astro-ph.HE",
    doi = "10.3847/1538-4357/ad4701",
    journal = "Astrophys. J.",
    volume = "968",
    number = "2",
    pages = "83",
    year = "2024"
}

@article{Biesdorf:2024dor,
    author = {Biesdorf, Carline and Schaffner-Bielich, J\"urgen and Tolos, Laura},
    title = "{Masquerading hybrid stars with dark matter}",
    eprint = "2412.05207",
    archivePrefix = "arXiv",
    primaryClass = "hep-ph",
    doi = "10.1103/PhysRevD.111.083038",
    journal = "Phys. Rev. D",
    volume = "111",
    number = "8",
    pages = "083038",
    year = "2025"
}

@article{Giangrandi:2022wht,
    author = "Giangrandi, Edoardo and Sagun, Violetta and Ivanytskyi, Oleksii and Provid\^encia, Constan\c{c}a and Dietrich, Tim",
    title = "{The Effects of Self-interacting Bosonic Dark Matter on Neutron Star Properties}",
    eprint = "2209.10905",
    archivePrefix = "arXiv",
    primaryClass = "astro-ph.HE",
    doi = "10.3847/1538-4357/ace104",
    journal = "Astrophys. J.",
    volume = "953",
    number = "1",
    pages = "115",
    year = "2023"
}

@article{Diedrichs:2023trk,
    author = {Diedrichs, Robin Fynn and Becker, Niklas and Jockel, C\'edric and Christian, Jan-Erik and Sagunski, Laura and Schaffner-Bielich, J\"urgen},
    title = "{Tidal deformability of fermion-boson stars: Neutron stars admixed with ultralight dark matter}",
    eprint = "2303.04089",
    archivePrefix = "arXiv",
    primaryClass = "gr-qc",
    doi = "10.1103/PhysRevD.108.064009",
    journal = "Phys. Rev. D",
    volume = "108",
    number = "6",
    pages = "064009",
    year = "2023"
}

@article{Sedrakian:2015krq,
    author = "Sedrakian, Armen",
    title = "{Axion cooling of neutron stars}",
    eprint = "1512.07828",
    archivePrefix = "arXiv",
    primaryClass = "astro-ph.HE",
    doi = "10.1103/PhysRevD.93.065044",
    journal = "Phys. Rev. D",
    volume = "93",
    number = "6",
    pages = "065044",
    year = "2016"
}

@article{Sedrakian:2018kdm,
    author = "Sedrakian, Armen",
    title = "{Axion cooling of neutron stars. II. Beyond hadronic axions}",
    eprint = "1810.00190",
    archivePrefix = "arXiv",
    primaryClass = "astro-ph.HE",
    doi = "10.1103/PhysRevD.99.043011",
    journal = "Phys. Rev. D",
    volume = "99",
    number = "4",
    pages = "043011",
    year = "2019"
}

@article{Avila:2023rzj,
    author = "\'Avila, Afonso and Giangrandi, Edoardo and Sagun, Violetta and Ivanytskyi, Oleksii and Provid\^encia, Constan\c{c}a",
    title = "{Rapid neutron star cooling triggered by dark matter}",
    eprint = "2309.03894",
    archivePrefix = "arXiv",
    primaryClass = "astro-ph.HE",
    doi = "10.1093/mnras/stae337",
    journal = "Mon. Not. Roy. Astron. Soc.",
    volume = "528",
    number = "4",
    pages = "6319--6328",
    year = "2024"
}

@article{Sagun:2023rzp,
    author = "Sagun, Violetta and Giangrandi, Edoardo and Dietrich, Tim and Ivanytskyi, Oleksii and Negreiros, Rodrigo and Provid\^encia, Constan\c{c}a",
    title = "{What Is the Nature of the HESS J1731-347 Compact Object?}",
    eprint = "2306.12326",
    archivePrefix = "arXiv",
    primaryClass = "astro-ph.HE",
    doi = "10.3847/1538-4357/acfc9e",
    journal = "Astrophys. J.",
    volume = "958",
    number = "1",
    pages = "49",
    year = "2023"
}

@article{Pitz:2024xvh,
    author = {Pitz, Sarah Louisa and Schaffner-Bielich, J\"urgen},
    title = "{Generating ultracompact neutron stars with bosonic dark matter}",
    eprint = "2408.13157",
    archivePrefix = "arXiv",
    primaryClass = "astro-ph.HE",
    doi = "10.1103/PhysRevD.111.043050",
    journal = "Phys. Rev. D",
    volume = "111",
    number = "4",
    pages = "043050",
    year = "2025"
}

@article{Yang:2024ycl,
    author = "Yang, Shu-Hua and Pi, Chun-Mei and Weber, Fridolin",
    title = "{Strange stars admixed with mirror dark matter: Confronting observations of XTE J1814-338}",
    eprint = "2409.15969",
    archivePrefix = "arXiv",
    primaryClass = "astro-ph.HE",
    doi = "10.1103/PhysRevD.111.043037",
    journal = "Phys. Rev. D",
    volume = "111",
    number = "4",
    pages = "043037",
    year = "2025"
}

@article{Lopes:2024ixl,
    author = "Lopes, Luiz L. and Issifu, Adamu",
    title = "{XTE J1814-338 as a dark matter admixed neutron star}",
    eprint = "2411.17105",
    archivePrefix = "arXiv",
    primaryClass = "astro-ph.HE",
    doi = "10.1016/j.dark.2025.101922",
    journal = "Phys. Dark Univ.",
    volume = "48",
    pages = "101922",
    year = "2025"
}

@article{Barbat:2024yvi,
    author = {Barbat, Mikel F. and Schaffner-Bielich, J\"urgen and Tolos, Laura},
    title = "{Comprehensive study of compact stars with dark matter}",
    eprint = "2404.12875",
    archivePrefix = "arXiv",
    primaryClass = "astro-ph.HE",
    doi = "10.1103/PhysRevD.110.023013",
    journal = "Phys. Rev. D",
    volume = "110",
    number = "2",
    pages = "023013",
    year = "2024"
}

@article{Shirke:2024ymc,
    author = {Shirke, Swarnim and Pradhan, Bikram Keshari and Chatterjee, Debarati and Sagunski, Laura and Schaffner-Bielich, J\"urgen},
    title = "{Effects of dark matter on f-mode oscillations of neutron stars}",
    eprint = "2403.18740",
    archivePrefix = "arXiv",
    primaryClass = "gr-qc",
    reportNumber = "LIGO-P2400135",
    doi = "10.1103/PhysRevD.110.063025",
    journal = "Phys. Rev. D",
    volume = "110",
    number = "6",
    pages = "063025",
    year = "2024"
}

@article{Thakur:2024btu,
    author = "Thakur, Prashant and Malik, Tuhin and Das, Arpan and Jha, T. K. and Sharma, B. K. and Provid\^encia, Constan\c{c}a",
    title = "{Feasibility of dark matter admixed neutron star based on recent observational constraints}",
    eprint = "2408.03780",
    archivePrefix = "arXiv",
    primaryClass = "nucl-th",
    month = "8",
    year = "2024"
}

@inbook{Blaschke:2022knl,
    author = "Blaschke, D. and Ivanytskyi, O. and Shahrbaf, M.",
    title = "{Quark deconfinement in compact stars through sexaquark condensation}",
    eprint = "2202.05061",
    archivePrefix = "arXiv",
    primaryClass = "nucl-th",
    doi = "10.1142/9789811220913_0008",
    month = "2",
    year = "2022"
}

@article{Issifu:2024htq,
    author = "Issifu, Adamu and Thakur, Prashant and da Silva, Franciele M. and Marquez, Kau D. and Menezes, D\'ebora P. and Dutra, M. and Louren\c{c}o, O. and Frederico, Tobias",
    title = "{Supernova remnants with mirror dark matter and hyperons}",
    eprint = "2412.17946",
    archivePrefix = "arXiv",
    primaryClass = "hep-ph",
    doi = "10.1103/PhysRevD.111.083026",
    journal = "Phys. Rev. D",
    volume = "111",
    number = "8",
    pages = "083026",
    year = "2025"
}

@article{Rutherford:2024bli,
    author = "Rutherford, Nathan and Prescod-Weinstein, Chanda and Watts, Anna",
    title = "{Probing fermionic asymmetric dark matter cores using global neutron star properties}",
    eprint = "2410.00140",
    archivePrefix = "arXiv",
    primaryClass = "astro-ph.HE",
    month = "9",
    year = "2024"
}

@article{Shawqi:2024jmk,
    author = "Shawqi, Shafayat and Morsink, Sharon M.",
    title = "{Interpreting Mass and Radius Measurements of Neutron Stars with Dark Matter Halos}",
    eprint = "2406.03332",
    archivePrefix = "arXiv",
    primaryClass = "astro-ph.HE",
    reportNumber = "INT-PUB-24-025",
    doi = "10.3847/1538-4357/ad77c1",
    journal = "Astrophys. J.",
    volume = "975",
    number = "1",
    pages = "123",
    year = "2024"
}

@article{Hajkarim:2024ecp,
    author = {Hajkarim, Fazlollah and Schaffner-Bielich, J\"urgen and Tolos, Laura},
    title = "{Thermodynamic Consistent Description of Compact Stars of Two Interacting Fluids: The Case of Neutron Stars with Higgs Portal Dark Matter}",
    eprint = "2412.04585",
    archivePrefix = "arXiv",
    primaryClass = "hep-ph",
    month = "12",
    year = "2024"
}

@article{Kumar:2025ytm,
    author = "Kumar, Ankit and Sotani, Hajime",
    title = "{Impact of dark matter distribution on neutron star properties}",
    eprint = "2501.07052",
    archivePrefix = "arXiv",
    primaryClass = "astro-ph.HE",
    reportNumber = "RIKEN-iTHEMS-Report-25",
    doi = "10.1103/PhysRevD.111.043016",
    journal = "Phys. Rev. D",
    volume = "111",
    number = "4",
    pages = "043016",
    year = "2025"
}

@article{Kodama:1994np,
    author = "Kodama, Nobuaki and Oka, Makoto and Hatsuda, Tetsuo",
    title = "{H dibaryon in the QCD sum rule}",
    eprint = "hep-ph/9404221",
    archivePrefix = "arXiv",
    reportNumber = "TIT-HEP-249-NP",
    doi = "10.1016/0375-9474(94)90908-3",
    journal = "Nucl. Phys. A",
    volume = "580",
    pages = "445--454",
    year = "1994"
}

@article{Azizi:2019xla,
    author = "Azizi, K. and Agaev, S. S. and Sundu, H.",
    title = "{The Scalar Hexaquark $uuddss$: a Candidate to Dark Matter?}",
    eprint = "1904.09913",
    archivePrefix = "arXiv",
    primaryClass = "hep-ph",
    doi = "10.1088/1361-6471/ab9a0e",
    journal = "J. Phys. G",
    volume = "47",
    number = "9",
    pages = "095001",
    year = "2020"
}

@article{Gross:2018ivp,
    author = "Gross, Christian and Polosa, Antonello and Strumia, Alessandro and Urbano, Alfredo and Xue, Wei",
    title = "{Dark Matter in the Standard Model?}",
    eprint = "1803.10242",
    archivePrefix = "arXiv",
    primaryClass = "hep-ph",
    reportNumber = "CERN-TH-2018-065",
    doi = "10.1103/PhysRevD.98.063005",
    journal = "Phys. Rev. D",
    volume = "98",
    number = "6",
    pages = "063005",
    year = "2018"
}

@article{Evans:2023zde,
    author = "Evans, Nick and Ward, Matthew",
    title = "{Running anomalous dimensions in holographic QCD: From the proton to the sexaquark}",
    eprint = "2304.10816",
    archivePrefix = "arXiv",
    primaryClass = "hep-ph",
    doi = "10.1103/PhysRevD.108.026018",
    journal = "Phys. Rev. D",
    volume = "108",
    number = "2",
    pages = "026018",
    year = "2023"
}

@article{Inoue:2010es,
    author = "Inoue, Takashi and Ishii, Noriyoshi and Aoki, Sinya and Doi, Takumi and Hatsuda, Tetsuo and Ikeda, Yoichi and Murano, Keiko and Nemura, Hidekatsu and Sasaki, Kenji",
    collaboration = "HAL QCD",
    title = "{Bound H-dibaryon in Flavor SU(3) Limit of Lattice QCD}",
    eprint = "1012.5928",
    archivePrefix = "arXiv",
    primaryClass = "hep-lat",
    doi = "10.1103/PhysRevLett.106.162002",
    journal = "Phys. Rev. Lett.",
    volume = "106",
    pages = "162002",
    year = "2011"
}

@article{NPLQCD:2010ocs,
    author = "Beane, S. R. and others",
    collaboration = "NPLQCD",
    title = "{Evidence for a Bound H-dibaryon from Lattice QCD}",
    eprint = "1012.3812",
    archivePrefix = "arXiv",
    primaryClass = "hep-lat",
    reportNumber = "JLAB-THY-10-1296",
    doi = "10.1103/PhysRevLett.106.162001",
    journal = "Phys. Rev. Lett.",
    volume = "106",
    pages = "162001",
    year = "2011"
}

@article{Green:2021qol,
    author = "Green, Jeremy R. and Hanlon, Andrew D. and Junnarkar, Parikshit M. and Wittig, Hartmut",
    title = "{Weakly bound $H$ dibaryon from SU(3)-flavor-symmetric QCD}",
    eprint = "2103.01054",
    archivePrefix = "arXiv",
    primaryClass = "hep-lat",
    reportNumber = "MITP-21-009, CERN-TH-2021-024",
    doi = "10.1103/PhysRevLett.127.242003",
    journal = "Phys. Rev. Lett.",
    volume = "127",
    number = "24",
    pages = "242003",
    year = "2021"
}

@article{Shanahan:2011su,
    author = "Shanahan, P. E. and Thomas, A. W. and Young, R. D.",
    title = "{Mass of the H-dibaryon}",
    eprint = "1106.2851",
    archivePrefix = "arXiv",
    primaryClass = "nucl-th",
    reportNumber = "ADP-10-22-T744",
    doi = "10.1103/PhysRevLett.107.092004",
    journal = "Phys. Rev. Lett.",
    volume = "107",
    pages = "092004",
    year = "2011"
}

@article{Takahashi:2001nm,
    author = "Takahashi, H. and others",
    title = "{Observation of a (Lambda Lambda)He-6 double hypernucleus}",
    doi = "10.1103/PhysRevLett.87.212502",
    journal = "Phys. Rev. Lett.",
    volume = "87",
    pages = "212502",
    year = "2001"
}

@article{Farrar:2023wta,
    author = "Farrar, Glennys R. and Wang, Zihui",
    title = "{Constraints on long-lived di-baryons and di-baryonic dark matter}",
    eprint = "2306.03123",
    archivePrefix = "arXiv",
    primaryClass = "hep-ph",
    month = "6",
    year = "2023"
}

@article{ALICE:2022sco,
    author = "Acharya, Shreyasi and others",
    collaboration = "ALICE",
    title = "{Measurement of the Lifetime and \ensuremath{\Lambda} Separation Energy of H\ensuremath{\Lambda}3}",
    eprint = "2209.07360",
    archivePrefix = "arXiv",
    primaryClass = "nucl-ex",
    reportNumber = "CERN-EP-2022-188",
    doi = "10.1103/PhysRevLett.131.102302",
    journal = "Phys. Rev. Lett.",
    volume = "131",
    number = "10",
    pages = "102302",
    year = "2023"
}

@article{ALargeIonColliderExperiment:2021puh,
    author = "Acharya, Shreyasi and others",
    collaboration = "A Large Ion Collider Experiment, ALICE",
    title = "{Hypertriton Production in p-Pb Collisions at $\sqrt {s_{NN}}$=5.02\,\,TeV}",
    eprint = "2107.10627",
    archivePrefix = "arXiv",
    primaryClass = "nucl-ex",
    reportNumber = "CERN-EP-2021-139",
    doi = "10.1103/PhysRevLett.128.252003",
    journal = "Phys. Rev. Lett.",
    volume = "128",
    number = "25",
    pages = "252003",
    year = "2022"
}

@article{STAR:2010gyg,
    author = "Abelev, B. I. and others",
    collaboration = "STAR",
    title = "{Observation of an Antimatter Hypernucleus}",
    eprint = "1003.2030",
    archivePrefix = "arXiv",
    primaryClass = "nucl-ex",
    doi = "10.1126/science.1183980",
    journal = "Science",
    volume = "328",
    pages = "58--62",
    year = "2010"
}

@article{Rappold:2015una,
    author = "Rappold, C. and others",
    title = "{Hypernuclear production cross section in the reaction of $^6Li$ + $^{12}C$ at 2A GeV}",
    doi = "10.1016/j.physletb.2015.05.059",
    journal = "Phys. Lett. B",
    volume = "747",
    pages = "129--134",
    year = "2015"
}

@article{Leung:2022flt,
    author = "Leung, Yue-Hang",
    title = "{Hypernuclei and Antihypernuclei Production in Heavy-Ion Collisions}",
    eprint = "2204.01393",
    archivePrefix = "arXiv",
    primaryClass = "nucl-ex",
    doi = "10.1051/epjconf/202225908001",
    journal = "EPJ Web Conf.",
    volume = "259",
    pages = "08001",
    year = "2022"
}

@article{Bauswein:2018bma,
    author = "Bauswein, Andreas and Bastian, Niels-Uwe F. and Blaschke, David B. and Chatziioannou, Katerina and Clark, James A. and Fischer, Tobias and Oertel, Micaela",
    title = "{Identifying a first-order phase transition in neutron star mergers through gravitational waves}",
    eprint = "1809.01116",
    archivePrefix = "arXiv",
    primaryClass = "astro-ph.HE",
    doi = "10.1103/PhysRevLett.122.061102",
    journal = "Phys. Rev. Lett.",
    volume = "122",
    number = "6",
    pages = "061102",
    year = "2019"
}

@article{Most:2018eaw,
    author = {Most, Elias R. and Papenfort, L. Jens and Dexheimer, Veronica and Hanauske, Matthias and Schramm, Stefan and St\"ocker, Horst and Rezzolla, Luciano},
    title = "{Signatures of quark-hadron phase transitions in general-relativistic neutron-star mergers}",
    eprint = "1807.03684",
    archivePrefix = "arXiv",
    primaryClass = "astro-ph.HE",
    doi = "10.1103/PhysRevLett.122.061101",
    journal = "Phys. Rev. Lett.",
    volume = "122",
    number = "6",
    pages = "061101",
    year = "2019"
}

@article{Weih:2019xvw,
    author = "Weih, Lukas R. and Hanauske, Matthias and Rezzolla, Luciano",
    title = "{Postmerger Gravitational-Wave Signatures of Phase Transitions in Binary Mergers}",
    eprint = "1912.09340",
    archivePrefix = "arXiv",
    primaryClass = "gr-qc",
    doi = "10.1103/PhysRevLett.124.171103",
    journal = "Phys. Rev. Lett.",
    volume = "124",
    number = "17",
    pages = "171103",
    year = "2020"
}

@inbook{Bauswein:2022vtq,
    author = "Bauswein, Andreas and Blaschke, David and Fischer, Tobias",
    title = "{Effects of a strong phase transition on supernova explosions, compact stars and their mergers}",
    eprint = "2203.17188",
    archivePrefix = "arXiv",
    primaryClass = "nucl-th",
    doi = "10.1142/9789811220944_0008",
    month = "3",
    year = "2022"
}

@article{Largani:2023kjx,
    author = "Largani, Noshad Khosravi and Fischer, Tobias and Shibagaki, Shota and Cerd\'a-Dur\'an, Pablo and Torres-Forn\'e, Alejandro",
    title = "{Neutron stars in accreting systems \textendash{} Signatures of the QCD phase transition}",
    eprint = "2311.15992",
    archivePrefix = "arXiv",
    primaryClass = "astro-ph.HE",
    doi = "10.1051/0004-6361/202348742",
    journal = "Astron. Astrophys.",
    volume = "687",
    pages = "A245",
    year = "2024"
}

@article{Satz:1998kg,
    author = "Satz, Helmut",
    editor = "Karsch, F. and Lombardo, M. P.",
    title = "{Deconfinement and percolation}",
    eprint = "hep-ph/9805418",
    archivePrefix = "arXiv",
    reportNumber = "BI-TP-98-11",
    doi = "10.1016/S0375-9474(98)00508-9",
    journal = "Nucl. Phys. A",
    volume = "642",
    pages = "130--142",
    year = "1998"
}

@article{Magas:2003wi,
    author = "Magas, V. and Satz, H.",
    title = "{Conditions for confinement and freezeout}",
    eprint = "hep-ph/0308155",
    archivePrefix = "arXiv",
    reportNumber = "BI-TP-2003-20",
    doi = "10.1140/epjc/s2003-01375-1",
    journal = "Eur. Phys. J. C",
    volume = "32",
    pages = "115--119",
    year = "2003"
}

@article{Castorina:2008vu,
    author = "Castorina, P. and Redlich, K. and Satz, H.",
    title = "{The Phase Diagram of Hadronic Matter}",
    eprint = "0807.4469",
    archivePrefix = "arXiv",
    primaryClass = "hep-ph",
    reportNumber = "BI-TP-2008-17",
    doi = "10.1140/epjc/s10052-008-0795-z",
    journal = "Eur. Phys. J. C",
    volume = "59",
    pages = "67--73",
    year = "2009"
}

@article{Braun-Munzinger:2014lba,
    author = "Braun-Munzinger, Peter and Kalweit, Alexander and Redlich, Krzysztof and Stachel, Johanna",
    title = "{Confronting fluctuations of conserved charges in central nuclear collisions at the LHC with predictions from Lattice QCD}",
    eprint = "1412.8614",
    archivePrefix = "arXiv",
    primaryClass = "hep-ph",
    doi = "10.1016/j.physletb.2015.05.077",
    journal = "Phys. Lett. B",
    volume = "747",
    pages = "292--298",
    year = "2015"
}

@article{Fukushima:2020cmk,
    author = "Fukushima, Kenji and Kojo, Toru and Weise, Wolfram",
    title = "{Hard-core deconfinement and soft-surface delocalization from nuclear to quark matter}",
    eprint = "2008.08436",
    archivePrefix = "arXiv",
    primaryClass = "hep-ph",
    doi = "10.1103/PhysRevD.102.096017",
    journal = "Phys. Rev. D",
    volume = "102",
    number = "9",
    pages = "096017",
    year = "2020"
}

@article{Marczenko:2022jhl,
    author = "Marczenko, Micha\l{} and McLerran, Larry and Redlich, Krzysztof and Sasaki, Chihiro",
    title = "{Reaching percolation and conformal limits in neutron stars}",
    eprint = "2207.13059",
    archivePrefix = "arXiv",
    primaryClass = "nucl-th",
    doi = "10.1103/PhysRevC.107.025802",
    journal = "Phys. Rev. C",
    volume = "107",
    number = "2",
    pages = "025802",
    year = "2023"
}

@article{Schafer:1998ef,
    author = {Sch\"afer, Thomas and Wilczek, Frank},
    title = "{Continuity of quark and hadron matter}",
    eprint = "hep-ph/9811473",
    archivePrefix = "arXiv",
    reportNumber = "IASSNS-HEP-98-100",
    doi = "10.1103/PhysRevLett.82.3956",
    journal = "Phys. Rev. Lett.",
    volume = "82",
    pages = "3956--3959",
    year = "1999"
}

@article{Hirono:2018fjr,
    author = "Hirono, Yuji and Tanizaki, Yuya",
    title = "{Quark-Hadron Continuity beyond the Ginzburg-Landau Paradigm}",
    eprint = "1811.10608",
    archivePrefix = "arXiv",
    primaryClass = "hep-th",
    reportNumber = "RBRC-1295",
    doi = "10.1103/PhysRevLett.122.212001",
    journal = "Phys. Rev. Lett.",
    volume = "122",
    number = "21",
    pages = "212001",
    year = "2019"
}

@article{Fujimoto:2019sxg,
    author = "Fujimoto, Yuki and Fukushima, Kenji and Weise, Wolfram",
    title = "{Continuity from neutron matter to two-flavor quark matter with $^1 S_0$ and $^3 P_2$ superfluidity}",
    eprint = "1908.09360",
    archivePrefix = "arXiv",
    primaryClass = "hep-ph",
    doi = "10.1103/PhysRevD.101.094009",
    journal = "Phys. Rev. D",
    volume = "101",
    number = "9",
    pages = "094009",
    year = "2020"
}

@article{Baym:2017whm,
    author = "Baym, Gordon and Hatsuda, Tetsuo and Kojo, Toru and Powell, Philip D. and Song, Yifan and Takatsuka, Tatsuyuki",
    title = "{From hadrons to quarks in neutron stars: a review}",
    eprint = "1707.04966",
    archivePrefix = "arXiv",
    primaryClass = "astro-ph.HE",
    reportNumber = "RIKEN-ITHEMS-REPORT-17, RIKEN-QHP-316, RIKEN-iTHEMS-Report-17",
    doi = "10.1088/1361-6633/aaae14",
    journal = "Rept. Prog. Phys.",
    volume = "81",
    number = "5",
    pages = "056902",
    year = "2018"
}

@article{Kojo:2021hqh,
    author = "Kojo, Toru and Suenaga, Daiki",
    title = "{Peaks of sound velocity in two color dense QCD: Quark saturation effects and semishort range correlations}",
    eprint = "2110.02100",
    archivePrefix = "arXiv",
    primaryClass = "hep-ph",
    doi = "10.1103/PhysRevD.105.076001",
    journal = "Phys. Rev. D",
    volume = "105",
    number = "7",
    pages = "076001",
    year = "2022"
}

@article{Kojo:2020krb,
    author = "Kojo, Toru",
    title = "{QCD equations of state and speed of sound in neutron stars}",
    eprint = "2011.10940",
    archivePrefix = "arXiv",
    primaryClass = "nucl-th",
    doi = "10.1007/s43673-021-00011-6",
    journal = "AAPPS Bull.",
    volume = "31",
    number = "1",
    pages = "11",
    year = "2021"
}

@article{Hensh:2024onv,
    author = "Hensh, Sudipta and Huang, Yong-Jia and Kojo, Toru and Baiotti, Luca and Takami, Kentaro and Nagataki, Shigehiro and Sotani, Hajime",
    title = "{Neutron-quark stars: Discerning viable alternatives for the higher-density part of the equation of state of compact stars}",
    eprint = "2407.09446",
    archivePrefix = "arXiv",
    primaryClass = "astro-ph.HE",
    reportNumber = "RIKEN-iTHEMS-Report-24",
    month = "7",
    year = "2024"
}

@article{Ayriyan:2017nby,
    author = "Ayriyan, A. and Bastian, N. -U. and Blaschke, D. and Grigorian, H. and Maslov, K. and Voskresensky, D. N.",
    title = "{Robustness of third family solutions for hybrid stars against mixed phase effects}",
    eprint = "1711.03926",
    archivePrefix = "arXiv",
    primaryClass = "nucl-th",
    doi = "10.1103/PhysRevC.97.045802",
    journal = "Phys. Rev. C",
    volume = "97",
    number = "4",
    pages = "045802",
    year = "2018"
}

@article{Ayriyan:2017tvl,
    author = "Ayriyan, Alexander and Grigorian, Hovik",
    editor = "Adam, G. and Bu\v{s}a, J. and Hnati\v{c}, Michal and Podgainy, D.",
    title = "{Model of the Phase Transition Mimicking the Pasta Phase in Cold and Dense Quark-Hadron Matter}",
    eprint = "1710.05637",
    archivePrefix = "arXiv",
    primaryClass = "astro-ph.HE",
    doi = "10.1051/epjconf/201817303003",
    journal = "EPJ Web Conf.",
    volume = "173",
    pages = "03003",
    year = "2018"
}

@article{Antoniadis:2013pzd,
    author = "Antoniadis, John and others",
    title = "{A Massive Pulsar in a Compact Relativistic Binary}",
    eprint = "1304.6875",
    archivePrefix = "arXiv",
    primaryClass = "astro-ph.HE",
    doi = "10.1126/science.1233232",
    journal = "Science",
    volume = "340",
    pages = "6131",
    year = "2013"
}

@article{Kojo:2024ejq,
    author = "Kojo, Toru",
    title = "{Stiffening of matter in quark-hadron continuity: a mini-review}",
    eprint = "2412.20442",
    archivePrefix = "arXiv",
    primaryClass = "nucl-th",
    month = "12",
    year = "2024"
}

@article{Fujimoto:2022xhv,
    author = "Fujimoto, Yuki and Fukushima, Kenji and Hotokezaka, Kenta and Kyutoku, Koutarou",
    title = "{Gravitational Wave Signal for Quark Matter with Realistic Phase Transition}",
    eprint = "2205.03882",
    archivePrefix = "arXiv",
    primaryClass = "astro-ph.HE",
    reportNumber = "INT-PUB-22-015",
    doi = "10.1103/PhysRevLett.130.091404",
    journal = "Phys. Rev. Lett.",
    volume = "130",
    number = "9",
    pages = "091404",
    year = "2023"
}

@article{Visinelli:2017ooc,
    author = "Visinelli, Luca and Baum, Sebastian and Redondo, Javier and Freese, Katherine and Wilczek, Frank",
    title = "{Dilute and dense axion stars}",
    eprint = "1710.08910",
    archivePrefix = "arXiv",
    primaryClass = "astro-ph.CO",
    reportNumber = "MCTP-17-20A, MIT-CTP-4949, NORDITA-2017-112",
    doi = "10.1016/j.physletb.2017.12.010",
    journal = "Phys. Lett. B",
    volume = "777",
    pages = "64--72",
    year = "2018"
}

@article{PhysRev.55.364,
  title = {Static Solutions of Einstein's Field Equations for Spheres of Fluid},
  author = {Tolman, Richard C.},
  journal = {Phys. Rev.},
  volume = {55},
  issue = {4},
  pages = {364--373},
  numpages = {0},
  year = {1939},
  month = {Feb},
  publisher = {American Physical Society},
  doi = {10.1103/PhysRev.55.364},
  url = {https://link.aps.org/doi/10.1103/PhysRev.55.364}
}

@article{PhysRev.55.374,
  title = {On Massive Neutron Cores},
  author = {Oppenheimer, J. R. and Volkoff, G. M.},
  journal = {Phys. Rev.},
  volume = {55},
  issue = {4},
  pages = {374--381},
  numpages = {0},
  year = {1939},
  month = {Feb},
  publisher = {American Physical Society},
  doi = {10.1103/PhysRev.55.374},
  url = {https://link.aps.org/doi/10.1103/PhysRev.55.374}
}

@article{Gal:2024nbr,
    author = "Gal, Avraham",
    title = "{Hypernuclear constraints on the existence and lifetime of a deeply bound H dibaryon}",
    eprint = "2404.12801",
    archivePrefix = "arXiv",
    primaryClass = "nucl-th",
    doi = "10.1016/j.physletb.2024.138973",
    journal = "Phys. Lett. B",
    volume = "857",
    pages = "138973",
    year = "2024"
}

@article{Alvarez-Castillo:2016oln,
    author = "Alvarez-Castillo, D. and Ayriyan, A. and Benic, S. and Blaschke, D. and Grigorian, H. and Typel, S.",
    title = "{New class of hybrid EoS and Bayesian M-R data analysis}",
    eprint = "1603.03457",
    archivePrefix = "arXiv",
    primaryClass = "nucl-th",
    doi = "10.1140/epja/i2016-16069-2",
    journal = "Eur. Phys. J. A",
    volume = "52",
    number = "3",
    pages = "69",
    year = "2016"
}

@article{Horvath:2023uwl,
    author = "Horvath, J. E. and Rocha, L. S. and de S\'a, L. M. and Moraes, P. H. R. S. and Bar\~ao, L. G. and de Avellar, M. G. B. and Bernardo, A. and Bachega, R. R. A.",
    title = "{A light strange star in the remnant HESS J1731\ensuremath{-}347: Minimal consistency checks}",
    eprint = "2303.10264",
    archivePrefix = "arXiv",
    primaryClass = "astro-ph.HE",
    doi = "10.1051/0004-6361/202345885",
    journal = "Astron. Astrophys.",
    volume = "672",
    pages = "L11",
    year = "2023"
}

@article{Yuan:2025mmn,
    author = "Yuan, Ya-Jing and Zhou, Xia",
    title = "{Thermal Evolution of the Central Compact Object in HESS J1731\ensuremath{-}347 as Evidence for a Color-flavor-locked Strange Star}",
    doi = "10.1088/1674-4527/adce4e",
    journal = "Res. Astron. Astrophys.",
    volume = "25",
    number = "5",
    pages = "055016",
    year = "2025"
}

@article{Alvarez-Castillo:2025yzu,
    author = "Alvarez-Castillo, David E.",
    title = "{Properties of the Object HESS J1731-347 as a Twin Compact Star}",
    eprint = "2504.00240",
    archivePrefix = "arXiv",
    primaryClass = "astro-ph.HE",
    month = "3",
    year = "2025"
}

@article{DiClemente:2022wqp,
    author = "Di Clemente, Francesco and Drago, Alessandro and Pagliara, Giuseppe",
    title = "{Is the Compact Object Associated with HESS J1731-347 a Strange Quark Star? A Possible Astrophysical Scenario for Its Formation}",
    eprint = "2211.07485",
    archivePrefix = "arXiv",
    primaryClass = "astro-ph.HE",
    doi = "10.3847/1538-4357/ad445b",
    journal = "Astrophys. J.",
    volume = "967",
    number = "2",
    pages = "159",
    year = "2024"
}

@article{Mauviard:2025dmd,
    author = "Mauviard, Lucien and others",
    title = "{A NICER view of the 1.4 solar-mass edge-on pulsar PSR J0614--3329}",
    eprint = "2506.14883",
    archivePrefix = "arXiv",
    primaryClass = "astro-ph.HE",
    month = "6",
    year = "2025"
}

@article{Typel:2024myq,
    author = "Typel, Stefan and Shlomo, Shalom",
    title = "{Improving relativistic energy density functionals with tensor couplings}",
    eprint = "2408.00425",
    archivePrefix = "arXiv",
    primaryClass = "nucl-th",
    doi = "10.1140/epja/s10050-024-01442-z",
    journal = "Eur. Phys. J. A",
    volume = "60",
    number = "11",
    pages = "236",
    year = "2024"
}

@article{Das:2025pjl,
    author = "Das, H. C. and Burgio, G. F.",
    title = "{Neutron Decay Anomaly and Its Effects on Neutron Star Properties}",
    eprint = "2505.09190",
    archivePrefix = "arXiv",
    primaryClass = "astro-ph.HE",
    doi = "10.3390/universe11050159",
    journal = "Universe",
    volume = "11",
    number = "5",
    year = "2025"
}

@article{Das:2025duq,
    author = "Das, Arijit and Jaikumar, Prashanth and Karekkat, Adarsh and Mandal, Tanumoy",
    title = "{Effects of BHF-corrected effective mass on sound speed, conformality and observables of dark matter admixed neutron stars}",
    eprint = "2506.15328",
    archivePrefix = "arXiv",
    primaryClass = "hep-ph",
    month = "6",
    year = "2025"
}

@article{Klangburam:2025rcb,
    author = "Klangburam, Tanech and Pongkitivanichkul, Chakrit",
    title = "{Axionlike particle mediated dark matter and neutron star properties in the quantum hadrodynamics model}",
    eprint = "2503.12430",
    archivePrefix = "arXiv",
    primaryClass = "astro-ph.HE",
    doi = "10.1103/PhysRevD.111.103031",
    journal = "Phys. Rev. D",
    volume = "111",
    number = "10",
    pages = "103031",
    year = "2025"
}

@article{Lenzi:2022ypb,
    author = "Lenzi, C\'esar H. and Dutra, Mariana and Louren\c{c}o, Odilon and Lopes, Luiz L. and Menezes, D\'ebora P.",
    title = "{Dark matter effects on hybrid star properties}",
    eprint = "2212.12615",
    archivePrefix = "arXiv",
    primaryClass = "hep-ph",
    doi = "10.1140/epjc/s10052-023-11416-y",
    journal = "Eur. Phys. J. C",
    volume = "83",
    number = "3",
    pages = "266",
    year = "2023"
}

@article{Routaray:2024fcq,
    author = "Routaray, Pinku and Parmar, Vishal and Das, H. C. and Kumar, Bharat and Burgio, G. F. and Schulze, H. -J.",
    title = "{Effects of asymmetric dark matter on a magnetized neutron star: A two-fluid approach}",
    eprint = "2412.21097",
    archivePrefix = "arXiv",
    primaryClass = "nucl-th",
    doi = "10.1103/PhysRevD.111.103045",
    journal = "Phys. Rev. D",
    volume = "111",
    number = "10",
    pages = "103045",
    year = "2025"
}

@article{Marzola:2024ame,
    author = "Marzola, Isabella and Rodrigues, Everson H. and Coelho, Anderson F. and Louren\c{c}o, Odilon",
    title = "{Strange stars admixed with dark matter: Equiparticle model in a two fluid approach}",
    eprint = "2408.16583",
    archivePrefix = "arXiv",
    primaryClass = "astro-ph.HE",
    doi = "10.1103/PhysRevD.111.063076",
    journal = "Phys. Rev. D",
    volume = "111",
    number = "6",
    pages = "063076",
    year = "2025"
}

@article{Cipriani:2025tga,
    author = "Cipriani, Lorenzo and Giangrandi, Edoardo and Sagun, Violetta and Doneva, Daniela D. and Yazadjiev, Stoytcho S.",
    title = "{Rapidly spinning dark matter-admixed neutron stars}",
    eprint = "2502.17948",
    archivePrefix = "arXiv",
    primaryClass = "astro-ph.HE",
    doi = "10.1103/qcl7-m5kf",
    journal = "Phys. Rev. D",
    volume = "111",
    number = "12",
    pages = "123005",
    year = "2025"
}

@article{Guha:2024pnn,
    author = "Guha, Atanu and Sen, Debashree",
    title = "{Constraining the mass of fermionic dark matter from its feeble interaction with hadronic matter via dark mediators in neutron stars}",
    eprint = "2401.14419",
    archivePrefix = "arXiv",
    primaryClass = "astro-ph.HE",
    doi = "10.1103/PhysRevD.109.043038",
    journal = "Phys. Rev. D",
    volume = "109",
    number = "4",
    pages = "043038",
    year = "2024"
}

@article{Dengler:2025ntz,
    author = "Dengler, Yannick and Kulkarni, Suchita and Maas, Axel and Radl, Kevin",
    title = "{Strongly Interacting Dark Matter admixed Neutron Stars}",
    eprint = "2503.19691",
    archivePrefix = "arXiv",
    primaryClass = "hep-ph",
    month = "3",
    year = "2025"
}

@article{Vinciguerra:2023qxq,
    author = "Vinciguerra, Serena and others",
    title = "{An Updated Mass\textendash{}Radius Analysis of the 2017\textendash{}2018 NICER Data Set of PSR J0030+0451}",
    eprint = "2308.09469",
    archivePrefix = "arXiv",
    primaryClass = "astro-ph.HE",
    doi = "10.3847/1538-4357/acfb83",
    journal = "Astrophys. J.",
    volume = "961",
    number = "1",
    pages = "62",
    year = "2024"
}

@article{Collier:2022cpr,
    author = "Collier, Michael and Croon, Djuna and Leane, Rebecca K.",
    title = "{Tidal Love numbers of novel and admixed celestial objects}",
    eprint = "2205.15337",
    archivePrefix = "arXiv",
    primaryClass = "gr-qc",
    reportNumber = "SLAC-PUB-17679, IPPP/22/35",
    doi = "10.1103/PhysRevD.106.123027",
    journal = "Phys. Rev. D",
    volume = "106",
    number = "12",
    pages = "123027",
    year = "2022"
}

@article{Liu:2025cwy,
    author = "Liu, Xue-Zhi and Mahapatra, Premachand and Huang, Chun and Hazarika, Ayush and Singha, Chiranjeeb and Das, Prasanta Kumar",
    title = "{Revealing Dark Matter's Role in Neutron Stars Anisotropy: A Bayesian Approach Using Multi-messenger Observations}",
    eprint = "2506.08376",
    archivePrefix = "arXiv",
    primaryClass = "astro-ph.HE",
    month = "6",
    year = "2025"
}

@article{Scordino:2024ehe,
    author = "Scordino, Domenico and Bombaci, Ignazio",
    title = "{Dark matter admixed neutron stars with a realistic nuclear equation of state from chiral nuclear interactions}",
    eprint = "2405.19251",
    archivePrefix = "arXiv",
    primaryClass = "astro-ph.HE",
    doi = "10.1016/j.jheap.2025.01.008",
    journal = "JHEAp",
    volume = "45",
    pages = "371--381",
    year = "2025"
}

@article{Liu:2025vwm,
    author = "Liu, Hong-Ming and Chu, Peng-Cheng and Liu, He and Li, Xiao-Hua and Li, Zeng-Hua",
    title = "{Dark matter effects on the properties of quark stars and the implications for the peculiar objects}",
    eprint = "2501.04382",
    archivePrefix = "arXiv",
    primaryClass = "nucl-th",
    month = "1",
    year = "2025"
}

@article{Vikiaris:2023vau,
    author = "Vikiaris, M. and Petousis, V. and Veselsky, M. and Moustakidis, Ch. C.",
    title = "{Supramassive dark objects with neutron star origin}",
    eprint = "2312.07412",
    archivePrefix = "arXiv",
    primaryClass = "astro-ph.HE",
    doi = "10.1103/PhysRevD.109.123006",
    journal = "Phys. Rev. D",
    volume = "109",
    number = "12",
    pages = "123006",
    year = "2024"
}

@dataset{vinciguerra_2023_8239000,
  author       = {Vinciguerra, Serena and
                  Salmi, Tuomo and
                  Watts, Anna L. and
                  Choudhury, Devarshi and
                  Riley, Thomas E. and
                  Ray, Paul S. and
                  Bogdanov, Slavko and
                  Kini, Yves and
                  Guillot, Sebastien and
                  Chakrabarty, Deepto and
                  Ho, Wynn C. G. and
                  Huppenkothen, Daniela and
                  Morsink, Sharon M. and
                  Wadiasingh, Zorawar and
                  Wolff, Micheal T.},
  title        = {An updated mass-radius analysis of the 2017-2018
                   NICER data set of PSR J0030+0451
                  },
  month        = nov,
  year         = 2023,
  publisher    = {Zenodo},
  version      = {v1.0.0},
  doi          = {10.5281/zenodo.8239000},
  url          = {https://doi.org/10.5281/zenodo.8239000},
}

@dataset{mauviard_2025_15603406,
  author       = {Mauviard, Lucien and
                  Guillot, Sebastien and
                  Salmi, Tuomo and
                  Choudhury, Devarshi and
                  Dorsman, Bas and
                  González-Caniulef, Denis and
                  Hoogkamer, Mariska and
                  Huppenkothen, Daniela and
                  Kazantsev, Christine and
                  Kini, Yves and
                  Olive, Jean-François and
                  Stammler, Pierre and
                  Watts, Anna and
                  Mendes, Melissa and
                  Rutherford, Nathan and
                  Schwenk, Achim and
                  Svensson, Isak and
                  Bogdanov, Slavko and
                  Kerr, Matthew and
                  Ray, Paul and
                  Guillemot, Lucas and
                  Cognard, Ismaël and
                  Theureau, Gilles},
  title        = {A NICER view of the 1.4 solar masses edge-on
                   pulsar PSR J0614-3329
                  },
  month        = jun,
  year         = 2025,
  publisher    = {Zenodo},
  version      = {1.0.0},
  doi          = {10.5281/zenodo.15603406},
  url          = {https://doi.org/10.5281/zenodo.15603406},
}

@article{Jangal:2025maa,
    author = "Jangal, F. Moradi and Moshfegh, H. R. and Azizi, K.",
    title = "{Impact of QCD sum rules coupling constants on neutron stars structure}",
    eprint = "2501.01234",
    archivePrefix = "arXiv",
    primaryClass = "hep-ph",
    doi = "10.1140/epjc/s10052-025-14417-1",
    journal = "Eur. Phys. J. C",
    volume = "85",
    number = "6",
    pages = "691",
    year = "2025"
}

@article{Tong:2025sui,
    author = "Tong, Hui and Elhatisari, Serdar and Mei{\ss}ner, Ulf-G.",
    title = "{Hyperneutron Stars from an Ab Initio Calculation}",
    eprint = "2502.14435",
    archivePrefix = "arXiv",
    primaryClass = "nucl-th",
    doi = "10.3847/1538-4357/adba47",
    journal = "Astrophys. J.",
    volume = "982",
    number = "2",
    pages = "164",
    year = "2025"
}

@article{Choi:2023qtk,
    author = "Choi, Soonchul and Hiyama, Emiko and Hyun, Chang Ho and Cheoun, Myung-Ki",
    title = "{$\Lambda\Lambda$ Interaction in a Nuclear Density Functional Theory and Hyperon Puzzle of the Neutron Star}",
    eprint = "2309.01348",
    archivePrefix = "arXiv",
    primaryClass = "nucl-th",
    month = "9",
    year = "2023"
}

@article{Fujimoto:2024doc,
    author = "Fujimoto, Yuki and Kojo, Toru and McLerran, Larry",
    title = "{Quarkyonic matter pieces together the hyperon puzzle}",
    eprint = "2410.22758",
    archivePrefix = "arXiv",
    primaryClass = "nucl-th",
    reportNumber = "INT-PUB-24-056, RIKEN-iTHEMS-Report-24",
    month = "10",
    year = "2024"
}

@article{Tsiopelas:2024ksy,
    author = "Tsiopelas, Stefanos and Sedrakian, Armen and Oertel, Micaela",
    title = "{Finite-temperature equations of state of compact stars with hyperons: three-dimensional tables}",
    eprint = "2406.00484",
    archivePrefix = "arXiv",
    primaryClass = "nucl-th",
    doi = "10.1140/epja/s10050-024-01351-1",
    journal = "Eur. Phys. J. A",
    volume = "60",
    number = "6",
    pages = "127",
    year = "2024"
}

@article{Sedrakian:2020kbi,
    author = "Sedrakian, Armen and Weber, Fridolin and Li, Jia Jie",
    title = "{Confronting GW190814 with hyperonization in dense matter and hypernuclear compact stars}",
    eprint = "2007.09683",
    archivePrefix = "arXiv",
    primaryClass = "astro-ph.HE",
    doi = "10.1103/PhysRevD.102.041301",
    journal = "Phys. Rev. D",
    volume = "102",
    number = "4",
    pages = "041301",
    year = "2020"
}

@article{Li:2023vso,
    author = "Li, Jia Jie and Sedrakian, Armen",
    title = "{Baryonic models of ultra-low-mass compact stars for the central compact object in HESS J1731-347}",
    eprint = "2306.14185",
    archivePrefix = "arXiv",
    primaryClass = "nucl-th",
    doi = "10.1016/j.physletb.2023.138062",
    journal = "Phys. Lett. B",
    volume = "844",
    pages = "138062",
    year = "2023"
}

@article{Tolos:2024adu,
    author = "Tolos, Laura",
    title = "{Strangeness in Astrophysics}",
    eprint = "2409.06461",
    archivePrefix = "arXiv",
    primaryClass = "nucl-th",
    doi = "10.1051/epjconf/202531601009",
    journal = "EPJ Web Conf.",
    volume = "316",
    pages = "01009",
    year = "2025"
}

@article{Maslov_2015,
   title={Solution of the hyperon puzzle within a relativistic mean-field model},
   volume={748},
   ISSN={0370-2693},
   url={http://dx.doi.org/10.1016/j.physletb.2015.07.032},
   DOI={10.1016/j.physletb.2015.07.032},
   journal={Physics Letters B},
   publisher={Elsevier BV},
   author={Maslov, K.A. and Kolomeitsev, E.E. and Voskresensky, D.N.},
   year={2015},
   month=sep, pages={369–375} }

@misc{cirelli2024darkmatter,
      title={Dark Matter}, 
      author={Marco Cirelli and Alessandro Strumia and Jure Zupan},
      year={2024},
      eprint={2406.01705},
      archivePrefix={arXiv},
      primaryClass={hep-ph},
      url={https://arxiv.org/abs/2406.01705}, 
}

@article{Yang:2024sxi,
    author = "Yang, S. -H. and Pi, C. -M.",
    title = "{Color-flavor locked strange stars admixed with mirror dark matter and the observations of compact stars}",
    eprint = "2402.14262",
    archivePrefix = "arXiv",
    primaryClass = "astro-ph.HE",
    doi = "10.1088/1475-7516/2024/09/052",
    journal = "JCAP",
    volume = "09",
    pages = "052",
    year = "2024"
}

@article{Ferreira:2022fjo,
    author = "Ferreira, Osvaldo and Fraga, Eduardo S.",
    title = "{Strange magnetars admixed with fermionic dark matter}",
    eprint = "2209.10959",
    archivePrefix = "arXiv",
    primaryClass = "hep-ph",
    doi = "10.1088/1475-7516/2023/04/012",
    journal = "JCAP",
    volume = "04",
    pages = "012",
    year = "2023"
}

@article{Arvikar:2025hej,
    author = "Arvikar, Payaswinee and Gautam, Sakshi and Venneti, Anagh and Banik, Sarmistha",
    title = "{Exploring Fermionic Dark Matter Admixed Neutron Stars in the Light of Astrophysical Observations}",
    eprint = "2506.20736",
    archivePrefix = "arXiv",
    primaryClass = "astro-ph.HE",
    month = "6",
    year = "2025"
}

@article{Mukherjee:2025omu,
    author = "Mukherjee, Samanwaya and Aswathi, P. S. and Singha, Chiranjeeb and Ganguly, Apratim",
    title = "{Bose-Einstein Condensate Dark Matter in the Core of Neutron Stars: Implications for Gravitational-wave Observations}",
    eprint = "2506.22353",
    archivePrefix = "arXiv",
    primaryClass = "gr-qc",
    month = "6",
    year = "2025"
}

@article{Pang:2022rzc,
    author = "Pang, Peter T. H. and others",
    title = "{An updated nuclear-physics and multi-messenger astrophysics framework for binary neutron star mergers}",
    eprint = "2205.08513",
    archivePrefix = "arXiv",
    primaryClass = "astro-ph.HE",
    reportNumber = "LA-UR-22-23872, LIGO-P2200150",
    doi = "10.1038/s41467-023-43932-6",
    journal = "Nature Commun.",
    volume = "14",
    number = "1",
    pages = "8352",
    year = "2023"
}

@article{Ecker:2024uqv,
    author = "Ecker, Christian and Gorda, Tyler and Kurkela, Aleksi and Rezzolla, Luciano",
    title = "{Constraining the equation of state in neutron-star cores via the long-ringdown signal}",
    eprint = "2403.03246",
    archivePrefix = "arXiv",
    primaryClass = "astro-ph.HE",
    doi = "10.1038/s41467-025-56500-x",
    journal = "Nature Commun.",
    volume = "16",
    number = "1",
    pages = "1320",
    year = "2025"
}

@article{Gholami:2024ety,
    author = {Gholami, Hosein and Rather, Ishfaq Ahmad and Hofmann, Marco and Buballa, Michael and Schaffner-Bielich, J{\"u}rgen},
    title = "{Astrophysical constraints on color-superconducting phases in compact stars within the RG-consistent NJL model}",
    eprint = "2411.04064",
    archivePrefix = "arXiv",
    primaryClass = "hep-ph",
    doi = "10.1103/PhysRevD.111.103034",
    journal = "Phys. Rev. D",
    volume = "111",
    number = "10",
    pages = "103034",
    year = "2025"
}

@article{Lindblom:2025wme,
    author = "Lindblom, Lee and Lewis, Steve M. and Weber, Fridolin",
    title = "{Representing equations of state with strong first-order phase transitions}",
    eprint = "2506.06201",
    archivePrefix = "arXiv",
    primaryClass = "nucl-th",
    doi = "10.1103/j1z7-jfc6",
    journal = "Phys. Rev. D",
    volume = "111",
    number = "12",
    pages = "123035",
    year = "2025"
}

@article{Li:2023zty,
    author = "Li, Jia Jie and Sedrakian, Armen and Alford, Mark",
    title = "{Relativistic Hybrid Stars with Sequential First-order Phase Transitions in Light of Multimessenger Constraints}",
    eprint = "2301.10940",
    archivePrefix = "arXiv",
    primaryClass = "astro-ph.HE",
    doi = "10.3847/1538-4357/acb688",
    journal = "Astrophys. J.",
    volume = "944",
    number = "2",
    pages = "206",
    year = "2023"
}

@article{Li:2024sft,
    author = "Li, Jia Jie and Sedrakian, Armen and Alford, Mark",
    title = "{Confronting new NICER mass-radius measurements with phase transition in dense matter and twin compact stars}",
    eprint = "2409.05322",
    archivePrefix = "arXiv",
    primaryClass = "astro-ph.HE",
    doi = "10.1088/1475-7516/2025/02/002",
    journal = "JCAP",
    volume = "02",
    pages = "002",
    year = "2025"
}

@article{Huang:2025vfl,
    author = "Huang, Chun and Sourav, Shashwat",
    title = "{Constraining First-order Phase Transition inside Neutron Stars with Application of Bayesian Techniques on PSR J0437{\textendash}4715 NICER Data}",
    eprint = "2502.11976",
    archivePrefix = "arXiv",
    primaryClass = "astro-ph.HE",
    doi = "10.3847/1538-4357/adbb67",
    journal = "Astrophys. J.",
    volume = "983",
    number = "1",
    pages = "17",
    year = "2025"
}

@article{Christian:2023hez,
    author = {Christian, Jan-Erik and Schaffner-Bielich, J{\"u}rgen and Rosswog, Stephan},
    title = "{Which first order phase transitions to quark matter are possible in neutron stars?}",
    eprint = "2312.10148",
    archivePrefix = "arXiv",
    primaryClass = "nucl-th",
    doi = "10.1103/PhysRevD.109.063035",
    journal = "Phys. Rev. D",
    volume = "109",
    number = "6",
    pages = "063035",
    year = "2024"
}

@article{Counsell:2025hcv,
    author = "Counsell, A. R. and Gittins, F. and Andersson, N. and Tews, I.",
    title = "{Interface modes in inspiralling neutron stars: A gravitational-wave probe of first-order phase transitions}",
    eprint = "2504.06181",
    archivePrefix = "arXiv",
    primaryClass = "gr-qc",
    reportNumber = "LA-UR-25-22199, INT-PUB-25-008",
    month = "4",
    year = "2025"
}

@article{Fujimoto:2024ymt,
    author = "Fujimoto, Yuki and Fukushima, Kenji and Hotokezaka, Kenta and Kyutoku, Koutarou",
    title = "{Signature of hadron-quark crossover in binary-neutron-star mergers}",
    eprint = "2408.10298",
    archivePrefix = "arXiv",
    primaryClass = "astro-ph.HE",
    reportNumber = "INT-PUB-24-041",
    doi = "10.1103/PhysRevD.111.063054",
    journal = "Phys. Rev. D",
    volume = "111",
    number = "6",
    pages = "063054",
    year = "2025"
}

@article{Salmi:2024aum,
    author = "Salmi, Tuomo and others",
    title = "{The Radius of the High-mass Pulsar PSR J0740+6620 with 3.6 yr of NICER Data}",
    eprint = "2406.14466",
    archivePrefix = "arXiv",
    primaryClass = "astro-ph.HE",
    doi = "10.3847/1538-4357/ad5f1f",
    journal = "Astrophys. J.",
    volume = "974",
    number = "2",
    pages = "294",
    year = "2024"
}

@article{WASA-at-COSY:2011bjg,
    author = "Adlarson, P. and others",
    collaboration = "WASA-at-COSY",
    title = "{ABC Effect in Basic Double-Pionic Fusion --- Observation of a new resonance?}",
    eprint = "1104.0123",
    archivePrefix = "arXiv",
    primaryClass = "nucl-ex",
    doi = "10.1103/PhysRevLett.106.242302",
    journal = "Phys. Rev. Lett.",
    volume = "106",
    pages = "242302",
    year = "2011"
}

@article{Vidana:2017qey,
    author = "Vida{\~n}a, I. and Bashkanov, M. and Watts, D. P. and Pastore, A.",
    title = "{The $d^*(2380)$ in neutron stars - a new degree of freedom?}",
    eprint = "1706.09701",
    archivePrefix = "arXiv",
    primaryClass = "nucl-th",
    doi = "10.1016/j.physletb.2018.03.052",
    journal = "Phys. Lett. B",
    volume = "781",
    pages = "112--116",
    year = "2018"
}

@article{Celi:2023gtj,
    author = "Celi, Marcos O. and Bashkanov, Mikhail and Mariani, Mauro and Orsaria, Milva G. and Pastore, Alessandro and Ranea-Sandoval, Ignacio F. and Weber, Fridolin",
    title = "{Destabilization of high-mass neutron stars by the emergence of d*-hexaquarks}",
    eprint = "2312.03880",
    archivePrefix = "arXiv",
    primaryClass = "nucl-th",
    doi = "10.1103/PhysRevD.109.023004",
    journal = "Phys. Rev. D",
    volume = "109",
    number = "2",
    pages = "023004",
    year = "2024"
}

@article{Celi:2025wnc,
    author = "Celi, Marcos O. and Mariani, Mauro and Kumar, Rajesh and Bashkanov, Mikhail and Orsaria, Milva G. and Pastore, Alessandro and Ranea-Sandoval, Ignacio F. and Dexheimer, Veronica",
    title = "{Exploring the role of d* hexaquarks on quark deconfinement and hybrid stars}",
    eprint = "2504.00981",
    archivePrefix = "arXiv",
    primaryClass = "nucl-th",
    doi = "10.1103/3lyv-45jp",
    journal = "Phys. Rev. D",
    volume = "112",
    number = "2",
    pages = "023027",
    year = "2025"
}

@article{Ayriyan:2021prr,
    author = "Ayriyan, A. and Blaschke, D. and Grunfeld, A. G. and Alvarez-Castillo, D. and Grigorian, H. and Abgaryan, V.",
    title = "{Bayesian analysis of multimessenger M-R data with interpolated hybrid EoS}",
    eprint = "2102.13485",
    archivePrefix = "arXiv",
    primaryClass = "astro-ph.HE",
    doi = "10.1140/epja/s10050-021-00619-0",
    journal = "Eur. Phys. J. A",
    volume = "57",
    number = "11",
    pages = "318",
    year = "2021"
}

@article{Zhang:2022pse,
    author = "Zhang, Chen and Ren, Jing",
    title = "{Hybrid stars may have an inverted structure}",
    eprint = "2211.12043",
    archivePrefix = "arXiv",
    primaryClass = "astro-ph.HE",
    doi = "10.1103/PhysRevD.108.063012",
    journal = "Phys. Rev. D",
    volume = "108",
    number = "6",
    pages = "063012",
    year = "2023"
}

@article{Zhang:2023zth,
    author = "Zhang, Chen and Luo, Yudong and Li, Hong-bo and Shao, Lijing and Xu, Renxin",
    title = "{Radial and nonradial oscillations of inverted hybrid stars}",
    eprint = "2306.08234",
    archivePrefix = "arXiv",
    primaryClass = "astro-ph.HE",
    doi = "10.1103/PhysRevD.109.063020",
    journal = "Phys. Rev. D",
    volume = "109",
    number = "6",
    pages = "063020",
    year = "2024"
}

@article{Negreiros:2024cvr,
    author = "Negreiros, Rodrigo and Zhang, Chen and Xu, Renxin",
    title = "{Rotational properties of inverted hybrid stars}",
    eprint = "2407.06410",
    archivePrefix = "arXiv",
    primaryClass = "astro-ph.HE",
    doi = "10.1103/PhysRevD.111.063026",
    journal = "Phys. Rev. D",
    volume = "111",
    number = "6",
    pages = "063026",
    year = "2025"
}

\clearpage

\end{document}